\documentclass[longauth]{aa}

\usepackage{graphicx}
\usepackage{natbib}
\usepackage{scalerel}
\usepackage{csquotes}

\usepackage[table]{xcolor}

\usepackage{txfonts}
\usepackage[pdfencoding=auto,psdextra]{hyperref}
\hypersetup{
    colorlinks=true,
    linkcolor=blue,
    filecolor=magenta,      
    urlcolor=blue,
    citecolor=blue
}
\makeatletter
\renewcommand*\aa@pageof{, page \thepage{} of \pageref*{LastPage}}
\makeatother

\usepackage[utf8]{inputenc}

\usepackage[switch, modulo]{lineno}

\usepackage{euclid}
\usepackage{orcidlink}

\begin{document}

\defcitealias{assefWISEAGNCatalog2018}{A18}
\defcitealias{Q1-SP015}{MB25}
\defcitealias{Q1-SP027}{MZ25}
\defcitealias{Q1-SP003}{RW25}
\defcitealias{Q1cite}{Q1}

%
%

\title{Euclid Quick Data Release (Q1)}
\subtitle{First \Euclid statistical study of galaxy mergers and their connection to active galactic nuclei}


\DeclareRobustCommand{\orcid}[1]{\orcidlink{#1}}

\author{
Euclid Collaboration: A.~La~Marca\orcid{0000-0002-7217-5120}\thanks{\email{a.la.marca@sron.nl}}\inst{\ref{aff1},\ref{aff2}}
\and L.~Wang\orcid{0000-0002-6736-9158}\inst{\ref{aff1},\ref{aff2}}
\and B.~Margalef-Bentabol\orcid{0000-0001-8702-7019}\inst{\ref{aff1}}
\and L.~Gabarra\orcid{0000-0002-8486-8856}\inst{\ref{aff3}}
\and Y.~Toba\orcid{0000-0002-3531-7863}\inst{\ref{aff4},\ref{aff5},\ref{aff6}}
\and M.~Mezcua\orcid{0000-0003-4440-259X}\inst{\ref{aff7},\ref{aff8}}
\and V.~Rodriguez-Gomez\orcid{0000-0002-9495-0079}\inst{\ref{aff9}}
\and F.~Ricci\orcid{0000-0001-5742-5980}\inst{\ref{aff10},\ref{aff11}}
\and S.~Fotopoulou\orcid{0000-0002-9686-254X}\inst{\ref{aff12}}
\and T.~Matamoro~Zatarain\orcid{0009-0007-2976-293X}\inst{\ref{aff12}}
\and V.~Allevato\orcid{0000-0001-7232-5152}\inst{\ref{aff13}}
\and F.~La~Franca\orcid{0000-0002-1239-2721}\inst{\ref{aff10}}
\and F.~Shankar\orcid{0000-0001-8973-5051}\inst{\ref{aff14}}
\and L.~Bisigello\orcid{0000-0003-0492-4924}\inst{\ref{aff15}}
\and G.~Stevens\orcid{0000-0002-8885-4443}\inst{\ref{aff12}}
\and M.~Siudek\orcid{0000-0002-2949-2155}\inst{\ref{aff16},\ref{aff7}}
\and W.~Roster\orcid{0000-0002-9149-6528}\inst{\ref{aff17}}
\and M.~Salvato\orcid{0000-0001-7116-9303}\inst{\ref{aff17}}
\and C.~Tortora\orcid{0000-0001-7958-6531}\inst{\ref{aff13}}
\and L.~Spinoglio\orcid{0000-0001-8840-1551}\inst{\ref{aff18}}
\and A.~W.~S.~Man\orcid{0000-0003-2475-124X}\inst{\ref{aff19}}
\and J.~H.~Knapen\orcid{0000-0003-1643-0024}\inst{\ref{aff20},\ref{aff21}}
\and M.~Baes\orcid{0000-0002-3930-2757}\inst{\ref{aff22}}
\and D.~O'Ryan\orcid{0000-0003-1217-4617}\inst{\ref{aff23}}
\and N.~Aghanim\orcid{0000-0002-6688-8992}\inst{\ref{aff24}}
\and B.~Altieri\orcid{0000-0003-3936-0284}\inst{\ref{aff23}}
\and A.~Amara\inst{\ref{aff25}}
\and S.~Andreon\orcid{0000-0002-2041-8784}\inst{\ref{aff26}}
\and N.~Auricchio\orcid{0000-0003-4444-8651}\inst{\ref{aff27}}
\and H.~Aussel\orcid{0000-0002-1371-5705}\inst{\ref{aff28}}
\and C.~Baccigalupi\orcid{0000-0002-8211-1630}\inst{\ref{aff29},\ref{aff30},\ref{aff31},\ref{aff32}}
\and M.~Baldi\orcid{0000-0003-4145-1943}\inst{\ref{aff33},\ref{aff27},\ref{aff34}}
\and S.~Bardelli\orcid{0000-0002-8900-0298}\inst{\ref{aff27}}
\and P.~Battaglia\orcid{0000-0002-7337-5909}\inst{\ref{aff27}}
\and A.~Biviano\orcid{0000-0002-0857-0732}\inst{\ref{aff30},\ref{aff29}}
\and A.~Bonchi\orcid{0000-0002-2667-5482}\inst{\ref{aff35}}
\and E.~Branchini\orcid{0000-0002-0808-6908}\inst{\ref{aff36},\ref{aff37},\ref{aff26}}
\and M.~Brescia\orcid{0000-0001-9506-5680}\inst{\ref{aff38},\ref{aff13}}
\and J.~Brinchmann\orcid{0000-0003-4359-8797}\inst{\ref{aff39},\ref{aff40}}
\and S.~Camera\orcid{0000-0003-3399-3574}\inst{\ref{aff41},\ref{aff42},\ref{aff43}}
\and G.~Ca\~nas-Herrera\orcid{0000-0003-2796-2149}\inst{\ref{aff44},\ref{aff45},\ref{aff46}}
\and V.~Capobianco\orcid{0000-0002-3309-7692}\inst{\ref{aff43}}
\and C.~Carbone\orcid{0000-0003-0125-3563}\inst{\ref{aff47}}
\and J.~Carretero\orcid{0000-0002-3130-0204}\inst{\ref{aff48},\ref{aff49}}
\and M.~Castellano\orcid{0000-0001-9875-8263}\inst{\ref{aff11}}
\and G.~Castignani\orcid{0000-0001-6831-0687}\inst{\ref{aff27}}
\and S.~Cavuoti\orcid{0000-0002-3787-4196}\inst{\ref{aff13},\ref{aff50}}
\and K.~C.~Chambers\orcid{0000-0001-6965-7789}\inst{\ref{aff51}}
\and A.~Cimatti\inst{\ref{aff52}}
\and C.~Colodro-Conde\inst{\ref{aff20}}
\and G.~Congedo\orcid{0000-0003-2508-0046}\inst{\ref{aff53}}
\and C.~J.~Conselice\orcid{0000-0003-1949-7638}\inst{\ref{aff54}}
\and L.~Conversi\orcid{0000-0002-6710-8476}\inst{\ref{aff55},\ref{aff23}}
\and Y.~Copin\orcid{0000-0002-5317-7518}\inst{\ref{aff56}}
\and A.~Costille\inst{\ref{aff57}}
\and F.~Courbin\orcid{0000-0003-0758-6510}\inst{\ref{aff58},\ref{aff59}}
\and H.~M.~Courtois\orcid{0000-0003-0509-1776}\inst{\ref{aff60}}
\and M.~Cropper\orcid{0000-0003-4571-9468}\inst{\ref{aff61}}
\and A.~Da~Silva\orcid{0000-0002-6385-1609}\inst{\ref{aff62},\ref{aff63}}
\and H.~Degaudenzi\orcid{0000-0002-5887-6799}\inst{\ref{aff64}}
\and G.~De~Lucia\orcid{0000-0002-6220-9104}\inst{\ref{aff30}}
\and A.~M.~Di~Giorgio\orcid{0000-0002-4767-2360}\inst{\ref{aff18}}
\and C.~Dolding\orcid{0009-0003-7199-6108}\inst{\ref{aff61}}
\and H.~Dole\orcid{0000-0002-9767-3839}\inst{\ref{aff24}}
\and F.~Dubath\orcid{0000-0002-6533-2810}\inst{\ref{aff64}}
\and C.~A.~J.~Duncan\orcid{0009-0003-3573-0791}\inst{\ref{aff54}}
\and X.~Dupac\inst{\ref{aff23}}
\and A.~Ealet\orcid{0000-0003-3070-014X}\inst{\ref{aff56}}
\and S.~Escoffier\orcid{0000-0002-2847-7498}\inst{\ref{aff65}}
\and M.~Fabricius\orcid{0000-0002-7025-6058}\inst{\ref{aff17},\ref{aff66}}
\and M.~Farina\orcid{0000-0002-3089-7846}\inst{\ref{aff18}}
\and R.~Farinelli\inst{\ref{aff27}}
\and F.~Faustini\orcid{0000-0001-6274-5145}\inst{\ref{aff35},\ref{aff11}}
\and S.~Ferriol\inst{\ref{aff56}}
\and F.~Finelli\orcid{0000-0002-6694-3269}\inst{\ref{aff27},\ref{aff67}}
\and M.~Frailis\orcid{0000-0002-7400-2135}\inst{\ref{aff30}}
\and E.~Franceschi\orcid{0000-0002-0585-6591}\inst{\ref{aff27}}
\and S.~Galeotta\orcid{0000-0002-3748-5115}\inst{\ref{aff30}}
\and K.~George\orcid{0000-0002-1734-8455}\inst{\ref{aff66}}
\and B.~Gillis\orcid{0000-0002-4478-1270}\inst{\ref{aff53}}
\and C.~Giocoli\orcid{0000-0002-9590-7961}\inst{\ref{aff27},\ref{aff34}}
\and P.~G\'omez-Alvarez\orcid{0000-0002-8594-5358}\inst{\ref{aff68},\ref{aff23}}
\and J.~Gracia-Carpio\inst{\ref{aff17}}
\and B.~R.~Granett\orcid{0000-0003-2694-9284}\inst{\ref{aff26}}
\and A.~Grazian\orcid{0000-0002-5688-0663}\inst{\ref{aff15}}
\and F.~Grupp\inst{\ref{aff17},\ref{aff66}}
\and L.~Guzzo\orcid{0000-0001-8264-5192}\inst{\ref{aff69},\ref{aff26},\ref{aff70}}
\and S.~Gwyn\orcid{0000-0001-8221-8406}\inst{\ref{aff71}}
\and S.~V.~H.~Haugan\orcid{0000-0001-9648-7260}\inst{\ref{aff72}}
\and W.~Holmes\inst{\ref{aff73}}
\and I.~M.~Hook\orcid{0000-0002-2960-978X}\inst{\ref{aff74}}
\and F.~Hormuth\inst{\ref{aff75}}
\and A.~Hornstrup\orcid{0000-0002-3363-0936}\inst{\ref{aff76},\ref{aff77}}
\and P.~Hudelot\inst{\ref{aff78}}
\and K.~Jahnke\orcid{0000-0003-3804-2137}\inst{\ref{aff79}}
\and M.~Jhabvala\inst{\ref{aff80}}
\and B.~Joachimi\orcid{0000-0001-7494-1303}\inst{\ref{aff81}}
\and E.~Keih\"anen\orcid{0000-0003-1804-7715}\inst{\ref{aff82}}
\and S.~Kermiche\orcid{0000-0002-0302-5735}\inst{\ref{aff65}}
\and A.~Kiessling\orcid{0000-0002-2590-1273}\inst{\ref{aff73}}
\and B.~Kubik\orcid{0009-0006-5823-4880}\inst{\ref{aff56}}
\and M.~K\"ummel\orcid{0000-0003-2791-2117}\inst{\ref{aff66}}
\and M.~Kunz\orcid{0000-0002-3052-7394}\inst{\ref{aff83}}
\and H.~Kurki-Suonio\orcid{0000-0002-4618-3063}\inst{\ref{aff84},\ref{aff85}}
\and Q.~Le~Boulc'h\inst{\ref{aff86}}
\and A.~M.~C.~Le~Brun\orcid{0000-0002-0936-4594}\inst{\ref{aff87}}
\and D.~Le~Mignant\orcid{0000-0002-5339-5515}\inst{\ref{aff57}}
\and S.~Ligori\orcid{0000-0003-4172-4606}\inst{\ref{aff43}}
\and P.~B.~Lilje\orcid{0000-0003-4324-7794}\inst{\ref{aff72}}
\and V.~Lindholm\orcid{0000-0003-2317-5471}\inst{\ref{aff84},\ref{aff85}}
\and I.~Lloro\orcid{0000-0001-5966-1434}\inst{\ref{aff88}}
\and G.~Mainetti\orcid{0000-0003-2384-2377}\inst{\ref{aff86}}
\and D.~Maino\inst{\ref{aff69},\ref{aff47},\ref{aff70}}
\and E.~Maiorano\orcid{0000-0003-2593-4355}\inst{\ref{aff27}}
\and O.~Mansutti\orcid{0000-0001-5758-4658}\inst{\ref{aff30}}
\and S.~Marcin\inst{\ref{aff89}}
\and O.~Marggraf\orcid{0000-0001-7242-3852}\inst{\ref{aff90}}
\and M.~Martinelli\orcid{0000-0002-6943-7732}\inst{\ref{aff11},\ref{aff91}}
\and N.~Martinet\orcid{0000-0003-2786-7790}\inst{\ref{aff57}}
\and F.~Marulli\orcid{0000-0002-8850-0303}\inst{\ref{aff92},\ref{aff27},\ref{aff34}}
\and R.~Massey\orcid{0000-0002-6085-3780}\inst{\ref{aff93}}
\and S.~Maurogordato\inst{\ref{aff94}}
\and E.~Medinaceli\orcid{0000-0002-4040-7783}\inst{\ref{aff27}}
\and S.~Mei\orcid{0000-0002-2849-559X}\inst{\ref{aff95},\ref{aff96}}
\and M.~Melchior\inst{\ref{aff97}}
\and Y.~Mellier\inst{\ref{aff98},\ref{aff78}}
\and M.~Meneghetti\orcid{0000-0003-1225-7084}\inst{\ref{aff27},\ref{aff34}}
\and E.~Merlin\orcid{0000-0001-6870-8900}\inst{\ref{aff11}}
\and G.~Meylan\inst{\ref{aff99}}
\and A.~Mora\orcid{0000-0002-1922-8529}\inst{\ref{aff100}}
\and M.~Moresco\orcid{0000-0002-7616-7136}\inst{\ref{aff92},\ref{aff27}}
\and L.~Moscardini\orcid{0000-0002-3473-6716}\inst{\ref{aff92},\ref{aff27},\ref{aff34}}
\and R.~Nakajima\orcid{0009-0009-1213-7040}\inst{\ref{aff90}}
\and C.~Neissner\orcid{0000-0001-8524-4968}\inst{\ref{aff101},\ref{aff49}}
\and S.-M.~Niemi\inst{\ref{aff44}}
\and J.~W.~Nightingale\orcid{0000-0002-8987-7401}\inst{\ref{aff102}}
\and C.~Padilla\orcid{0000-0001-7951-0166}\inst{\ref{aff101}}
\and S.~Paltani\orcid{0000-0002-8108-9179}\inst{\ref{aff64}}
\and F.~Pasian\orcid{0000-0002-4869-3227}\inst{\ref{aff30}}
\and K.~Pedersen\inst{\ref{aff103}}
\and W.~J.~Percival\orcid{0000-0002-0644-5727}\inst{\ref{aff104},\ref{aff105},\ref{aff106}}
\and V.~Pettorino\inst{\ref{aff44}}
\and S.~Pires\orcid{0000-0002-0249-2104}\inst{\ref{aff28}}
\and G.~Polenta\orcid{0000-0003-4067-9196}\inst{\ref{aff35}}
\and M.~Poncet\inst{\ref{aff107}}
\and L.~A.~Popa\inst{\ref{aff108}}
\and L.~Pozzetti\orcid{0000-0001-7085-0412}\inst{\ref{aff27}}
\and F.~Raison\orcid{0000-0002-7819-6918}\inst{\ref{aff17}}
\and R.~Rebolo\orcid{0000-0003-3767-7085}\inst{\ref{aff20},\ref{aff109},\ref{aff21}}
\and A.~Renzi\orcid{0000-0001-9856-1970}\inst{\ref{aff110},\ref{aff111}}
\and J.~Rhodes\orcid{0000-0002-4485-8549}\inst{\ref{aff73}}
\and G.~Riccio\inst{\ref{aff13}}
\and E.~Romelli\orcid{0000-0003-3069-9222}\inst{\ref{aff30}}
\and M.~Roncarelli\orcid{0000-0001-9587-7822}\inst{\ref{aff27}}
\and B.~Rusholme\orcid{0000-0001-7648-4142}\inst{\ref{aff112}}
\and R.~Saglia\orcid{0000-0003-0378-7032}\inst{\ref{aff66},\ref{aff17}}
\and Z.~Sakr\orcid{0000-0002-4823-3757}\inst{\ref{aff113},\ref{aff114},\ref{aff115}}
\and D.~Sapone\orcid{0000-0001-7089-4503}\inst{\ref{aff116}}
\and B.~Sartoris\orcid{0000-0003-1337-5269}\inst{\ref{aff66},\ref{aff30}}
\and J.~A.~Schewtschenko\orcid{0000-0002-4913-6393}\inst{\ref{aff53}}
\and P.~Schneider\orcid{0000-0001-8561-2679}\inst{\ref{aff90}}
\and T.~Schrabback\orcid{0000-0002-6987-7834}\inst{\ref{aff117}}
\and M.~Scodeggio\inst{\ref{aff47}}
\and A.~Secroun\orcid{0000-0003-0505-3710}\inst{\ref{aff65}}
\and G.~Seidel\orcid{0000-0003-2907-353X}\inst{\ref{aff79}}
\and M.~Seiffert\orcid{0000-0002-7536-9393}\inst{\ref{aff73}}
\and S.~Serrano\orcid{0000-0002-0211-2861}\inst{\ref{aff8},\ref{aff118},\ref{aff7}}
\and P.~Simon\inst{\ref{aff90}}
\and C.~Sirignano\orcid{0000-0002-0995-7146}\inst{\ref{aff110},\ref{aff111}}
\and G.~Sirri\orcid{0000-0003-2626-2853}\inst{\ref{aff34}}
\and L.~Stanco\orcid{0000-0002-9706-5104}\inst{\ref{aff111}}
\and J.~Steinwagner\orcid{0000-0001-7443-1047}\inst{\ref{aff17}}
\and P.~Tallada-Cresp\'{i}\orcid{0000-0002-1336-8328}\inst{\ref{aff48},\ref{aff49}}
\and A.~N.~Taylor\inst{\ref{aff53}}
\and H.~I.~Teplitz\orcid{0000-0002-7064-5424}\inst{\ref{aff119}}
\and I.~Tereno\inst{\ref{aff62},\ref{aff120}}
\and N.~Tessore\orcid{0000-0002-9696-7931}\inst{\ref{aff81}}
\and S.~Toft\orcid{0000-0003-3631-7176}\inst{\ref{aff121},\ref{aff122}}
\and R.~Toledo-Moreo\orcid{0000-0002-2997-4859}\inst{\ref{aff123}}
\and F.~Torradeflot\orcid{0000-0003-1160-1517}\inst{\ref{aff49},\ref{aff48}}
\and I.~Tutusaus\orcid{0000-0002-3199-0399}\inst{\ref{aff114}}
\and L.~Valenziano\orcid{0000-0002-1170-0104}\inst{\ref{aff27},\ref{aff67}}
\and J.~Valiviita\orcid{0000-0001-6225-3693}\inst{\ref{aff84},\ref{aff85}}
\and T.~Vassallo\orcid{0000-0001-6512-6358}\inst{\ref{aff66},\ref{aff30}}
\and G.~Verdoes~Kleijn\orcid{0000-0001-5803-2580}\inst{\ref{aff2}}
\and A.~Veropalumbo\orcid{0000-0003-2387-1194}\inst{\ref{aff26},\ref{aff37},\ref{aff36}}
\and Y.~Wang\orcid{0000-0002-4749-2984}\inst{\ref{aff119}}
\and J.~Weller\orcid{0000-0002-8282-2010}\inst{\ref{aff66},\ref{aff17}}
\and A.~Zacchei\orcid{0000-0003-0396-1192}\inst{\ref{aff30},\ref{aff29}}
\and G.~Zamorani\orcid{0000-0002-2318-301X}\inst{\ref{aff27}}
\and F.~M.~Zerbi\inst{\ref{aff26}}
\and I.~A.~Zinchenko\orcid{0000-0002-2944-2449}\inst{\ref{aff66}}
\and E.~Zucca\orcid{0000-0002-5845-8132}\inst{\ref{aff27}}
\and M.~Ballardini\orcid{0000-0003-4481-3559}\inst{\ref{aff124},\ref{aff125},\ref{aff27}}
\and M.~Bolzonella\orcid{0000-0003-3278-4607}\inst{\ref{aff27}}
\and E.~Bozzo\orcid{0000-0002-8201-1525}\inst{\ref{aff64}}
\and C.~Burigana\orcid{0000-0002-3005-5796}\inst{\ref{aff126},\ref{aff67}}
\and R.~Cabanac\orcid{0000-0001-6679-2600}\inst{\ref{aff114}}
\and A.~Cappi\inst{\ref{aff27},\ref{aff94}}
\and D.~Di~Ferdinando\inst{\ref{aff34}}
\and J.~A.~Escartin~Vigo\inst{\ref{aff17}}
\and M.~Huertas-Company\orcid{0000-0002-1416-8483}\inst{\ref{aff20},\ref{aff16},\ref{aff127},\ref{aff128}}
\and J.~Mart\'{i}n-Fleitas\orcid{0000-0002-8594-569X}\inst{\ref{aff100}}
\and S.~Matthew\orcid{0000-0001-8448-1697}\inst{\ref{aff53}}
\and N.~Mauri\orcid{0000-0001-8196-1548}\inst{\ref{aff52},\ref{aff34}}
\and R.~B.~Metcalf\orcid{0000-0003-3167-2574}\inst{\ref{aff92},\ref{aff27}}
\and A.~Pezzotta\orcid{0000-0003-0726-2268}\inst{\ref{aff129},\ref{aff17}}
\and M.~P\"ontinen\orcid{0000-0001-5442-2530}\inst{\ref{aff84}}
\and C.~Porciani\orcid{0000-0002-7797-2508}\inst{\ref{aff90}}
\and I.~Risso\orcid{0000-0003-2525-7761}\inst{\ref{aff130}}
\and V.~Scottez\inst{\ref{aff98},\ref{aff131}}
\and M.~Sereno\orcid{0000-0003-0302-0325}\inst{\ref{aff27},\ref{aff34}}
\and M.~Tenti\orcid{0000-0002-4254-5901}\inst{\ref{aff34}}
\and M.~Viel\orcid{0000-0002-2642-5707}\inst{\ref{aff29},\ref{aff30},\ref{aff32},\ref{aff31},\ref{aff132}}
\and M.~Wiesmann\orcid{0009-0000-8199-5860}\inst{\ref{aff72}}
\and Y.~Akrami\orcid{0000-0002-2407-7956}\inst{\ref{aff133},\ref{aff134}}
\and S.~Alvi\orcid{0000-0001-5779-8568}\inst{\ref{aff124}}
\and I.~T.~Andika\orcid{0000-0001-6102-9526}\inst{\ref{aff135},\ref{aff136}}
\and S.~Anselmi\orcid{0000-0002-3579-9583}\inst{\ref{aff111},\ref{aff110},\ref{aff137}}
\and M.~Archidiacono\orcid{0000-0003-4952-9012}\inst{\ref{aff69},\ref{aff70}}
\and F.~Atrio-Barandela\orcid{0000-0002-2130-2513}\inst{\ref{aff138}}
\and C.~Benoist\inst{\ref{aff94}}
\and K.~Benson\inst{\ref{aff61}}
\and D.~Bertacca\orcid{0000-0002-2490-7139}\inst{\ref{aff110},\ref{aff15},\ref{aff111}}
\and M.~Bethermin\orcid{0000-0002-3915-2015}\inst{\ref{aff139}}
\and A.~Blanchard\orcid{0000-0001-8555-9003}\inst{\ref{aff114}}
\and L.~Blot\orcid{0000-0002-9622-7167}\inst{\ref{aff140},\ref{aff137}}
\and H.~B\"ohringer\orcid{0000-0001-8241-4204}\inst{\ref{aff17},\ref{aff141},\ref{aff142}}
\and S.~Borgani\orcid{0000-0001-6151-6439}\inst{\ref{aff143},\ref{aff29},\ref{aff30},\ref{aff31},\ref{aff132}}
\and M.~L.~Brown\orcid{0000-0002-0370-8077}\inst{\ref{aff54}}
\and S.~Bruton\orcid{0000-0002-6503-5218}\inst{\ref{aff144}}
\and A.~Calabro\orcid{0000-0003-2536-1614}\inst{\ref{aff11}}
\and B.~Camacho~Quevedo\orcid{0000-0002-8789-4232}\inst{\ref{aff8},\ref{aff7}}
\and F.~Caro\inst{\ref{aff11}}
\and C.~S.~Carvalho\inst{\ref{aff120}}
\and T.~Castro\orcid{0000-0002-6292-3228}\inst{\ref{aff30},\ref{aff31},\ref{aff29},\ref{aff132}}
\and F.~Cogato\orcid{0000-0003-4632-6113}\inst{\ref{aff92},\ref{aff27}}
\and S.~Conseil\orcid{0000-0002-3657-4191}\inst{\ref{aff56}}
\and T.~Contini\orcid{0000-0003-0275-938X}\inst{\ref{aff114}}
\and A.~R.~Cooray\orcid{0000-0002-3892-0190}\inst{\ref{aff145}}
\and O.~Cucciati\orcid{0000-0002-9336-7551}\inst{\ref{aff27}}
\and S.~Davini\orcid{0000-0003-3269-1718}\inst{\ref{aff37}}
\and F.~De~Paolis\orcid{0000-0001-6460-7563}\inst{\ref{aff146},\ref{aff147},\ref{aff148}}
\and G.~Desprez\orcid{0000-0001-8325-1742}\inst{\ref{aff2}}
\and A.~D\'iaz-S\'anchez\orcid{0000-0003-0748-4768}\inst{\ref{aff149}}
\and J.~J.~Diaz\inst{\ref{aff20}}
\and S.~Di~Domizio\orcid{0000-0003-2863-5895}\inst{\ref{aff36},\ref{aff37}}
\and J.~M.~Diego\orcid{0000-0001-9065-3926}\inst{\ref{aff150}}
\and P.-A.~Duc\orcid{0000-0003-3343-6284}\inst{\ref{aff139}}
\and A.~Enia\orcid{0000-0002-0200-2857}\inst{\ref{aff33},\ref{aff27}}
\and Y.~Fang\inst{\ref{aff66}}
\and A.~G.~Ferrari\orcid{0009-0005-5266-4110}\inst{\ref{aff34}}
\and A.~Finoguenov\orcid{0000-0002-4606-5403}\inst{\ref{aff84}}
\and A.~Fontana\orcid{0000-0003-3820-2823}\inst{\ref{aff11}}
\and F.~Fontanot\orcid{0000-0003-4744-0188}\inst{\ref{aff30},\ref{aff29}}
\and A.~Franco\orcid{0000-0002-4761-366X}\inst{\ref{aff147},\ref{aff146},\ref{aff148}}
\and K.~Ganga\orcid{0000-0001-8159-8208}\inst{\ref{aff95}}
\and J.~Garc\'ia-Bellido\orcid{0000-0002-9370-8360}\inst{\ref{aff133}}
\and T.~Gasparetto\orcid{0000-0002-7913-4866}\inst{\ref{aff30}}
\and V.~Gautard\inst{\ref{aff151}}
\and E.~Gaztanaga\orcid{0000-0001-9632-0815}\inst{\ref{aff7},\ref{aff8},\ref{aff152}}
\and F.~Giacomini\orcid{0000-0002-3129-2814}\inst{\ref{aff34}}
\and F.~Gianotti\orcid{0000-0003-4666-119X}\inst{\ref{aff27}}
\and G.~Gozaliasl\orcid{0000-0002-0236-919X}\inst{\ref{aff153},\ref{aff84}}
\and M.~Guidi\orcid{0000-0001-9408-1101}\inst{\ref{aff33},\ref{aff27}}
\and C.~M.~Gutierrez\orcid{0000-0001-7854-783X}\inst{\ref{aff154}}
\and A.~Hall\orcid{0000-0002-3139-8651}\inst{\ref{aff53}}
\and W.~G.~Hartley\inst{\ref{aff64}}
\and C.~Hern\'andez-Monteagudo\orcid{0000-0001-5471-9166}\inst{\ref{aff21},\ref{aff20}}
\and H.~Hildebrandt\orcid{0000-0002-9814-3338}\inst{\ref{aff155}}
\and J.~Hjorth\orcid{0000-0002-4571-2306}\inst{\ref{aff103}}
\and J.~J.~E.~Kajava\orcid{0000-0002-3010-8333}\inst{\ref{aff156},\ref{aff157}}
\and Y.~Kang\orcid{0009-0000-8588-7250}\inst{\ref{aff64}}
\and V.~Kansal\orcid{0000-0002-4008-6078}\inst{\ref{aff158},\ref{aff159}}
\and D.~Karagiannis\orcid{0000-0002-4927-0816}\inst{\ref{aff124},\ref{aff160}}
\and K.~Kiiveri\inst{\ref{aff82}}
\and C.~C.~Kirkpatrick\inst{\ref{aff82}}
\and S.~Kruk\orcid{0000-0001-8010-8879}\inst{\ref{aff23}}
\and J.~Le~Graet\orcid{0000-0001-6523-7971}\inst{\ref{aff65}}
\and L.~Legrand\orcid{0000-0003-0610-5252}\inst{\ref{aff161},\ref{aff162}}
\and M.~Lembo\orcid{0000-0002-5271-5070}\inst{\ref{aff124},\ref{aff125}}
\and F.~Lepori\orcid{0009-0000-5061-7138}\inst{\ref{aff163}}
\and G.~Leroy\orcid{0009-0004-2523-4425}\inst{\ref{aff164},\ref{aff93}}
\and G.~F.~Lesci\orcid{0000-0002-4607-2830}\inst{\ref{aff92},\ref{aff27}}
\and J.~Lesgourgues\orcid{0000-0001-7627-353X}\inst{\ref{aff165}}
\and L.~Leuzzi\orcid{0009-0006-4479-7017}\inst{\ref{aff92},\ref{aff27}}
\and T.~I.~Liaudat\orcid{0000-0002-9104-314X}\inst{\ref{aff166}}
\and A.~Loureiro\orcid{0000-0002-4371-0876}\inst{\ref{aff167},\ref{aff168}}
\and J.~Macias-Perez\orcid{0000-0002-5385-2763}\inst{\ref{aff169}}
\and G.~Maggio\orcid{0000-0003-4020-4836}\inst{\ref{aff30}}
\and M.~Magliocchetti\orcid{0000-0001-9158-4838}\inst{\ref{aff18}}
\and E.~A.~Magnier\orcid{0000-0002-7965-2815}\inst{\ref{aff51}}
\and F.~Mannucci\orcid{0000-0002-4803-2381}\inst{\ref{aff170}}
\and R.~Maoli\orcid{0000-0002-6065-3025}\inst{\ref{aff171},\ref{aff11}}
\and C.~J.~A.~P.~Martins\orcid{0000-0002-4886-9261}\inst{\ref{aff172},\ref{aff39}}
\and L.~Maurin\orcid{0000-0002-8406-0857}\inst{\ref{aff24}}
\and M.~Miluzio\inst{\ref{aff23},\ref{aff173}}
\and P.~Monaco\orcid{0000-0003-2083-7564}\inst{\ref{aff143},\ref{aff30},\ref{aff31},\ref{aff29}}
\and C.~Moretti\orcid{0000-0003-3314-8936}\inst{\ref{aff32},\ref{aff132},\ref{aff30},\ref{aff29},\ref{aff31}}
\and G.~Morgante\inst{\ref{aff27}}
\and K.~Naidoo\orcid{0000-0002-9182-1802}\inst{\ref{aff152}}
\and A.~Navarro-Alsina\orcid{0000-0002-3173-2592}\inst{\ref{aff90}}
\and S.~Nesseris\orcid{0000-0002-0567-0324}\inst{\ref{aff133}}
\and F.~Passalacqua\orcid{0000-0002-8606-4093}\inst{\ref{aff110},\ref{aff111}}
\and K.~Paterson\orcid{0000-0001-8340-3486}\inst{\ref{aff79}}
\and L.~Patrizii\inst{\ref{aff34}}
\and A.~Pisani\orcid{0000-0002-6146-4437}\inst{\ref{aff65},\ref{aff174}}
\and D.~Potter\orcid{0000-0002-0757-5195}\inst{\ref{aff163}}
\and S.~Quai\orcid{0000-0002-0449-8163}\inst{\ref{aff92},\ref{aff27}}
\and M.~Radovich\orcid{0000-0002-3585-866X}\inst{\ref{aff15}}
\and P.-F.~Rocci\inst{\ref{aff24}}
\and S.~Sacquegna\orcid{0000-0002-8433-6630}\inst{\ref{aff146},\ref{aff147},\ref{aff148}}
\and M.~Sahl\'en\orcid{0000-0003-0973-4804}\inst{\ref{aff175}}
\and D.~B.~Sanders\orcid{0000-0002-1233-9998}\inst{\ref{aff51}}
\and E.~Sarpa\orcid{0000-0002-1256-655X}\inst{\ref{aff32},\ref{aff132},\ref{aff31}}
\and C.~Scarlata\orcid{0000-0002-9136-8876}\inst{\ref{aff176}}
\and J.~Schaye\orcid{0000-0002-0668-5560}\inst{\ref{aff46}}
\and A.~Schneider\orcid{0000-0001-7055-8104}\inst{\ref{aff163}}
\and D.~Sciotti\orcid{0009-0008-4519-2620}\inst{\ref{aff11},\ref{aff91}}
\and E.~Sellentin\inst{\ref{aff177},\ref{aff46}}
\and L.~C.~Smith\orcid{0000-0002-3259-2771}\inst{\ref{aff178}}
\and S.~A.~Stanford\orcid{0000-0003-0122-0841}\inst{\ref{aff179}}
\and K.~Tanidis\orcid{0000-0001-9843-5130}\inst{\ref{aff3}}
\and G.~Testera\inst{\ref{aff37}}
\and R.~Teyssier\orcid{0000-0001-7689-0933}\inst{\ref{aff174}}
\and S.~Tosi\orcid{0000-0002-7275-9193}\inst{\ref{aff36},\ref{aff37},\ref{aff26}}
\and A.~Troja\orcid{0000-0003-0239-4595}\inst{\ref{aff110},\ref{aff111}}
\and M.~Tucci\inst{\ref{aff64}}
\and C.~Valieri\inst{\ref{aff34}}
\and A.~Venhola\orcid{0000-0001-6071-4564}\inst{\ref{aff180}}
\and D.~Vergani\orcid{0000-0003-0898-2216}\inst{\ref{aff27}}
\and G.~Verza\orcid{0000-0002-1886-8348}\inst{\ref{aff181}}
\and P.~Vielzeuf\orcid{0000-0003-2035-9339}\inst{\ref{aff65}}
\and N.~A.~Walton\orcid{0000-0003-3983-8778}\inst{\ref{aff178}}
\and E.~Soubrie\orcid{0000-0001-9295-1863}\inst{\ref{aff24}}
\and D.~Scott\orcid{0000-0002-6878-9840}\inst{\ref{aff19}}
}
										   
\institute{SRON Netherlands Institute for Space Research, Landleven 12, 9747 AD, Groningen, The Netherlands\label{aff1}
\and
Kapteyn Astronomical Institute, University of Groningen, PO Box 800, 9700 AV Groningen, The Netherlands\label{aff2}
\and
Department of Physics, Oxford University, Keble Road, Oxford OX1 3RH, UK\label{aff3}
\and
Department of Physical Sciences, Ritsumeikan University, Kusatsu, Shiga 525-8577, Japan\label{aff4}
\and
National Astronomical Observatory of Japan, 2-21-1 Osawa, Mitaka, Tokyo 181-8588, Japan\label{aff5}
\and
Academia Sinica Institute of Astronomy and Astrophysics (ASIAA), 11F of ASMAB, No.~1, Section 4, Roosevelt Road, Taipei 10617, Taiwan\label{aff6}
\and
Institute of Space Sciences (ICE, CSIC), Campus UAB, Carrer de Can Magrans, s/n, 08193 Barcelona, Spain\label{aff7}
\and
Institut d'Estudis Espacials de Catalunya (IEEC),  Edifici RDIT, Campus UPC, 08860 Castelldefels, Barcelona, Spain\label{aff8}
\and
Instituto de Radioastronom\'ia y Astrof\'isica, Universidad Nacional Aut\'onoma de M\'exico, A.P. 72-3, 58089 Morelia, Mexico\label{aff9}
\and
Department of Mathematics and Physics, Roma Tre University, Via della Vasca Navale 84, 00146 Rome, Italy\label{aff10}
\and
INAF-Osservatorio Astronomico di Roma, Via Frascati 33, 00078 Monteporzio Catone, Italy\label{aff11}
\and
School of Physics, HH Wills Physics Laboratory, University of Bristol, Tyndall Avenue, Bristol, BS8 1TL, UK\label{aff12}
\and
INAF-Osservatorio Astronomico di Capodimonte, Via Moiariello 16, 80131 Napoli, Italy\label{aff13}
\and
School of Physics \& Astronomy, University of Southampton, Highfield Campus, Southampton SO17 1BJ, UK\label{aff14}
\and
INAF-Osservatorio Astronomico di Padova, Via dell'Osservatorio 5, 35122 Padova, Italy\label{aff15}
\and
Instituto de Astrof\'isica de Canarias (IAC); Departamento de Astrof\'isica, Universidad de La Laguna (ULL), 38200, La Laguna, Tenerife, Spain\label{aff16}
\and
Max Planck Institute for Extraterrestrial Physics, Giessenbachstr. 1, 85748 Garching, Germany\label{aff17}
\and
INAF-Istituto di Astrofisica e Planetologia Spaziali, via del Fosso del Cavaliere, 100, 00100 Roma, Italy\label{aff18}
\and
Department of Physics and Astronomy, University of British Columbia, Vancouver, BC V6T 1Z1, Canada\label{aff19}
\and
Instituto de Astrof\'{\i}sica de Canarias, V\'{\i}a L\'actea, 38205 La Laguna, Tenerife, Spain\label{aff20}
\and
Universidad de La Laguna, Departamento de Astrof\'{\i}sica, 38206 La Laguna, Tenerife, Spain\label{aff21}
\and
Sterrenkundig Observatorium, Universiteit Gent, Krijgslaan 281 S9, 9000 Gent, Belgium\label{aff22}
\and
ESAC/ESA, Camino Bajo del Castillo, s/n., Urb. Villafranca del Castillo, 28692 Villanueva de la Ca\~nada, Madrid, Spain\label{aff23}
\and
Universit\'e Paris-Saclay, CNRS, Institut d'astrophysique spatiale, 91405, Orsay, France\label{aff24}
\and
School of Mathematics and Physics, University of Surrey, Guildford, Surrey, GU2 7XH, UK\label{aff25}
\and
INAF-Osservatorio Astronomico di Brera, Via Brera 28, 20122 Milano, Italy\label{aff26}
\and
INAF-Osservatorio di Astrofisica e Scienza dello Spazio di Bologna, Via Piero Gobetti 93/3, 40129 Bologna, Italy\label{aff27}
\and
Universit\'e Paris-Saclay, Universit\'e Paris Cit\'e, CEA, CNRS, AIM, 91191, Gif-sur-Yvette, France\label{aff28}
\and
IFPU, Institute for Fundamental Physics of the Universe, via Beirut 2, 34151 Trieste, Italy\label{aff29}
\and
INAF-Osservatorio Astronomico di Trieste, Via G. B. Tiepolo 11, 34143 Trieste, Italy\label{aff30}
\and
INFN, Sezione di Trieste, Via Valerio 2, 34127 Trieste TS, Italy\label{aff31}
\and
SISSA, International School for Advanced Studies, Via Bonomea 265, 34136 Trieste TS, Italy\label{aff32}
\and
Dipartimento di Fisica e Astronomia, Universit\`a di Bologna, Via Gobetti 93/2, 40129 Bologna, Italy\label{aff33}
\and
INFN-Sezione di Bologna, Viale Berti Pichat 6/2, 40127 Bologna, Italy\label{aff34}
\and
Space Science Data Center, Italian Space Agency, via del Politecnico snc, 00133 Roma, Italy\label{aff35}
\and
Dipartimento di Fisica, Universit\`a di Genova, Via Dodecaneso 33, 16146, Genova, Italy\label{aff36}
\and
INFN-Sezione di Genova, Via Dodecaneso 33, 16146, Genova, Italy\label{aff37}
\and
Department of Physics "E. Pancini", University Federico II, Via Cinthia 6, 80126, Napoli, Italy\label{aff38}
\and
Instituto de Astrof\'isica e Ci\^encias do Espa\c{c}o, Universidade do Porto, CAUP, Rua das Estrelas, PT4150-762 Porto, Portugal\label{aff39}
\and
Faculdade de Ci\^encias da Universidade do Porto, Rua do Campo de Alegre, 4150-007 Porto, Portugal\label{aff40}
\and
Dipartimento di Fisica, Universit\`a degli Studi di Torino, Via P. Giuria 1, 10125 Torino, Italy\label{aff41}
\and
INFN-Sezione di Torino, Via P. Giuria 1, 10125 Torino, Italy\label{aff42}
\and
INAF-Osservatorio Astrofisico di Torino, Via Osservatorio 20, 10025 Pino Torinese (TO), Italy\label{aff43}
\and
European Space Agency/ESTEC, Keplerlaan 1, 2201 AZ Noordwijk, The Netherlands\label{aff44}
\and
Institute Lorentz, Leiden University, Niels Bohrweg 2, 2333 CA Leiden, The Netherlands\label{aff45}
\and
Leiden Observatory, Leiden University, Einsteinweg 55, 2333 CC Leiden, The Netherlands\label{aff46}
\and
INAF-IASF Milano, Via Alfonso Corti 12, 20133 Milano, Italy\label{aff47}
\and
Centro de Investigaciones Energ\'eticas, Medioambientales y Tecnol\'ogicas (CIEMAT), Avenida Complutense 40, 28040 Madrid, Spain\label{aff48}
\and
Port d'Informaci\'{o} Cient\'{i}fica, Campus UAB, C. Albareda s/n, 08193 Bellaterra (Barcelona), Spain\label{aff49}
\and
INFN section of Naples, Via Cinthia 6, 80126, Napoli, Italy\label{aff50}
\and
Institute for Astronomy, University of Hawaii, 2680 Woodlawn Drive, Honolulu, HI 96822, USA\label{aff51}
\and
Dipartimento di Fisica e Astronomia "Augusto Righi" - Alma Mater Studiorum Universit\`a di Bologna, Viale Berti Pichat 6/2, 40127 Bologna, Italy\label{aff52}
\and
Institute for Astronomy, University of Edinburgh, Royal Observatory, Blackford Hill, Edinburgh EH9 3HJ, UK\label{aff53}
\and
Jodrell Bank Centre for Astrophysics, Department of Physics and Astronomy, University of Manchester, Oxford Road, Manchester M13 9PL, UK\label{aff54}
\and
European Space Agency/ESRIN, Largo Galileo Galilei 1, 00044 Frascati, Roma, Italy\label{aff55}
\and
Universit\'e Claude Bernard Lyon 1, CNRS/IN2P3, IP2I Lyon, UMR 5822, Villeurbanne, F-69100, France\label{aff56}
\and
Aix-Marseille Universit\'e, CNRS, CNES, LAM, Marseille, France\label{aff57}
\and
Institut de Ci\`{e}ncies del Cosmos (ICCUB), Universitat de Barcelona (IEEC-UB), Mart\'{i} i Franqu\`{e}s 1, 08028 Barcelona, Spain\label{aff58}
\and
Instituci\'o Catalana de Recerca i Estudis Avan\c{c}ats (ICREA), Passeig de Llu\'{\i}s Companys 23, 08010 Barcelona, Spain\label{aff59}
\and
UCB Lyon 1, CNRS/IN2P3, IUF, IP2I Lyon, 4 rue Enrico Fermi, 69622 Villeurbanne, France\label{aff60}
\and
Mullard Space Science Laboratory, University College London, Holmbury St Mary, Dorking, Surrey RH5 6NT, UK\label{aff61}
\and
Departamento de F\'isica, Faculdade de Ci\^encias, Universidade de Lisboa, Edif\'icio C8, Campo Grande, PT1749-016 Lisboa, Portugal\label{aff62}
\and
Instituto de Astrof\'isica e Ci\^encias do Espa\c{c}o, Faculdade de Ci\^encias, Universidade de Lisboa, Campo Grande, 1749-016 Lisboa, Portugal\label{aff63}
\and
Department of Astronomy, University of Geneva, ch. d'Ecogia 16, 1290 Versoix, Switzerland\label{aff64}
\and
Aix-Marseille Universit\'e, CNRS/IN2P3, CPPM, Marseille, France\label{aff65}
\and
Universit\"ats-Sternwarte M\"unchen, Fakult\"at f\"ur Physik, Ludwig-Maximilians-Universit\"at M\"unchen, Scheinerstrasse 1, 81679 M\"unchen, Germany\label{aff66}
\and
INFN-Bologna, Via Irnerio 46, 40126 Bologna, Italy\label{aff67}
\and
FRACTAL S.L.N.E., calle Tulip\'an 2, Portal 13 1A, 28231, Las Rozas de Madrid, Spain\label{aff68}
\and
Dipartimento di Fisica "Aldo Pontremoli", Universit\`a degli Studi di Milano, Via Celoria 16, 20133 Milano, Italy\label{aff69}
\and
INFN-Sezione di Milano, Via Celoria 16, 20133 Milano, Italy\label{aff70}
\and
NRC Herzberg, 5071 West Saanich Rd, Victoria, BC V9E 2E7, Canada\label{aff71}
\and
Institute of Theoretical Astrophysics, University of Oslo, P.O. Box 1029 Blindern, 0315 Oslo, Norway\label{aff72}
\and
Jet Propulsion Laboratory, California Institute of Technology, 4800 Oak Grove Drive, Pasadena, CA, 91109, USA\label{aff73}
\and
Department of Physics, Lancaster University, Lancaster, LA1 4YB, UK\label{aff74}
\and
Felix Hormuth Engineering, Goethestr. 17, 69181 Leimen, Germany\label{aff75}
\and
Technical University of Denmark, Elektrovej 327, 2800 Kgs. Lyngby, Denmark\label{aff76}
\and
Cosmic Dawn Center (DAWN), Denmark\label{aff77}
\and
Institut d'Astrophysique de Paris, UMR 7095, CNRS, and Sorbonne Universit\'e, 98 bis boulevard Arago, 75014 Paris, France\label{aff78}
\and
Max-Planck-Institut f\"ur Astronomie, K\"onigstuhl 17, 69117 Heidelberg, Germany\label{aff79}
\and
NASA Goddard Space Flight Center, Greenbelt, MD 20771, USA\label{aff80}
\and
Department of Physics and Astronomy, University College London, Gower Street, London WC1E 6BT, UK\label{aff81}
\and
Department of Physics and Helsinki Institute of Physics, Gustaf H\"allstr\"omin katu 2, 00014 University of Helsinki, Finland\label{aff82}
\and
Universit\'e de Gen\`eve, D\'epartement de Physique Th\'eorique and Centre for Astroparticle Physics, 24 quai Ernest-Ansermet, CH-1211 Gen\`eve 4, Switzerland\label{aff83}
\and
Department of Physics, P.O. Box 64, 00014 University of Helsinki, Finland\label{aff84}
\and
Helsinki Institute of Physics, Gustaf H{\"a}llstr{\"o}min katu 2, University of Helsinki, Helsinki, Finland\label{aff85}
\and
Centre de Calcul de l'IN2P3/CNRS, 21 avenue Pierre de Coubertin 69627 Villeurbanne Cedex, France\label{aff86}
\and
Laboratoire d'etude de l'Univers et des phenomenes eXtremes, Observatoire de Paris, Universit\'e PSL, Sorbonne Universit\'e, CNRS, 92190 Meudon, France\label{aff87}
\and
SKA Observatory, Jodrell Bank, Lower Withington, Macclesfield, Cheshire SK11 9FT, UK\label{aff88}
\and
University of Applied Sciences and Arts of Northwestern Switzerland, School of Computer Science, 5210 Windisch, Switzerland\label{aff89}
\and
Universit\"at Bonn, Argelander-Institut f\"ur Astronomie, Auf dem H\"ugel 71, 53121 Bonn, Germany\label{aff90}
\and
INFN-Sezione di Roma, Piazzale Aldo Moro, 2 - c/o Dipartimento di Fisica, Edificio G. Marconi, 00185 Roma, Italy\label{aff91}
\and
Dipartimento di Fisica e Astronomia "Augusto Righi" - Alma Mater Studiorum Universit\`a di Bologna, via Piero Gobetti 93/2, 40129 Bologna, Italy\label{aff92}
\and
Department of Physics, Institute for Computational Cosmology, Durham University, South Road, Durham, DH1 3LE, UK\label{aff93}
\and
Universit\'e C\^{o}te d'Azur, Observatoire de la C\^{o}te d'Azur, CNRS, Laboratoire Lagrange, Bd de l'Observatoire, CS 34229, 06304 Nice cedex 4, France\label{aff94}
\and
Universit\'e Paris Cit\'e, CNRS, Astroparticule et Cosmologie, 75013 Paris, France\label{aff95}
\and
CNRS-UCB International Research Laboratory, Centre Pierre Binetruy, IRL2007, CPB-IN2P3, Berkeley, USA\label{aff96}
\and
University of Applied Sciences and Arts of Northwestern Switzerland, School of Engineering, 5210 Windisch, Switzerland\label{aff97}
\and
Institut d'Astrophysique de Paris, 98bis Boulevard Arago, 75014, Paris, France\label{aff98}
\and
Institute of Physics, Laboratory of Astrophysics, Ecole Polytechnique F\'ed\'erale de Lausanne (EPFL), Observatoire de Sauverny, 1290 Versoix, Switzerland\label{aff99}
\and
Aurora Technology for European Space Agency (ESA), Camino bajo del Castillo, s/n, Urbanizacion Villafranca del Castillo, Villanueva de la Ca\~nada, 28692 Madrid, Spain\label{aff100}
\and
Institut de F\'{i}sica d'Altes Energies (IFAE), The Barcelona Institute of Science and Technology, Campus UAB, 08193 Bellaterra (Barcelona), Spain\label{aff101}
\and
School of Mathematics, Statistics and Physics, Newcastle University, Herschel Building, Newcastle-upon-Tyne, NE1 7RU, UK\label{aff102}
\and
DARK, Niels Bohr Institute, University of Copenhagen, Jagtvej 155, 2200 Copenhagen, Denmark\label{aff103}
\and
Waterloo Centre for Astrophysics, University of Waterloo, Waterloo, Ontario N2L 3G1, Canada\label{aff104}
\and
Department of Physics and Astronomy, University of Waterloo, Waterloo, Ontario N2L 3G1, Canada\label{aff105}
\and
Perimeter Institute for Theoretical Physics, Waterloo, Ontario N2L 2Y5, Canada\label{aff106}
\and
Centre National d'Etudes Spatiales -- Centre spatial de Toulouse, 18 avenue Edouard Belin, 31401 Toulouse Cedex 9, France\label{aff107}
\and
Institute of Space Science, Str. Atomistilor, nr. 409 M\u{a}gurele, Ilfov, 077125, Romania\label{aff108}
\and
Consejo Superior de Investigaciones Cientificas, Calle Serrano 117, 28006 Madrid, Spain\label{aff109}
\and
Dipartimento di Fisica e Astronomia "G. Galilei", Universit\`a di Padova, Via Marzolo 8, 35131 Padova, Italy\label{aff110}
\and
INFN-Padova, Via Marzolo 8, 35131 Padova, Italy\label{aff111}
\and
Caltech/IPAC, 1200 E. California Blvd., Pasadena, CA 91125, USA\label{aff112}
\and
Institut f\"ur Theoretische Physik, University of Heidelberg, Philosophenweg 16, 69120 Heidelberg, Germany\label{aff113}
\and
Institut de Recherche en Astrophysique et Plan\'etologie (IRAP), Universit\'e de Toulouse, CNRS, UPS, CNES, 14 Av. Edouard Belin, 31400 Toulouse, France\label{aff114}
\and
Universit\'e St Joseph; Faculty of Sciences, Beirut, Lebanon\label{aff115}
\and
Departamento de F\'isica, FCFM, Universidad de Chile, Blanco Encalada 2008, Santiago, Chile\label{aff116}
\and
Universit\"at Innsbruck, Institut f\"ur Astro- und Teilchenphysik, Technikerstr. 25/8, 6020 Innsbruck, Austria\label{aff117}
\and
Satlantis, University Science Park, Sede Bld 48940, Leioa-Bilbao, Spain\label{aff118}
\and
Infrared Processing and Analysis Center, California Institute of Technology, Pasadena, CA 91125, USA\label{aff119}
\and
Instituto de Astrof\'isica e Ci\^encias do Espa\c{c}o, Faculdade de Ci\^encias, Universidade de Lisboa, Tapada da Ajuda, 1349-018 Lisboa, Portugal\label{aff120}
\and
Cosmic Dawn Center (DAWN)\label{aff121}
\and
Niels Bohr Institute, University of Copenhagen, Jagtvej 128, 2200 Copenhagen, Denmark\label{aff122}
\and
Universidad Polit\'ecnica de Cartagena, Departamento de Electr\'onica y Tecnolog\'ia de Computadoras,  Plaza del Hospital 1, 30202 Cartagena, Spain\label{aff123}
\and
Dipartimento di Fisica e Scienze della Terra, Universit\`a degli Studi di Ferrara, Via Giuseppe Saragat 1, 44122 Ferrara, Italy\label{aff124}
\and
Istituto Nazionale di Fisica Nucleare, Sezione di Ferrara, Via Giuseppe Saragat 1, 44122 Ferrara, Italy\label{aff125}
\and
INAF, Istituto di Radioastronomia, Via Piero Gobetti 101, 40129 Bologna, Italy\label{aff126}
\and
Universit\'e PSL, Observatoire de Paris, Sorbonne Universit\'e, CNRS, LERMA, 75014, Paris, France\label{aff127}
\and
Universit\'e Paris-Cit\'e, 5 Rue Thomas Mann, 75013, Paris, France\label{aff128}
\and
INAF - Osservatorio Astronomico di Brera, via Emilio Bianchi 46, 23807 Merate, Italy\label{aff129}
\and
INAF-Osservatorio Astronomico di Brera, Via Brera 28, 20122 Milano, Italy, and INFN-Sezione di Genova, Via Dodecaneso 33, 16146, Genova, Italy\label{aff130}
\and
ICL, Junia, Universit\'e Catholique de Lille, LITL, 59000 Lille, France\label{aff131}
\and
ICSC - Centro Nazionale di Ricerca in High Performance Computing, Big Data e Quantum Computing, Via Magnanelli 2, Bologna, Italy\label{aff132}
\and
Instituto de F\'isica Te\'orica UAM-CSIC, Campus de Cantoblanco, 28049 Madrid, Spain\label{aff133}
\and
CERCA/ISO, Department of Physics, Case Western Reserve University, 10900 Euclid Avenue, Cleveland, OH 44106, USA\label{aff134}
\and
Technical University of Munich, TUM School of Natural Sciences, Physics Department, James-Franck-Str.~1, 85748 Garching, Germany\label{aff135}
\and
Max-Planck-Institut f\"ur Astrophysik, Karl-Schwarzschild-Str.~1, 85748 Garching, Germany\label{aff136}
\and
Laboratoire Univers et Th\'eorie, Observatoire de Paris, Universit\'e PSL, Universit\'e Paris Cit\'e, CNRS, 92190 Meudon, France\label{aff137}
\and
Departamento de F{\'\i}sica Fundamental. Universidad de Salamanca. Plaza de la Merced s/n. 37008 Salamanca, Spain\label{aff138}
\and
Universit\'e de Strasbourg, CNRS, Observatoire astronomique de Strasbourg, UMR 7550, 67000 Strasbourg, France\label{aff139}
\and
Center for Data-Driven Discovery, Kavli IPMU (WPI), UTIAS, The University of Tokyo, Kashiwa, Chiba 277-8583, Japan\label{aff140}
\and
Ludwig-Maximilians-University, Schellingstrasse 4, 80799 Munich, Germany\label{aff141}
\and
Max-Planck-Institut f\"ur Physik, Boltzmannstr. 8, 85748 Garching, Germany\label{aff142}
\and
Dipartimento di Fisica - Sezione di Astronomia, Universit\`a di Trieste, Via Tiepolo 11, 34131 Trieste, Italy\label{aff143}
\and
California Institute of Technology, 1200 E California Blvd, Pasadena, CA 91125, USA\label{aff144}
\and
Department of Physics \& Astronomy, University of California Irvine, Irvine CA 92697, USA\label{aff145}
\and
Department of Mathematics and Physics E. De Giorgi, University of Salento, Via per Arnesano, CP-I93, 73100, Lecce, Italy\label{aff146}
\and
INFN, Sezione di Lecce, Via per Arnesano, CP-193, 73100, Lecce, Italy\label{aff147}
\and
INAF-Sezione di Lecce, c/o Dipartimento Matematica e Fisica, Via per Arnesano, 73100, Lecce, Italy\label{aff148}
\and
Departamento F\'isica Aplicada, Universidad Polit\'ecnica de Cartagena, Campus Muralla del Mar, 30202 Cartagena, Murcia, Spain\label{aff149}
\and
Instituto de F\'isica de Cantabria, Edificio Juan Jord\'a, Avenida de los Castros, 39005 Santander, Spain\label{aff150}
\and
CEA Saclay, DFR/IRFU, Service d'Astrophysique, Bat. 709, 91191 Gif-sur-Yvette, France\label{aff151}
\and
Institute of Cosmology and Gravitation, University of Portsmouth, Portsmouth PO1 3FX, UK\label{aff152}
\and
Department of Computer Science, Aalto University, PO Box 15400, Espoo, FI-00 076, Finland\label{aff153}
\and
Instituto de Astrof\'\i sica de Canarias, c/ Via Lactea s/n, La Laguna 38200, Spain. Departamento de Astrof\'\i sica de la Universidad de La Laguna, Avda. Francisco Sanchez, La Laguna, 38200, Spain\label{aff154}
\and
Ruhr University Bochum, Faculty of Physics and Astronomy, Astronomical Institute (AIRUB), German Centre for Cosmological Lensing (GCCL), 44780 Bochum, Germany\label{aff155}
\and
Department of Physics and Astronomy, Vesilinnantie 5, 20014 University of Turku, Finland\label{aff156}
\and
Serco for European Space Agency (ESA), Camino bajo del Castillo, s/n, Urbanizacion Villafranca del Castillo, Villanueva de la Ca\~nada, 28692 Madrid, Spain\label{aff157}
\and
ARC Centre of Excellence for Dark Matter Particle Physics, Melbourne, Australia\label{aff158}
\and
Centre for Astrophysics \& Supercomputing, Swinburne University of Technology,  Hawthorn, Victoria 3122, Australia\label{aff159}
\and
Department of Physics and Astronomy, University of the Western Cape, Bellville, Cape Town, 7535, South Africa\label{aff160}
\and
DAMTP, Centre for Mathematical Sciences, Wilberforce Road, Cambridge CB3 0WA, UK\label{aff161}
\and
Kavli Institute for Cosmology Cambridge, Madingley Road, Cambridge, CB3 0HA, UK\label{aff162}
\and
Department of Astrophysics, University of Zurich, Winterthurerstrasse 190, 8057 Zurich, Switzerland\label{aff163}
\and
Department of Physics, Centre for Extragalactic Astronomy, Durham University, South Road, Durham, DH1 3LE, UK\label{aff164}
\and
Institute for Theoretical Particle Physics and Cosmology (TTK), RWTH Aachen University, 52056 Aachen, Germany\label{aff165}
\and
IRFU, CEA, Universit\'e Paris-Saclay 91191 Gif-sur-Yvette Cedex, France\label{aff166}
\and
Oskar Klein Centre for Cosmoparticle Physics, Department of Physics, Stockholm University, Stockholm, SE-106 91, Sweden\label{aff167}
\and
Astrophysics Group, Blackett Laboratory, Imperial College London, London SW7 2AZ, UK\label{aff168}
\and
Univ. Grenoble Alpes, CNRS, Grenoble INP, LPSC-IN2P3, 53, Avenue des Martyrs, 38000, Grenoble, France\label{aff169}
\and
INAF-Osservatorio Astrofisico di Arcetri, Largo E. Fermi 5, 50125, Firenze, Italy\label{aff170}
\and
Dipartimento di Fisica, Sapienza Universit\`a di Roma, Piazzale Aldo Moro 2, 00185 Roma, Italy\label{aff171}
\and
Centro de Astrof\'{\i}sica da Universidade do Porto, Rua das Estrelas, 4150-762 Porto, Portugal\label{aff172}
\and
HE Space for European Space Agency (ESA), Camino bajo del Castillo, s/n, Urbanizacion Villafranca del Castillo, Villanueva de la Ca\~nada, 28692 Madrid, Spain\label{aff173}
\and
Department of Astrophysical Sciences, Peyton Hall, Princeton University, Princeton, NJ 08544, USA\label{aff174}
\and
Theoretical astrophysics, Department of Physics and Astronomy, Uppsala University, Box 515, 751 20 Uppsala, Sweden\label{aff175}
\and
Minnesota Institute for Astrophysics, University of Minnesota, 116 Church St SE, Minneapolis, MN 55455, USA\label{aff176}
\and
Mathematical Institute, University of Leiden, Einsteinweg 55, 2333 CA Leiden, The Netherlands\label{aff177}
\and
Institute of Astronomy, University of Cambridge, Madingley Road, Cambridge CB3 0HA, UK\label{aff178}
\and
Department of Physics and Astronomy, University of California, Davis, CA 95616, USA\label{aff179}
\and
Space physics and astronomy research unit, University of Oulu, Pentti Kaiteran katu 1, FI-90014 Oulu, Finland\label{aff180}
\and
Center for Computational Astrophysics, Flatiron Institute, 162 5th Avenue, 10010, New York, NY, USA\label{aff181}
}

%
%
\abstract
{
Galaxy major mergers are indicated as one of the principal pathways to trigger active galactic nuclei (AGN). We present the first {statistical analysis of the major merger and AGN connection} in the Euclid Deep Fields, and showcase the statistical power of the \Euclid data. 
We constructed a stellar-mass-complete {($M_{\star}>10^{9.8}\,M_{\odot}$)} sample of galaxies from the quick data release (Q1) in the redshift range $z=0.5$--2.
We selected AGN using X-ray detections, optical spectroscopy, and mid-infrared (MIR) colours, and by processing \IE observations with an image decomposition algorithm. We used convolutional neural networks trained on cosmological hydrodynamic simulations to classify galaxies as mergers and non-mergers. 
We found a larger fraction of AGN in mergers compared to the non-merger controls for all AGN selections, with AGN excess factors ranging from two to six. The largest excess we observed was in the MIR AGN. Likewise, a generally larger merger fraction ($f_{\rm merg}$) was seen in active galaxies than in the non-active controls, with the excess depending on the AGN selection method. Furthermore, we analysed $f_{\rm merg}$ as a function of the AGN bolometric luminosity ($L_{\rm bol}$) and the contribution of the point-source component to the total galaxy light {in the \IE-band} ($f_{\rm{PSF}}$) as a proxy for the relative AGN contribution fraction. We uncovered a rising $f_{\rm merg}$, with an increasing $f_{\rm{PSF}}$ up to $f_{\rm{PSF}}\simeq 0.55$, after which we observed a decreasing trend. In the range $f_{\rm{PSF}} = 0.3$--0.7, mergers appear to be the dominant AGN fuelling mechanism. We then derived the point-source luminosity ($L_{\rm{PSF}}$) and showed that $f_{\rm merg}$ monotonically increases as a function of $L_{\rm{PSF}}$ at $z<0.9$, with $f_{\rm merg}\geq50\%$ for $L_{\rm{PSF}}\simeq 2\times10^{43}\,{\rm erg\,s^{-1}}$. Similarly, at $0.9\leq z \leq 2$, $f_{\rm merg}$ rises as a function of $L_{\rm{PSF}}$, though mergers do not dominate until $L_{\rm{PSF}} \simeq 10^{45}\,{\rm erg\,s^{-1}}$. 
For the X-ray and spectroscopically detected AGN, we derived the bolometric luminosity, $L_{\rm bol}$, which has a positive correlation with $f_{\rm merg}$ for X-ray AGN, while there is a less pronounced trend for spectroscopically selected AGN due to the smaller sample size.
At $L_{\rm bol} > 10^{45}\,{\rm erg\,s^{-1}}$, AGN mostly reside in mergers. We conclude that mergers are most strongly associated with the most powerful and dust-obscured AGN, which are typically linked to a fast-growing phase of the supermassive black hole, while other mechanisms, such as secular processes, might be the trigger of less luminous and dominant AGN.
}
%
%

%
%
   \titlerunning{First \Euclid statistical study of galaxy mergers}
   \authorrunning{Euclid Collaboration: A.~La~Marca et al.}
   
   \maketitle
%
%
%
%
   
\section{\label{sc:Intro} Introduction}

Galaxy mergers have long been considered a key driver of galaxy evolution, as they have the potential to significantly influence the growth and properties of both host galaxies and their central supermassive black holes \citep[SMBHs;][]{sandersWarmUltraluminousGalaxies1988,marconiLocalSupermassiveBlack2004}. During such encounters, tidal forces can lead to gas inflows towards central regions \citep{barnesTransformationsGalaxiesII1996}, that feed intense nuclear star formation and active galactic nucleus (AGN) activity \citep[e.g.,][]{springelModellingFeedbackStars2005,somervillePhysicalModelsGalaxy2015,blumenthalGoFlowUnderstanding2018}. Consequently, this process can trigger AGN feedback, which can severely affect the evolution of a galaxy, for example, by driving galactic-scale outflows and suppressing or enhancing star formation \citep[e.g.,][]{fabianObservationalEvidenceActive2012,harrisonAGNOutflowsFeedback2018}. Understanding the connection between mergers and AGN is therefore crucial for advancing our knowledge of galaxy evolution and the formation of large-scale structures \citep{alexanderWhatDrivesGrowth2012, heckmanCoevolutionGalaxiesSupermassive2014}.

Previous studies have generally shown that mergers can trigger AGN activation; however, the exact mechanisms driving this process remain poorly understood. While many simulation-based studies have suggested that mergers fuel SMBH accretion and initiate the AGN phase \citep[e.g.,][]{hopkinsCosmologicalFrameworkCoEvolution2008,blechaPowerInfraredAGN2018}, other simulations propose that mergers serve only as a secondary fuelling mechanism \citep[e.g.,][]{dimatteoBlackHoleGrowth2003, martinNormalBlackHoles2018,byrne-mamahitInteractingGalaxiesIllustrisTNG2023}. Similarly, mixed results have also emerged from observations. For example, multiple observational studies have reported a clear link between mergers and AGN triggering 
\citep{lacknerLateStageGalaxyMergers2014, kocevskiAreComptonthickAGNs2015, gouldingGalaxyInteractionsTrigger2018, ellisonDefinitiveMergerAGNConnection2019, gaoMergersTriggerActive2020, tobaOpticalIFUObservations2022, tanakaGalaxyCruiseDeep2023, bickleyXrayAGNsSRG2024}, with a possible dependency on AGN luminosity \citep{treisterMajorGalaxyMergers2012,weigelFractionAGNsMajor2018,pierceAGNTriggeringMechanisms2022,lamarcaDustPowerUnravelling2024}, dust obscuration \citep[][]{ricciGrowingSupermassiveBlack2017, ricciHardXrayView2021,donleyEvidenceMergerdrivenGrowth2018}, and environment \citep[][]{koulouridisLocalLargeScaleEnvironment2006, koulouridisAGNsMassiveGalaxy2024}. However, other studies have highlighted that mergers are a less significant mechanism, being outnumbered by secular processes  \citep{groginAGNHostGalaxies2005, allevatoXMMNewtonWideField2011, draperTaleTwoPopulations2012, marianMajorMergersAre2019, silvaGalaxyMergers252021, smethurstEvidenceNonmergerCoevolution2024, garlandMostLuminousMergerfree2023, villforthCompleteCatalogueMerger2023, bichangaPropertiesAGNDwarf2024}. Additionally, in several studies, there has been no observed dependence on AGN luminosity \citep{hewlettRedshiftEvolutionMajor2017, villforthHostGalaxiesLuminous2017, comerfordExcessActiveGalactic2024}.

The AGN triggering debate could arise from several factors. First, there are various methods to identify mergers, each with its advantages and limitations. Among the different methods, visual classification \citep{dargGalaxyZooProperties2010}, close spectroscopic pairs \citep{knapenInteractingGalaxiesNearby2015}, and non-parametric morphological statistics \citep{nevinAccurateIdentificationGalaxy2019} have been widely employed in the past. 
More recently, several studies have favoured machine learning (ML), in particular deep learning (DL), techniques \citep[e.g.,][]{wangConsistentFrameworkComparing2020}. These methods are reproducible, and once trained, they can process large samples efficiently \citep[for a review, see][]{margalef-bentabolGalaxyMergerChallenge2024}. However, their performance depends on the specific task and is constrained by the quality of the training labels. 
Second, as in merger detections, there is no unique method to identify AGN. Since AGN exhibit a diverse range of observational signatures and different characteristics of ongoing activity, they can be selected through a multitude of techniques, including X-ray detections, optical emission line ratios, variability, mid-infrared (MIR) colours, and radio emission \citep[for a review,
see][]{heckmanCoevolutionGalaxiesSupermassive2014}. As a result, different selection methods can lead to AGN and host galaxy samples with very different characteristics \citep{silvermanEvolutionAGNHost2008, hickoxObscuredActiveGalactic2018}.
For these reasons, a panchromatic approach has emerged in order to properly investigate the merger and AGN connection, accounting for different AGN types \citep{liMultiwavelengthStudyActive2023}.

\citet{lamarcaDustPowerUnravelling2024} exploited a large multi-wavelength dataset at $z<1$ and estimated the AGN contribution fraction parameter, which measures the AGN light contribution to the total galaxy light, through spectral energy distribution (SED) modelling in the rest-frame wavelength range 3--30\,\micron. The AGN were selected with multiple diagnostics, that is, X-ray, MIR, and SED modelling and a relation was proposed between the merger fraction and the AGN fraction relation, which revealed two distinct regimes. When the AGN is not dominant (low AGN fraction), the fraction of mergers stays roughly constant, with mergers representing only a secondary AGN triggering mechanism. However, for very dominant AGN, where the AGN fraction exceeds 0.8, the merger fraction rises rapidly towards 100\%. A similar picture was observed in the merger fraction as a function of the AGN bolometric luminosity. These findings could explain some of the conflicting results in the literature. Secular processes may be the principal fuelling mechanisms in non-dominant and relatively faint AGN, while major mergers are the main or only viable channel to trigger the most powerful and dominant AGN. 

So far, a lack of large survey data at high redshift has limited our understanding of the merger and AGN relation and its evolution. To improve our knowledge, particularly at earlier epochs, several key ingredients are needed, including deep imaging data with high spatial resolution to perform morphological classification, large volumes to construct large statistical samples, and multi-wavelength coverage to reliably select a diverse sample of AGN and derive physical properties of AGN and their host galaxies. The advent of \Euclid and the associated ancillary data finally offers the opportunity to investigate this problem throughout cosmic history up to `cosmic noon'.
\Euclid is a European Space Agency (ESA) mission \citep{Laureijs11} whose aim is to observe almost all of the extra-Galactic sky with two surveys. Its scientific objectives are outlined in \citet{EuclidSkyOverview}.
\Euclid operates in the optical and near-IR in four bands (\IE, \YE, \JE, and \HE), covering wavelengths from 0.53\,\micron\ to 2.02\,\micron, \citep{EuclidSkyVIS, EuclidSkyNISP}. 
Although \Euclid was designed as a cosmology mission, it will be able to detect billions of sources, of which at least $10$ million are expected to be AGN detected in \IE \citep{EP-Bisigello,EP-Selwood} and hundreds of thousands in the near-IR bands \citep{EP-Lusso}. This will dramatically increase the number of known AGN with high-resolution imaging, and provide an unprecedented opportunity to study the role of mergers in the evolution of AGN. 

The aim of this study is to investigate the connection between mergers and AGN using the first quick release of \Euclid data \citep[][hereafter \citetalias{Q1cite}]{Q1cite}. We constructed a stellar mass-complete sample of galaxies across the redshift range $0.5\leq z \leq 2$, with multi-wavelength data ranging from the X-ray to the MIR. We revisited two facets of the merger and AGN connection: {i)} Using a binary active--non-active AGN classification, we analysed whether mergers are a viable path to trigger AGN and assessed their significance, and {ii)} exploring continuous AGN properties, we studied how the fraction of mergers varies with AGN dominance and absolute power. Specifically, we explored whether galaxies hosting the most dominant and luminous AGN are more likely to be mergers. We developed a convolutional neural network (CNN) to identify mergers in \Euclid \IE images. To mitigate issues with visual classifications, we trained the CNN on mock \Euclid observations generated from cosmological hydrodynamic simulations, which include galaxy merger histories. We used four different diagnostics to select AGN and characterised each AGN based on the central point source luminosity relative to the host galaxy and, when possible, its bolometric luminosity. 

The paper is organised as follows. In Sect.~\ref{sc:Data}, we first introduce the \Euclid data products we use in this work, the ancillary multi-wavelength data, and our galaxy sample selection. Then, we describe the mock \Euclid observations generated from cosmological hydro-dynamical simulations to train our DL classifier. In Sect.~\ref{sc:Method}, we present our galaxy merger classifier and the various AGN selection methods adopted. In Sect.~\ref{sc:Results}, we first explore the merger and AGN connection using a binary classification of AGN and non-AGN. Next, we analyse this connection using continuous parameters that characterise the relative and absolute AGN power. 
We discuss possible caveats in our analysis in Sect.~\ref{sc:caveats}.
Finally, we summarise our main findings in Sect.~\ref{sc:Conc}. Throughout the paper, unless otherwise stated, we assume a flat $\Lambda$CDM Universe with $\Omega_{\rm{m}}=0.3$, $\Omega_{\Lambda}=0.7$, and $H_0=70\,{\rm km}\,{\rm s}^{-1}\,{\rm Mpc}^{-1}$ and express magnitudes in the AB system \citep{okeSecondaryStandardStars1983}.


\section{\label{sc:Data} Data}

In this section, we first describe the \Euclid data. Then, we present a brief description of the multi-wavelength ancillary data, from the X-ray to the MIR. Finally, we introduce the mock \Euclid VIS imaging data generated from the simulations. 

\subsection{\Euclid data}

This work focuses on exploring \citetalias{Q1cite}, comprising data from a single visit of the Euclid Deep Fields (EDFs), namely the EDF North (EDF-N), the EDF South (EDF-S), and the EDF Fornax (EDF-F), covering a total area of $\sim63\,{\rm deg}^2$. All EDFs have been observed in all four \Euclid photometric bands, that is \IE, \YE, \JE, and \HE.
These observations have been complemented by ground-based optical photometry taken with various instruments across the wavelength range 0.3--1.8\,\micron. Q1 includes imaging, spectroscopic data, and value-added catalogues, including photometric redshifts \citep[photo-$z$;][]{Q1-TP005}. 
Further details on Q1 can be found in \citet{Q1-TP001}, \citet{Q1-TP002}, \citet{Q1-TP003}, and \citet{Q1-TP004}.
All \Euclid data used in this work, catalogues and images, have been retrieved using the ESA Datalabs facility \citep{navarroESADatalabsDigital2024}.

\textbf{Catalogues}.
We selected a sample of galaxies from the \Euclid MER catalogue  (Euclid Collaboration: Altieri et al. in prep.) 
removing possible contaminants using the available columns as follows. First, we required a VIS \IE detection by imposing \verb|VIS_DET| $= 1$. Then, we applied the condition \verb|DET_QUALITY_FLAG| $< 4$ to filter out contaminants such as bad pixels, saturation, proximity to image borders, location within VIS or NIR bright star masks, presence within extended object areas, or omission by the deblending algorithm due to large pixel size. 
Additional flags can be used to filter out further contaminants, such as the \verb|SPURIOUS_FLAG|, which identifies spurious sources. We set this flag to $0$ to exclude such sources from our sample. Finally, we applied constraints on source flux and size by imposing \verb|MUMAX_MINUS_MAG| $>-2.6$ to filter out point-like sources and $23.9 - 2.5\,\logten($\verb|FLUX_DETECTION_TOTAL|$) <\,${ 23.5} to exclude faint objects. 

In addition to the photometric catalogue, we queried the official pipeline photo-$z$ and stellar masses for each source \citep{Q1-TP005}. We excluded objects flagged (greater than or equal to one) by \verb|PHZ_FLAGS|, \verb|PHYS_PARAM_FLAGS|, or \verb|QUALITY_FLAG|. When available, we used photometric redshifts and stellar masses estimated by \citet{Q1-SP031}, which complemented the \Euclid data with public IRAC observations for an improved quality of the recovered parameters. Finally, we limited our selection to galaxies within the redshift range $0.5\leq$ photo-$z\leq 2.0$ and with stellar mass $M_{\star}/M_{\odot}>10^{9.8}$. These galaxy parameters are given with their probability distributions and a set of possible values, from which we adopted the median values when applying these selections. The cut on stellar mass is motivated by our requirement of selecting a stellar mass complete sample, since \citet{Q1-SP031} showed that at $z=2$ the \Euclid galaxy sample is $90\%$ complete at $M_{\star}/M_{\odot}\gtrsim10^{9.8}$, based on the \citet{pozzettiZCOSMOS10kbrightSpectroscopic2010} methodology.  

\textbf{Images}. For our task of identifying mergers, we utilised VIS imaging data, which have a pixel resolution of \ang{;;0.1} and a signal-to-noise ratio ${\rm S/N}\geq10$ at $\IE \leq24.5$ \citep{EuclidSkyVIS}. For each galaxy in the selected sample, we made an $8\arcsec \times 8\arcsec$ (corresponding to a $80\times 80$ pixel grid) thumbnail centred on the source. This size approximately corresponds to a physical scale of $50\,\rm{kpc}\times50\,\rm{kpc}$ in the redshift range considered. We excluded sources for which generating an $8\arcsec \times 8\arcsec$ cutout was not feasible because they are at the edge of the field. Additionally, we retrieved the VIS point spread function (PSF), which was used when constructing the training dataset.

\textbf{Ancillary data}.
Q1 is complemented by ancillary multi-wavelength datasets from photometric surveys, including X-ray data from XMM-{\it Newton}, {\it Chandra} and eROSITA \citep[][hereafter \citetalias{Q1-SP003}, and references therein]{Q1-SP003}, GALEX \citep[Galaxy Evolution Explorer;][]{bianchiRevisedCatalogGALEX2017}, Hyper Suprime-Cam \citep[HSC;][]{miyazakiHyperSuprimeCamSystem2018}, \textit{Gaia} \citep{gaiacollaborationGaiaMission2016}, UNIONS (Ultraviolet Near-Infrared Optical Northern Survey; Gwyn et al. in prep.), DES \citep[Dark Energy Survey;][]{thedarkenergysurveycollaborationDarkEnergySurvey2005}, the Dark Energy Spectroscopic Instrument \citep[DESI;][]{desicollaborationDESIExperimentPart2016, desicollaborationOverviewInstrumentationDark2022} Legacy Imaging Surveys \citep{deyOverviewDESILegacy2019} and spectroscopic survey, and the WISE-AllWISE DR6 survey data \citep{wrightWidefieldInfraredSurvey2010a}. To create multi-wavelength catalogues for each one of the EDFs, \citet[][hereafter \citetalias{Q1-SP027}]{Q1-SP027} performed positional matches with the external surveys using the software \texttt{STILTS}, version 3.5--1 \citep{taylorSTILTSPackageCommandLine2006}. The matching process was customised for each EDF to account for differences in sky coverage. For detailed information on the matching procedure, including the radii used and selection criteria applied, we refer the reader to \citetalias{Q1-SP027}.

\subsection{The IllustrisTNG galaxy sample}

\begin{figure}
    \centering
    \includegraphics[width=0.49\textwidth]{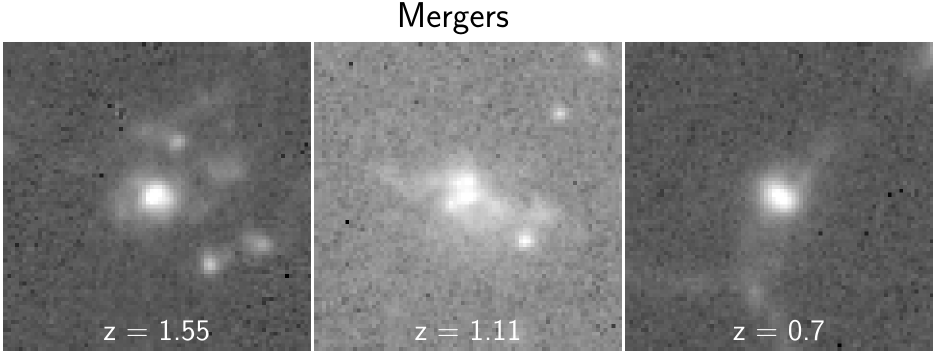}
    \includegraphics[width=0.49\textwidth]{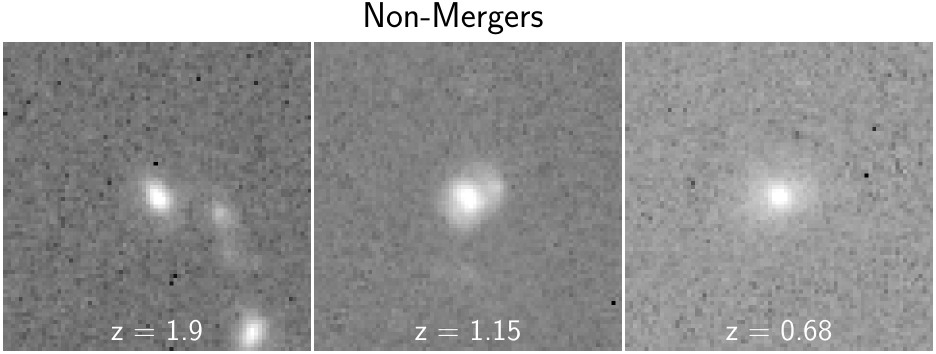}
    \caption{Examples of mock \Euclid VIS \IE-band observations of TNG galaxies. Galaxies were randomly selected among the TNG100 galaxies available. The images are $8\arcsec\times8\arcsec$ wide and log-scaled in the $1{\rm st}$--$99{\rm th}$ percentile range. The redshift of each galaxy is reported in each panel.
    }
    \label{fig:tng_examples}
\end{figure}

To train our merger identification model, we used simulated galaxies from the IllustrisTNG (hereafter TNG) cosmological hydrodynamical simulations, which provide detailed merger histories in large cosmological volumes, ensuring a large sample of galaxies. 
The TNG simulation consists of three different volumes varying in physical size and mass resolution \citep{marinacciFirstResultsIllustrisTNG2018a, naimanFirstResultsIllustrisTNG2018, nelsonFirstResultsIllustrisTNG2018, pillepichFirstResultsIllustrisTNG2018, springelFirstResultsIllustrisTNG2018a}. We used the TNG100 and TNG300 boxes (hereafter referred to as the TNG simulation), with their box size corresponding to $110.7$\,Mpc and $302.6$\,Mpc, respectively. 
The baryonic matter resolution is $1.4\times10^6\,M_{\odot}$ in TNG100 and $1.1\times10^7\,M_{\odot}$ in TNG300. Using both TNG suites allows us to confidently select galaxies down to $M_{\star}/M_{\odot}=10^9$ in TNG100, and to have a large sample of galaxies thanks to the TNG300 size. We required a minimum of about $1000$ baryonic particles, which, for TNG100, correspond to galaxies with stellar mass $M_{*}/M_{\odot}\geq 10^9$, while for TNG300 the lower mass limit is $M_{*}/M_{\odot}>8\times10^9$.

We selected galaxies within the redshift range $z=0.5$--$2$, corresponding to simulation snapshot numbers $67$--$33$. The time step between each snapshot is  $150\, {\rm Myr}$. For each galaxy, the TNG simulation provides a complete merger history \citep{rodriguez-gomezMergerRateGalaxies2015a} identified through the {\tt Subfind} algorithm \citep{springelPopulatingClusterGalaxies2001b}. We then define a subhalo as a merger if a merger event occurred in the previous 300\,Myr or will occur within the next 800\,Myr. Otherwise, the subhalo is considered a non-merger. This time window is motivated by simulation studies \citep[e.g.,][]{morenoInteractingGalaxiesFIRE22019}, which show that during this period, the majority of gas is transferred between galaxies, leading to enhanced star formation and nuclear activity. Here we considered only major merger events, with a stellar mass ratio $M_1/M_2 \leq 4$.
This selection includes pre-mergers (close galaxy pairs expected to merge within 800\,Myr), ongoing mergers, and recent post-mergers. As such, we sample a significant fraction of the merger timescale, covering the late stages of the dynamical interaction and the immediate aftermath of coalescence, but excluding wide pairs that are still in the early interaction phase. It is important to note that the merger rate and its evolution in hydrodynamical simulations such as TNG depend on the underlying galaxy-halo connection implemented in the simulation. Different simulations with distinct prescriptions for galaxy formation physics can yield different merger rates \citep[e.g.,][]{gryllsSignificantEffectsStellar2020}.

\begin{figure}
    \centering
    \includegraphics[width=0.49\textwidth]{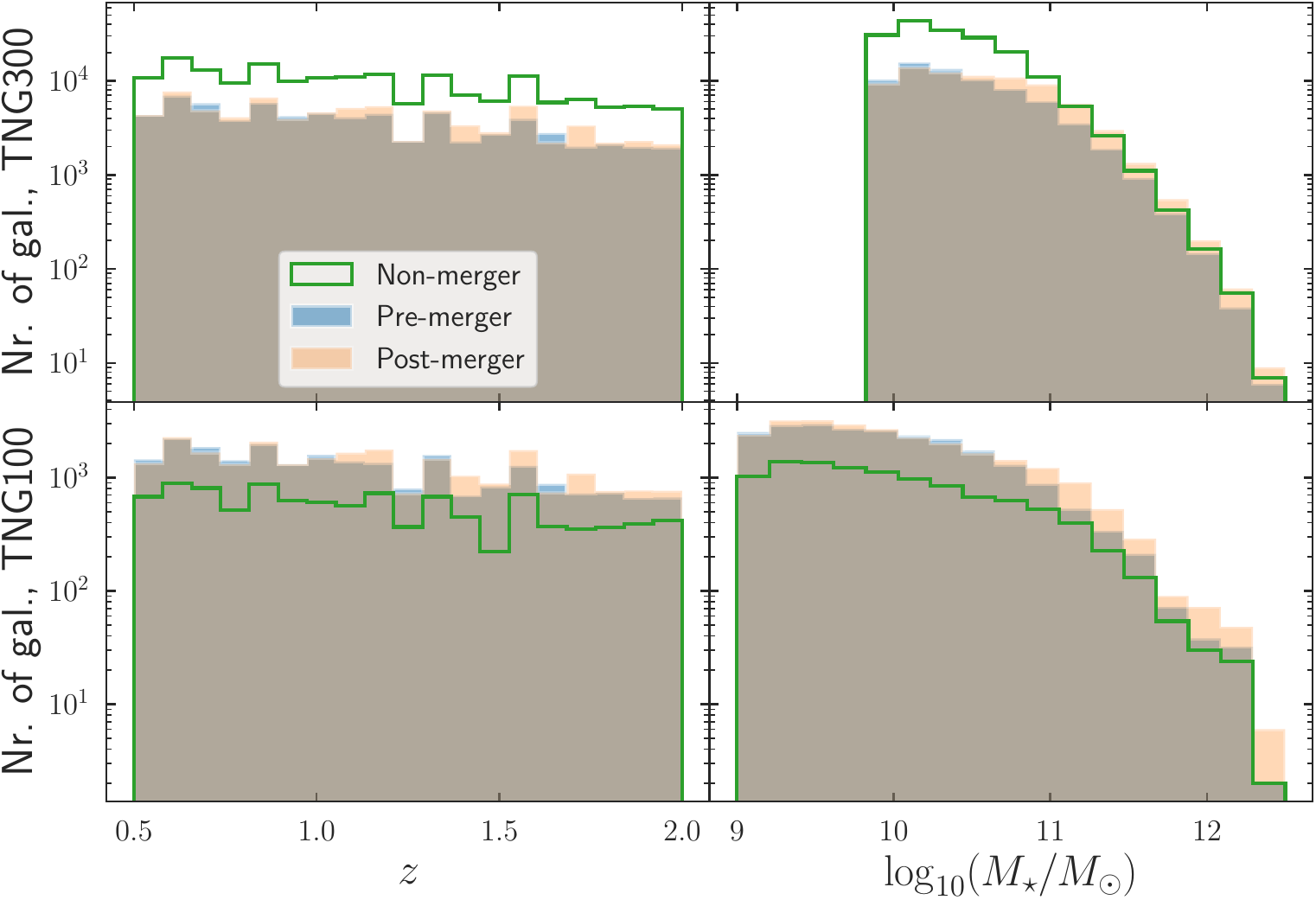}
    \caption{ Redshift (left column) and $M_{\star}$ (right column) distributions for pre-, post-, and non-merging galaxies in the TNG300 (top row) and TNG100 (bottom row) training sets.}
    \label{fig:tng_dist}
\end{figure}

The dataset was divided into training, validation, and testing sets, corresponding to 80\%, 10\%, and 10\% of the total sample, respectively. We ensured that galaxies involved in the same merger sequence were included in only one subset. These datasets result in 499\,523 galaxies for the training sample (427\,577 from TNG300 and 71\,946 from TNG100), 54\,911 for the validation set (46\,660 from TNG300 and 8251 from TNG100), 61\,697 for the testing set (52\,875 from TNG300 and 8822 from TNG100). Of these, 250\,142, 26\,999, and 30\,986 are mergers, respectively. In each set, we balanced the number of non-mergers with that of mergers. 
Figure~\ref{fig:tng_dist} displays the $z$ and $M_{\star}$ distributions of the training samples, split in pre-, post-, and non-mergers. 

We generated mock observations for the TNG galaxies following the \citet{margalef-bentabolGalaxyMergerChallenge2024} methodology. We prepared the mock VIS \IE-band observations, at the same pixel resolution of \ang{;;0.1}, as follows:
\begin{enumerate}
    \item Each stellar particle contributes to the galaxy's SED, determined by its mass, age, and metallicity. These SEDs are derived from the stellar population synthesis models of \citet{bruzualStellarPopulationSynthesis2003}. The summed SED was passed through the VIS filter to generate a smoothed 2D projected map \citep{rodriguez-gomezMergerRateGalaxies2015a}. The image is then cropped to $8\arcsec \times 8\arcsec$, corresponding to approximately $50\,\mathrm{kpc} \times 50\,\mathrm{kpc}$ in the relevant redshift range, matching the size used for the \Euclid galaxy images.
    
    \item Each image was convolved with a VIS PSF, randomly chosen to account for the spatial variation of the PSF across the field of view. 
    \item Poisson noise was added to each image to simulate the statistical variation in photon emission from sources over time.
    
    \item To ensure realistic merger classifications, it was essential to include observational effects \citep[e.g.,][]{huertas-companyHubbleSequence02019, rodriguez-gomezMergerRateGalaxies2015a}. We injected the TNG galaxies into actual \Euclid sky cutouts of $8\arcsec \times 8\arcsec$. To prepare the sky cutouts, we generated random coordinates within the area covered by \citetalias{Q1cite} data. We controlled the segmentation map for each coordinate, ensuring that within a $9\arcsec \times 9\arcsec$ box, all pixels were set to zero\footnote{This constraint ensures that there are no detected sources or artefacts in each image pixel. The $9\arcsec$ radius is derived from the estimated source density of the EDFs.}. When creating the cutouts, we controlled that there were no invalid pixels ({\tt NaN} values) and that the selected coordinates allowed for a perfectly square cutout without intersecting the edge of a tile. 
\end{enumerate}

To create a training sample that accounts for possible AGN, we added a central point source to the host galaxies. The central source intensity can be defined in relation to the host galaxy flux, given the PSF fraction ($f_{\rm{PSF}}$):
\begin{equation}
    f_{\rm{PSF}} = \frac{F_{\rm{PSF}}}{F_{\rm{host}} + F_{\rm{PSF}}} \;,
\end{equation}
where $F_{\rm{PSF}}$ and $F_{\rm{host}}$ are the fluxes within a \ang{;;0.5} aperture of the central source and the host galaxy, respectively. The observed VIS PSF models were used as the central point source. The $f_{\rm{PSF}}$ values were uniformly chosen in the range $0$--1. This operation was performed for a randomly selected 20\% of the TNG sample. We show examples of the final mock observations in Fig.~\ref{fig:tng_examples}. To see the effect of each step in the mock observations generation, we refer the reader to \citet{margalef-bentabolGalaxyMergerChallenge2024}.

Our mock images do not include dust attenuation, which may affect morphologies at $z=2$, where the \IE-band probes the rest-frame ultraviolet. However, previous studies have shown that including dust attenuation via full radiative transfer calculations yields only modest changes in the overall classification performance \citep{bottrellDeepLearningPredictions2019, rodriguez-gomezOpticalMorphologiesGalaxies2019, wangConsistentFrameworkComparing2020}, while being computationally prohibitive at the scale of our training sample. 
Moreover, modelling the effect of dust involves many assumptions (e.g. on dust composition and distribution), whose validity remains to be tested \citep{Zanisi2021DeepLearning}. We therefore followed previous works and used dust-free mock observations.

We normalised each image following \citet{bottrellDeepLearningPredictions2019}.
This normalisation ensures that all images are in a hyperbolic arcsin scale within the range 0--1, maximising the contrast of the central target. A summary of the main steps { applied} is provided below \citep[see][for a detailed description]{bottrellDeepLearningPredictions2019}.
\begin{enumerate}
  \renewcommand{\labelenumi}{\roman{enumi})}
  \item We took the hyperbolic arcsin of the images. Values below $-7$ were converted to NaNs. 
  \item We computed the median of each image, $a_{\rm min}$, and the 99th percentile, $a_{\rm max}$, considering a central box of side 25 pixels. 
  \item All values below $a_{\rm min}$ were set to $a_{\rm min}$, including the NaNs. Values above $a_{\rm max}$ were set to $a_{\rm max}$. The resulting clipped images were normalised by subtracting $a_{\rm min}$ and dividing by $a_{\rm max}-a_{\rm min}$.
\end{enumerate}
This mock \Euclid dataset was used to train, validate, and test our merger classifier, as described in the next section.


\section{\label{sc:Method} Methodology}

Here, we present the DL classifier developed to identify mergers in \Euclid images. Then, we describe the diagnostics used to select AGN.

\subsection{Merger classification using CNNs}\label{sc:CNN}

\begin{table}[]
\caption{Convolutional neural network architecture.}
\small
\begin{tabular}{lccc}
\hline\hline \\[-7pt]
Layer type & No. param. & Output shape & Properties \\ \hline\\[-7pt]
Input & 0 & (1,80,80) & \\
\hline\\[-7pt]
\begin{tabular}[l]{@{}l@{}}Convolutional\\ 32 filters (7,7)\end{tabular} & 1600 & (32,80,80) & \begin{tabular}[c]{@{}c@{}}1 pixels stride, \\ \enquote{same} padding, \\ ReLU act.\end{tabular} \\
Max Pooling & 0 & (32,40,40) & pool size 2 \\
Dropout & 0 & (32,40,40) & 30\% \\
\hline\\[-7pt]

\begin{tabular}[l]{@{}l@{}}Convolutional\\ 64 filters (7,7)\end{tabular} & 100\,416 & (64,40,40) & \begin{tabular}[c]{@{}c@{}}1 pixels stride, \\ \enquote{same} padding, \\ ReLU act.\end{tabular} \\
Max Pooling & 0 & (64,20,20) & pool size 2 \\
Dropout & 0 & (64,20,20) & 30\% \\
Batch Norm. & 256 & (64,20,20) & \\
\hline\\[-7pt]

\begin{tabular}[l]{@{}l@{}}Convolutional\\ 128 filters (7,7)\end{tabular} & 401\,536 & (128,20,20) & \begin{tabular}[c]{@{}c@{}}1 pixels stride, \\ \enquote{same} padding, \\ ReLU act.\end{tabular} \\
Max Pooling & 0 & (128,10,10) & pool size 2 \\
Dropout & 0 & (128,10,10) & 30\% \\
\hline\\[-7pt]

\begin{tabular}[l]{@{}l@{}}Convolutional\\ 128 filters (7,7)\end{tabular} & 802\,944 & (128,10,10) & \begin{tabular}[c]{@{}c@{}}1 pixels stride, \\ \enquote{same} padding, \\ ReLU act.\end{tabular} \\
Max Pooling & 0 & (128,5,5) & pool size 2 \\
Dropout & 0 & (128,5,5) & 30\% \\
\hline\\[-7pt]

Flatten & 0 & (32\,000) & \\

Dense & 819\,456 & (256) & \begin{tabular}[c]{@{}c@{}}256 units, \\ ReLU act.\end{tabular} \\
Dropout & 0 & (256) & 30\% \\
Dense & 32\,896 & (128) & \begin{tabular}[c]{@{}c@{}}128 units, \\ ReLU act.\end{tabular} \\
Dropout & 0 & (128) & 30\% \\
Dense & 129 & (1) & \begin{tabular}[c]{@{}c@{}}1 unit, \\ sigmoid act.\end{tabular} \\
\hline
\end{tabular}
\label{tab:CNN}
\tablefoot{
The columns are the name of the Keras layer (and the filters for the convolutional layers), the number of trainable parameters, the output shape, and the hyper-parameters for each layer. 
}
\end{table}

We developed a CNN \citep{lecunGradientbasedLearningApplied1998} to classify mergers and non-mergers. CNNs consist of multiple layers that apply learnable filters to an input image to capture features such as edges and textures. The later layers of the network are typically fully connected, combining the features from earlier layers to calculate a classification for the input image.
The architecture developed in this work is presented in Table~\ref{tab:CNN}, for which we utilised the Keras framework for the {\tt TensorFlow} platform \citep{cholletKerasDeepLearning2023, abadiTensorFlowLargeScaleMachine2016}. The CNN consists of four convolutional layers and three fully connected layers. For all layers, we adopted a rectified linear unit (ReLU) as an activation function, except for the final layer, where a sigmoid activation function was used. A stride of one pixel was used for the convolutional layers. We introduced dropout layers after each processing layer to prevent overfitting. These dropout layers randomly set input units to zero at a specified rate. To further prevent overfitting, early stopping in the training phase was used. The specific hyper-parameters, listed in Table \ref{tab:CNN}, include filter numbers and sizes, dropout rates, and strides, chosen based on a grid search. 

\begin{table}[h]
    \centering
    \caption{Overall performance of the CNN on the TNG test set.}
    \label{tab:performance}
    \begin{tabular}{lccc}
    \hline \hline\\[-7pt]
    Class & Precision & Recall & F1-score  \\
    \hline\\[-7pt]
    Mergers & 0.80 & 0.68 & 0.74 \\
    Non-mergers & 0.72 & 0.83 & 0.77 \\
    \hline
    \end{tabular}
    \tablefoot{Two thresholds were used for classifying galaxies as mergers (score $\geq 0.59$) and non-mergers (score $< 0.35$). }
\end{table}

\begin{figure*}
    \centering
    \includegraphics[width=0.99\textwidth]{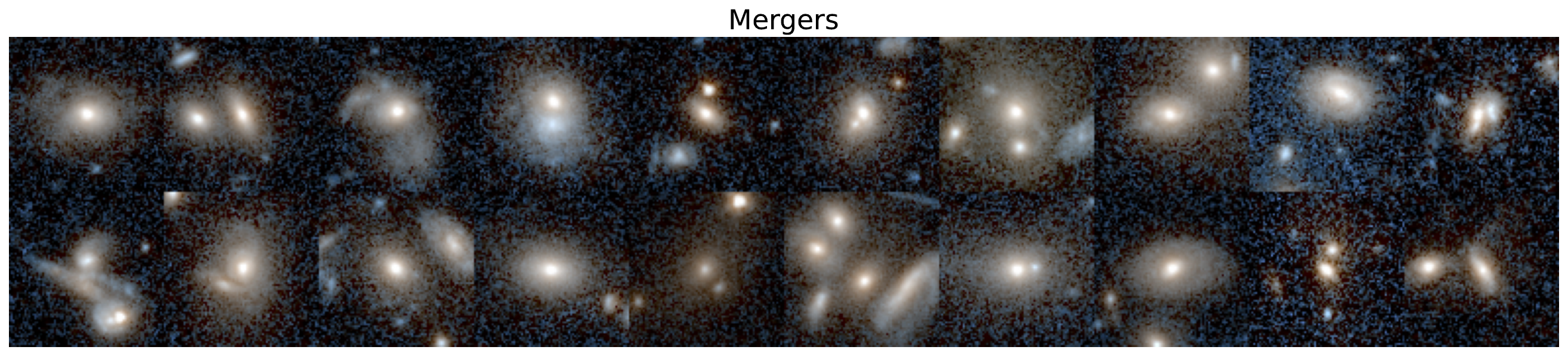}
    \includegraphics[width=0.99\textwidth]{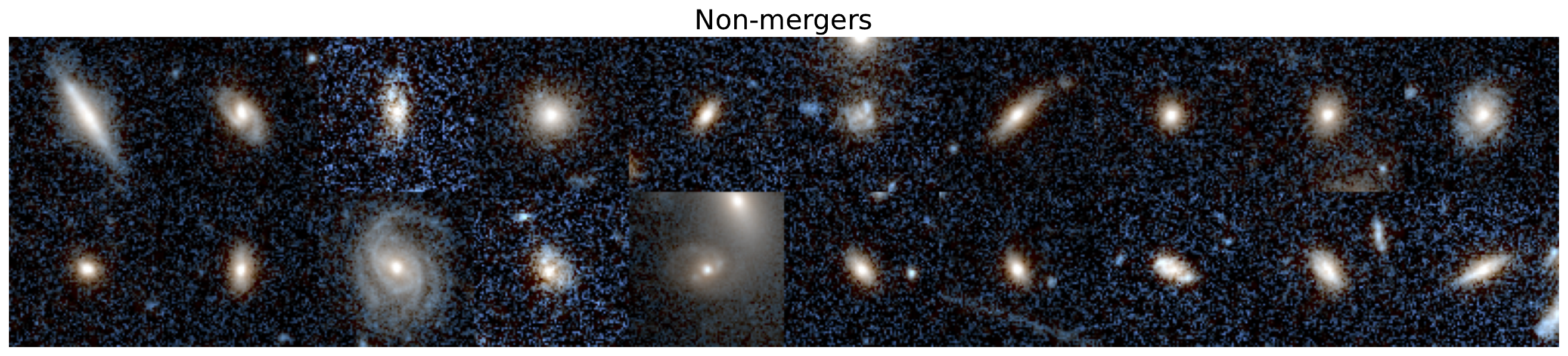}
    \caption{
    Examples of galaxies classified as mergers (\emph{top panel}) and as non-mergers (\emph{bottom panel}) by our algorithm. The cutouts are \Euclid composite RGB images where the R channel is \YE, the B channel is \IE, and the G channel is the mean, following a 99.85th percentile clip and an arcsinh stretch \citep[$x' = $~arcsinh$\,Qx$, with $Q=100$;][]{Q1-SP047}.
    }
    \label{fig:examples}
\end{figure*}

We evaluated the model performance on the test set from the TNG simulations, using common metrics such as `precision', `recall', and `F1-score'. Precision measures how often the model correctly predicts a given class, while recall focuses on how complete the model is at finding objects in a given class. 
In other words, precision is the number of objects correctly recovered for a class divided by the total number of objects predicted in that class. Recall is the number of objects correctly recovered for a class divided by the total number of objects in that class.
F1-score is the harmonic mean of precision and recall. All metrics are calculated on a balanced sample (50\% mergers and 50\% non-mergers). 

In this work, we are interested in selecting a pure sample of mergers. Hence, we followed \citet{lamarcaDustPowerUnravelling2024} to have a higher classification precision for mergers. We searched for a threshold which maximises the F1-score to identify mergers while maintaining a precision greater than or equal to $0.80$ for the merger class. According to these prescriptions, the best threshold for mergers is $0.59$. Thus, all galaxies with a predicted score greater than $0.59$ are classified as mergers. 
Similarly, to select a sample of non-mergers with low contamination levels, we searched for a threshold that ensures at least 0.70 precision for non-mergers, maximising the F1-score. We lowered the precision to 0.70, given the poorer precision of our classifier for the non-merger class. It is important to highlight that the expected number of non-mergers is much larger than that of interacting galaxies \citep[e.g.,][]{ferreiraGalaxyMergerRates2020}. Therefore, in real galaxies, we expect a much lower contamination of mergers in predicted non-mergers. The set threshold is 0.35, meaning all galaxies with a predicted score below 0.35 are labelled as non-mergers. Performance metrics for mergers and non-mergers are shown in Table~\ref{tab:performance}. 
Setting a larger threshold for mergers and a lower one for non-mergers will further improve the purity of both classes, but also strongly affect their completeness, limiting the sample size of both classes.

The model performance is comparable { to} the performance of other recent studies. \citet{margalef-bentabolGalaxyMergerChallenge2024} benchmarked several state-of-the-art methods to identify major mergers in astronomical images out to $z=1$. Each model was trained on mock observations from cosmological hydrodynamical simulations, where mergers have been defined in a similar fashion to this work. Based on the performance metrics on the TNG-test set, the best model in \citet[][Table~3]{margalef-bentabolGalaxyMergerChallenge2024} obtained a precision of 0.80 and a recall of 0.74 (F1-score 0.77) for the merger class, which is consistent with the performance for mergers we report in Table~\ref{tab:performance}. 
While comparable to similar contemporary works, the model performance metrics (Table~\ref{tab:performance}) indicate non-negligible levels of sample contamination and incompleteness inherent to automated classification. We quantitatively assess the impact of these classification uncertainties on our key scientific results regarding the merger-AGN connection in Section~\ref{sect:MC}.

The \Euclid collaboration also provides detailed morphologies \citep{Q1-SP047}, including possible companions and merger features, based on predictions from the Bayesian DL classifier {\tt Zoobot} \citep{walmsleyPracticalGalaxyMorphology2022}. However, these predictions are limited to the top $1\%$ brightest and most extended galaxies, with the selection criteria being \texttt{segmentation\_area $>1200\,\rm{pixels}$} OR \IE$<20.5\,\rm{mag}$ AND \texttt{segmentation\_area $>200\,\rm{pixels}$}. In comparison, our stellar mass-complete sample goes down to \IE$\simeq23.5$ mag. Therefore, we developed our own classifier. We compare our model predictions with the {\tt Zoobot} classification for the common galaxies in Appendix~\ref{app:Zoobot}. 

The double threshold approach has the side effect of producing unclassified galaxies, defined as those with a predicted score between 0.35 and 0.59, inclusive. However, given the large sample size of Q1, this does not affect the analysis carried out in this work. The \Euclid galaxy sample constructed contains $105\,037$ sources classified as mergers, $254\,564$ as non-mergers, and $204\,082$ unclassified objects. These values correspond to $18.6\%$, $45.2\%$, and $36.2\%$ shares of the whole sample, respectively. The catalogue with the merger classification is available from Zenodo\footnote{\url{https://doi.org/10.5281/zenodo.17087033}}.
Hereafter, we focus on the classified galaxies and calculate merger fractions as 
\begin{align}
        & f_{\rm merg} = \frac{N_{\rm merger}}{N_{\rm classified}}=\frac{N_{\rm merger}}{N_{\rm merger} + N_{\rm non-merger}}\, ,
        \label{eq:f_merg}
\end{align} 
unless differently stated.
We show some randomly selected \Euclid merger and non-merger examples in Fig.~\ref{fig:examples}. Examples of unclassified galaxies are provided in Appendix~\ref{app:uncl}. Most galaxies classified as mergers are pair galaxies, with close companions clearly visible in the images. In comparison, non-merger galaxies appear to be quite regular and isolated. Here, we point out that two close galaxies, if both detected, were considered individually rather than as a single system.

\subsection{AGN identification}

\begin{table*}[h]
    \centering
    \caption{Active Galactic Nucleus counts for each selection used in this paper.}
    \begin{tabular}{lllr}
        \hline \hline\\[-7pt]
        AGN selection method & Reference & Description & No. \\
        \hline\\[-7pt]
        X-ray & \citetalias{Q1-SP003} & Extragalactic point-like X-rays sources from & 437 \\
        & & 4XMM-DR13, CSC2, and eROSITA surveys & \\
        \hline \\[-7pt]
        DESI, spectroscopy & \citet{siudekValueaddedCatalogPhysical2024} & Spectroscopic diagnostics based on emission & 160 \\
        & & lines for the matched DESI sources (see Sect.~\ref{sc:Method}) & \\
        \hline\\[-7pt]
        DL-based & \citetalias{Q1-SP015} & Galaxies with a predicted $f_{\rm{PSF}}>0.2$, based on  & $23\,338$ \\
        & & VIS images & \\
        \hline\\[-7pt]
        C75, AllWISE & \citetalias{assefWISEAGNCatalog2018} & 75\% Completeness-optimised MIR diagnostic & $5712$ \\
        & & Eq.~(\ref{eq:C75}) applied to the AllWISE data & \\
        \hline\\[-7pt]
        R90, AllWISE & \citetalias{assefWISEAGNCatalog2018} & 90\% Reliability-optimised MIR diagnostic & 556 \\
        & & Eq.~(\ref{eq:R90}) applied to the AllWISE data & \\
        \hline
    \end{tabular}
    \label{tab:agn_counts}
\end{table*}

\begin{figure}[h]
    \centering
    \includegraphics[width=0.49\textwidth]{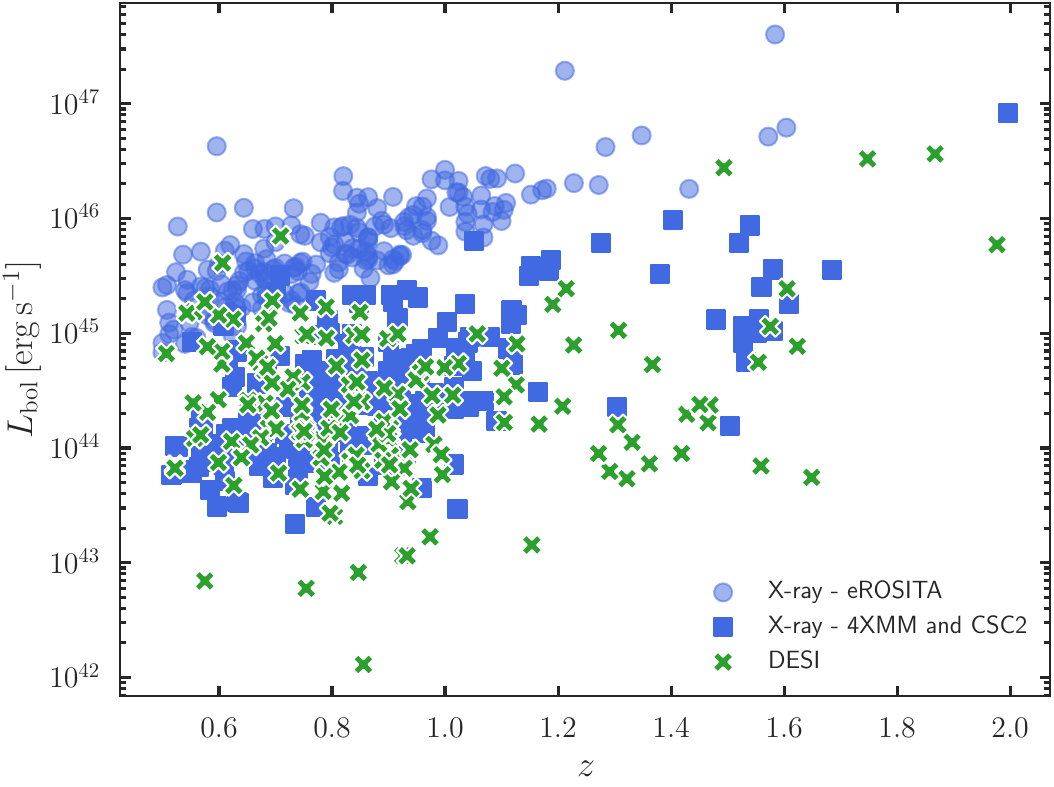}
    \caption{AGN bolometric luminosity ($L_{\rm bol}$) versus redshift for the X-ray and DESI-selected AGN.
    }
    \label{fig:Lbol_vs_z}
\end{figure}

The Q1 data are accompanied by a set of multi-wavelength catalogues that allow for multiple AGN selections. This work focuses on four main AGN detection techniques: X-ray, optical spectroscopy, DL-based image decomposition, and MIR colours. The first \Euclid AGN catalogue is presented in \citetalias{Q1-SP027} and includes all these AGN selections, except the DL-based image decomposition method, described in \citet[][hereafter \citetalias{Q1-SP015}]{Q1-SP015}. The AGN-selection techniques are detailed in these two papers. Here, we summarise the main aspects of the criteria used.

MIR colour selections defined in \citet[][hereafter \citetalias{assefWISEAGNCatalog2018}]{assefWISEAGNCatalog2018}. \citetalias{Q1-SP027} used two different diagnostics, C75 and R90, to select MIR AGN among the sources with AllWISE fluxes. The C75 selection, focusing on achieving 75\% completeness, is defined as
\begin{equation}
    W_1 - W_2 > 0.71\,\rm{Vega\,mag}\;,
    \label{eq:C75}
\end{equation}
while the R90 diagnostic, optimised for obtaining 90\% reliability, is
\begin{equation}
    W_1 - W_2 > \begin{cases}
      0.65\,{\rm e}^{0.153\,(W_2-13.86)^2}\;, & W_2>13.86\;,\\
      0.65\;, & W_2 \leq 13.86 \;.\\
    \end{cases}
    \label{eq:R90}
\end{equation}
These criteria were accompanied by some extra conditions. We only considered sources with $W_1$ and $W_2$ magnitudes fainter than the saturation limits of the survey set as $W_1>8$ and $W_2>7$ (Vega magnitudes), with S/N$_{W_2}>5$, and not flagged as either artefacts or affected by artefacts, meaning that the \verb|cc_flags| are equal to zero \citepalias{Q1-SP027}. 
In the case of EDF-F and EDF-S, multiple WISE fluxes are available, including the AllWISE and the LegacyDR10 WISE fluxes. The main difference is that the latter are obtained through forced photometry at the locations of the Legacy Surveys optical sources, resulting in a larger number of matches with \Euclid counterparts. Moreover, the extra conditions of the \citetalias{assefWISEAGNCatalog2018} diagnostics are not easily applicable to the LegacyDR10 WISE data. Therefore, considering that the EDF-N has only AllWISE data, we decided to work with the AllWISE MIR data also for EDF-F and EDF-S. 
    
Sources with an X-ray counterpart identified by \citetalias{Q1-SP003}. Several X-ray surveys observed the EDFs, such as the XMM-{\it Newton} 4XMM-DR13 survey \citep{webbXMMNewtonSerendipitousSurvey2020}, the {\it Chandra} Source Catalogue v.2.0 \citep[CSC2;][]{evansChandraSourceCatalog2024}, and the eROSITA  DR1 Main sample \citep{predehlEROSITAXrayTelescope2021a, merloniSRGEROSITAAllsky2024}. \citetalias{Q1-SP003} identified Q1 counterparts from these X-ray surveys using the Bayesian algorithm {\tt NWAY} \citep{salvatoFindingCounterpartsAllsky2018}. The final product is a catalogue of Q1 sources matched with several X-ray point-like sources. This catalogue also includes spectroscopic redshift, if available, otherwise photo-$z$, X-ray luminosities ($L_{\rm X}$), and a galactic or extragalactic probability (\texttt{Gal\_proba}). We refer the reader to \citetalias{Q1-SP003} for more details about the optical-X-ray matching procedure and the catalogue generation. To select a pure sample of X-ray AGN, we selected only sources with \texttt{match\_flag} $=1$,  \texttt{Gal\_proba} $<0.5$, optical signal-to-noise ${\rm S/N}\geq2$, and $L_{\rm X}\geq 10^{42}\,{\rm erg}\,{\rm s}^{-1}$. This soft X-ray luminosity threshold is generally sufficient for isolating AGN from other X-ray sources \citep{airdXraysGalaxyPopulation2017}.

The \citetalias{Q1-SP027} multi-wavelength catalogue provided DESI spectroscopic counterparts for 42\,706 galaxies, and thus allowed for spectroscopic AGN detection. We ran several diagnostics to identify quasars (QSOs) and AGN candidates based on these spectroscopic data. To select QSOs, we utilised the DESI spectral-type classification \citep[\texttt{SPECTYPE=QSO};][]{desicollaborationEarlyDataRelease2024}. For sources classified as galaxies (\texttt{SPECTYPE=GALAXY}), we used several methods to identify AGN based on emission line fluxes, widths, and equivalent widths measured with \verb|FastSpecFit| \citep{moustakas2023ascl.soft08005M}. \citetalias[][]{Q1-SP027} reports the details of these measurements, available for 40\,274 of the DESI EDR Q1 sources. 
This sample was accompanied by SED fitting performed by \citet{siudekValueaddedCatalogPhysical2024}, which provided stellar masses and AGN properties. They only kept sources with an SED fit with a reduced $\chi^2<17$. With these criteria, we found 160 counterparts in our stellar-mass-limited sample. 
This threshold for $\chi^2$ was adopted from \citet{siudekValueaddedCatalogPhysical2024}, who, based on extensive visual inspections (their Appendix~D.2), defined it as optimal for ensuring reliable SED fits in their value-added catalogue. For our DESI-selected AGN sample, the vast majority (139 out of 160) have even higher quality fits, with $\chi^2<5$, and excluding the remaining 21 sources with $\chi^2$ between 5 and 17 does not qualitatively affect our results.
    
\citetalias[][]{Q1-SP015} trained a DL-based algorithm to perform image decomposition in the VIS imaging data and provide an estimate of the PSF contribution ($f_{\rm{PSF}}$) with respect to the total galaxy light in the observed flux. 
Following the same technique as in \citet{margalef-bentabolAGNHost2024}, \citetalias{Q1-SP015} fine-tuned the pre-trained DL-architecture \texttt{Zoobot} \citep{walmsleyZoobotAdaptableDeep2023} to predict $f_{\rm PSF}$ from galaxy images. The training set consisted of mock \Euclid observations of TNG galaxies, where central point sources were injected with randomly chosen $f_{\rm PSF}$ values in the range $[0,\,1)$. The resulting DL model achieves high accuracy and precision in recovering $f_{\rm PSF}$, with a mean bias $\langle f_{\rm PSF}{\rm [injected]} - f_{\rm PSF}{\rm [{\tt Zoobot}]}\rangle=-0.0078$, and a root mean square error of $0.051$. Only about 5\% of galaxies with $f_{\rm PSF}{\rm [injected]}<0.05$ have a predicted $f_{\rm PSF}{\rm [{\tt Zoobot}]}>0.2$. 
Here, we labelled as DL-based AGN those galaxies with $f_{\rm{PSF}}>0.2$, which corresponds to a 4$\sigma$ cut given the mean uncertainty in the $f_{\rm{PSF}}$ estimates, and ensures a high-purity sample. 
Although this $f_{\rm PSF}$ threshold aims for a high-purity selection of dominant central point sources, we acknowledge that the derived $f_{\rm PSF}$ may include contributions from nuclear star formation in addition to AGN, and that its completeness could be affected by heavy nuclear obscuration. A more detailed discussion of the method, its performance, validation, and comparison with traditional selection methods, is presented in \citet{margalef-bentabolAGNHost2024} and \citetalias{Q1-SP015}.
We also estimate the AGN luminosity in the VIS band ($L_{\rm PSF}$) by multiplying the total galaxy flux by the predicted $f_{\rm PSF}$ and converting it into a luminosity using the photometric redshift.

\begin{figure*}
    \centering
    \includegraphics[width=0.95\textwidth]{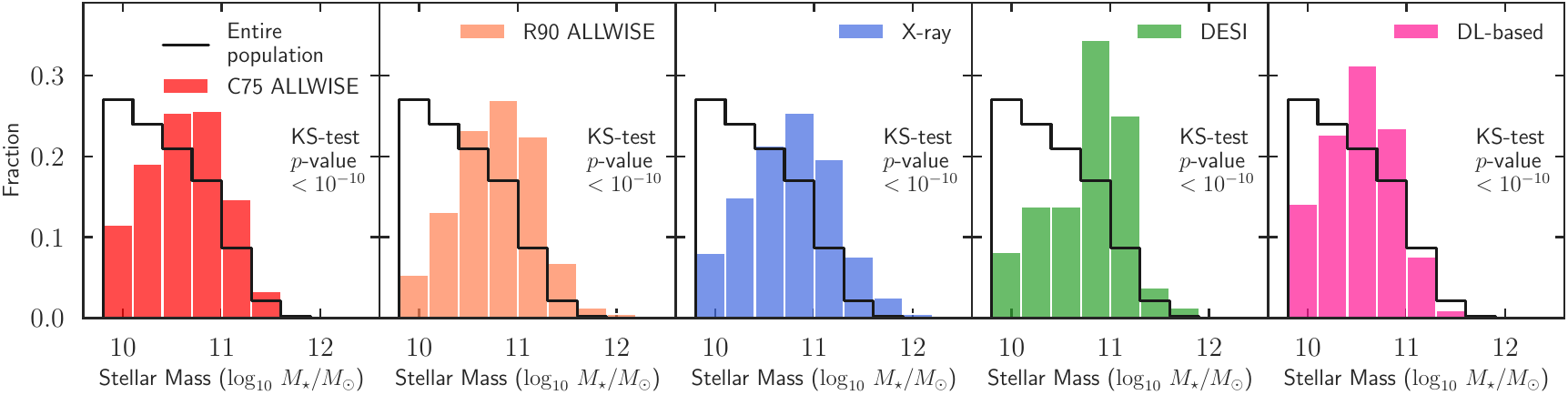}
    \includegraphics[width=0.95\textwidth]{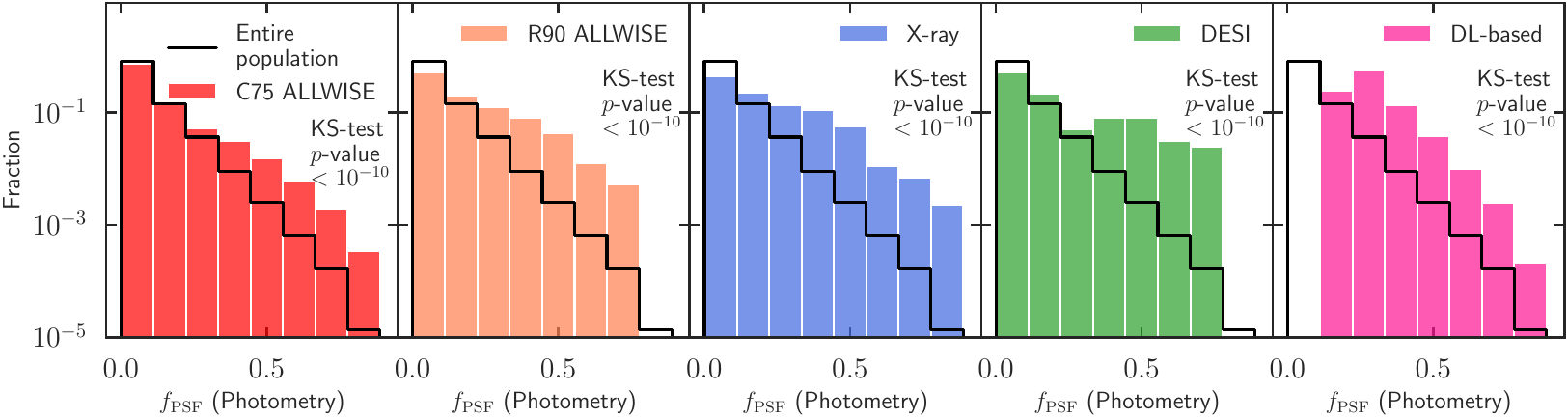}
    \caption{
    Stellar mass (upper row) and point-source contribution, estimated by the PSF fraction ($f_{\rm{PSF}}$, lower row) distributions for each AGN population. As a reference, we overlay the distribution for the entire sample of classified galaxies, including the selected AGN, in each panel. The distribution areas are normalised to unity. In each panel, we report the results of a two-sample KS test between the selected AGN and the entire galaxy sample.
    }
    \label{fig:agn_hist}
\end{figure*}

In total, our sample includes $28\,670$ classified galaxies hosting an AGN identified by at least one of the selection methods above.\footnote{Considering the whole sample, i.e. classified plus unclassified galaxies, about 39\% of AGN are labelled as unclassified, similar to the non-AGN population, where the unclassified fraction is 36\%.} The number of AGN identified per selection method is reported in Table~\ref{tab:agn_counts}. Many AGN have multiple detections. We show the intersection of the AGN selections in Appendix~\ref{app:AGN_sel}. 
The \citetalias[][]{Q1-SP027} Q1 AGN catalogue includes a variety of AGN selections, among which an AGN identification diagnostic based on \Euclid's photometry alone \citep{EP-Bisigello}. However, we did not consider this selection because its purity and completeness are poor in the absence of $u$-band observations. Moreover, this methodology is oriented towards obtaining a clean sample of quasars, which requires constraints on the point-like morphology of the source, which will bias against the detection of potential merging features. 

For DESI- and X-ray-selected AGN, we computed the AGN bolometric luminosity. In the first case, we utilised the $L_{\rm bol}$ estimated through SED fitting by \citet{siudekValueaddedCatalogPhysical2024}. The SED fitting was performed using Code Investigating GALaxy Emission \citep[{\tt CIGALE} v2022.1;][]{boquienCIGALEPythonCode2019, yangXCIGALEFittingAGN2020, yangFittingAGNGalaxy2022}, assuming for the AGN contribution the \citet{fritzRevisitingInfraredSpectra2006} templates. \citet{siudekValueaddedCatalogPhysical2024} found a close agreement between the $L_{\rm bol}$ derived from {\it Chandra} and SED fitting, with a median difference of $L_{\rm bol\, , SED}-L_{\rm bol\, , Chandra}\simeq -0.1\,\rm{dex}$.  
For the X-ray AGN, we used the X-ray luminosities ($L_{\rm X}$) from \citetalias{Q1-SP003}, which we converted into bolometric luminosities using the conversion factors from \citet{shenBolometricQuasarLuminosity2020}. Specifically, we used the double power law
\begin{equation}
    \frac{L_{\rm bol}}{L_{\rm X}} = c_1 \left( \frac{L_{\rm bol}}{10^{10}L_{\odot}} \right)^{k_1} + c_2 \left( \frac{L_{\rm bol}}{10^{10}L_{\odot}} \right)^{k_2} \;,
\end{equation}
where $c_1=5.712$, $k_1 = -0.026$, $c_2 = 12.60$, and $k_2=0.278$. We show how $L_{\rm bol}$ evolves with redshift for DESI and X-ray AGN in Fig.~\ref{fig:Lbol_vs_z}. X-ray sources are shown separately to highlight the difference in the survey characteristics \citepalias{Q1-SP003}. We note that 4XMM and CSC2 are deeper surveys compared to eROSITA, which covers a larger area but is biased towards brighter AGN ($L_{\rm bol}\gtrsim10^{45}\,{\rm erg}\,{\rm s}^{-1}$).

Different AGN selections correspond to different host galaxy properties \citep[e.g.,][]{silvermanEvolutionAGNHost2008}. We compare the stellar mass of the AGN candidates hosts for the different AGN selections in Fig.~\ref{fig:agn_hist}, top panels. Compared to the entire galaxy sample (active and non-active galaxies), AGN candidates reside in more massive galaxies, with their $M_{\star}$ distribution peaking at $10^{10.5}$--$10^{11}\,M_{\odot}$. 
Nevertheless, we should bear in mind that the stellar masses derived by \citet{Q1-SP031} or the official \Euclid pipeline do not consider the AGN component, which could bias the estimates of the $M_{\star}$ of AGN host galaxies. For example, the \citet{Q1-SP031} $M_{\star}$ are systematically lower by 0.07\,dex compared to those derived for the DESI sample, which included an AGN component in their SED fitting \citep{siudekValueaddedCatalogPhysical2024}.

To assess the statistical difference among the AGN populations and the entire galaxy sample, we ran a two-sample Kolmogorov-Smirnov test \citep[KS test;][]{hodgesSignificanceProbabilitySmirnov1958}. The KS test determines whether two samples come from the same parent distribution (null hypothesis). The $p$-value measures the probability of obtaining the observed difference between distributions, assuming the null hypothesis is true. If the $p$-value is below the significance level (here we take it to be 0.05), the difference between the two samples is statistically significant.
In each panel of Fig.~\ref{fig:agn_hist}, we report the resulting $p$-value between each AGN selection and the entire galaxy sample. The results confirm that the $M_{\star}$ distributions of AGN candidates are statistically different from the $M_{\star}$ distribution of the entire sample (active plus non-active galaxies). 

There are also some differences among the different AGN selections. DESI AGN live in extremely massive galaxies, with more than $70\%$ of these galaxies having $M_{\star}/M_{\odot}\geq 10^{11}$. X-ray AGN and R90 MIR AGN tend to be in slightly less massive galaxies, with average $M_{\star}/M_{\odot}\simeq10^{10.8}$, in agreement with previous studies \citep{bongiornoAccretingSupermassiveBlack2012, mountrichasGalaxyPropertiesType2021}. 
A KS test run to compare the $M_{\star}$ distributions of AllWISE R90, X-ray, and DESI AGN with each other, confirms this similarity ($p$-value$<0.05$).
DL-based and C75 MIR AGN inhabit the least massive galaxies, with average $M_{\star}/M_{\odot}\simeq10^{10.5}$. Also in this case, we found agreement with previous studies in the literature \citep[e.g.,][]{bornanciniPropertiesIRselectedActive2022}.
These differences might be due to selection biases. DESI AGN are spectroscopically selected and so naturally more likely to be in brighter, hence more massive, galaxies. Similarly, the difference between the stellar masses of R90 and C75 MIR AGN candidates hosts is expected because redder colours and brighter magnitudes are required \citepalias{assefWISEAGNCatalog2018} to select more reliable samples of AGN. The DL algorithm used to identify AGN components was trained using galaxies down to $10^9 M_{\odot}$. Thus, it is not surprising that this method allowed us to select AGN in less massive galaxies than the other methods. 

In Fig.~\ref{fig:agn_hist}, bottom panels, we compare the $f_{\rm{PSF}}$ distribution for the different AGN selections. As expected, all AGN types show a larger fraction of galaxies with higher $f_{\rm{PSF}}$ values compared to the entire galaxy sample. It is not surprising that the largest fraction of $f_{\rm{PSF}}\geq0.5$ galaxies is observed in DESI AGN, these being optically selected spectroscopic AGN. However, we might be missing the extremely dominant point sources ($f_{\rm PSF}>0.8$) because {\tt CIGALE} fails to estimate the stellar mass correctly when the AGN outshines the host galaxy. 
The KS tests confirm the difference between AGN candidate hosts and the entire galaxy population. 
This statistical difference confirms that the $f_{\rm PSF}$ parameter effectively isolates galaxies with a prominent central luminous component, characteristic of AGN activity, across all selection methods. It also provides an additional validation for the statistical reliability of our DL-based method in quantifying the AGN contribution.
In Appendix~\ref{app:AGN_sel}, we also compared the redshift distributions of each AGN selection.

\section{\label{sc:Results} Results}

In this section, we first construct control samples of mergers and non-mergers and AGN and non-AGN galaxies. Then, we investigate the merger and AGN relation by adopting a binary AGN--non-AGN classification and exploring continuous AGN parameters. All experiments are divided into two redshift bins, which are $0.5\leq z<0.9$, and $0.9\leq z \leq2.0$, with roughly equal numbers of AGN.

\subsection{Control pools}\label{sect:controls}
 
Proper control samples are crucial as AGN occurrence and the merger rate can depend on host galaxy properties such as stellar mass and redshift \citep[e.g.,][]{airdPRIMUSDependenceAGN2012,ferreiraGalaxyMergerRates2020}. Specifically, the merger and AGN control samples satisfy the following conditions:
\begin{align}\label{control_eq}
    &|z_{\rm control}-z_{\rm sample}| \leq 0.04\;z_{\rm sample}\, ,\\
    &|\logten(M_{\star, \rm control} / M_{\star, \rm sample}) | \leq 0.2\,\rm{dex}\, .
\end{align}
We chose these values according to the estimated normalised median absolute deviations for photo$-z$ and $M_{\star}$ \citep{Q1-SP031}.
These two conditions ensure that each galaxy (AGN) is compared with a sample of galaxies with similar redshift and stellar mass.
For each galaxy (AGN) in the original sample, we required at least ten counterparts that satisfy these criteria. When more than ten controls were found, we randomly picked ten of them. If there were fewer than ten controls, we iteratively increased our tolerances by a factor of 1.5 for each parameter. This operation was performed up to three times; otherwise, we rejected the galaxy (AGN). When constructing controls for AGN galaxies, we sampled from all galaxies that do not host any detected AGN, a pool of $330\,931$ possible galaxies. 
Non-AGN controls were constructed independently for each AGN selection. 

While matching in other physical parameters, such as star formation rate, could further refine the control samples, we have not included these in the current analysis. Specifically, star formation rate estimates available in \citetalias{Q1cite} are subject to considerable uncertainties at this stage \citep[e.g., normalised median absolute deviations of $\sim 0.45-0.64\,{\rm dex}$, as detailed in ][]{Q1-SP031}. Given these large uncertainties, attempting to match the star formation rate would not improve the control.

\subsection{Merger and AGN relation using a binary AGN classification}\label{sc:binary}

In the first set of experiments, we investigated whether mergers can trigger AGN by examining the incidence rate of AGN in mergers and non-merger controls, and whether they are the primary trigger, by comparing the merger fraction in AGN and non-AGN controls.

\subsubsection{AGN frequency in mergers and non-mergers}

\begin{figure}[h]
    \centering
    \includegraphics[width=0.49\textwidth]{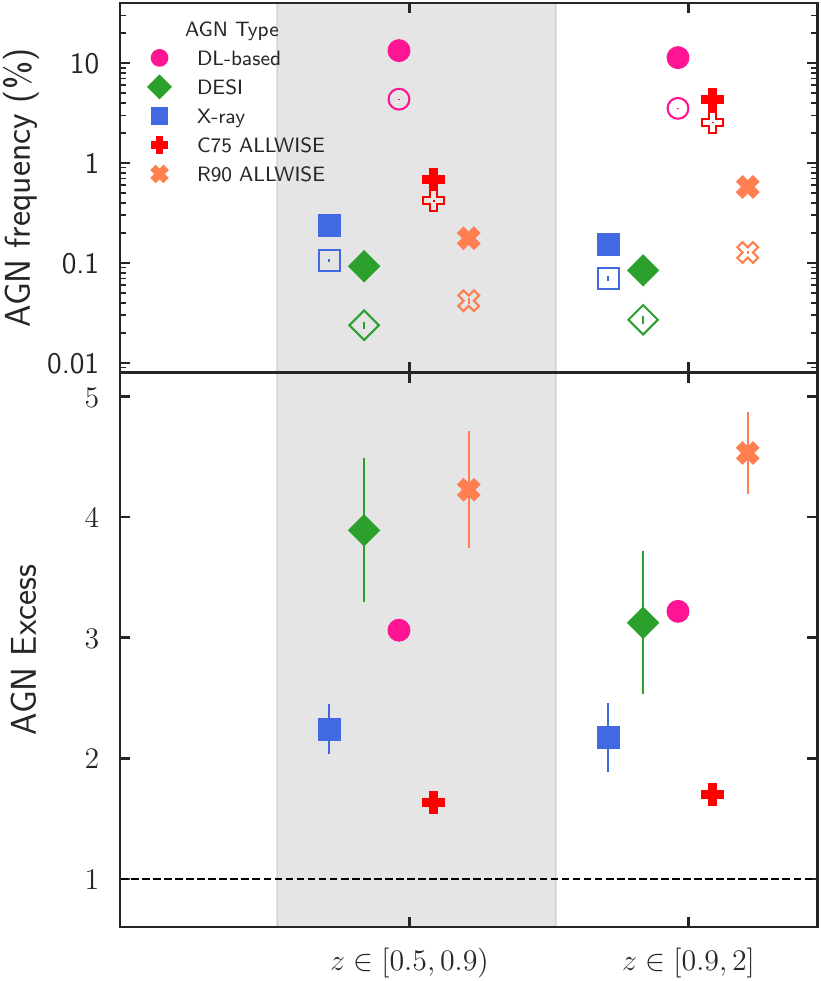}
    \caption{Active galactic nuclei frequency (Eq.~\ref{eq:agn_freq}) in mergers and non-merger controls in two redshift bins. 
    \emph{Top}: Frequency of AGN in mergers (filled symbols) and non-merger controls (empty symbols). 
    \emph{Bottom}: AGN excess in mergers compared to non-merger controls. The excess is the AGN frequency in mergers divided by that in the relative non-mergers. 
    }
    \label{fig:agn_freq}
\end{figure}

\begin{table*}[h]
\caption{Frequency of selected AGN types in mergers and non-merger controls (Eq.~\ref{eq:agn_freq}), divided into two redshift bins. }
\small
  \centering
  \begin{tabular}{lcccccc}
  \hline\hline \\[-7pt]
  & \multicolumn{3}{c}{$0.5\leq z < 0.9$} & \multicolumn{3}{c}{$0.9\leq z \leq 2.0$} \\
  AGN type & M & NM (control) & Excess & M & NM (control) & Excess \\
  \hline\\[-7pt]
  X-ray & $0.24\pm0.02\%$ & $0.106\pm0.004\%$ & $2.2\pm0.2$ & $0.15\pm0.02\%$ & $0.070\pm 0.004\%$ & $2.2\pm 0.3$ \\
  \citetalias{Q1-SP003} & (146/61\,393) & (652/614\,411) & & (66/43\,644) & (319/435\,959) & \\
  \hline\\[-7pt]
  DESI & $0.09\pm 0.01\%$ & $0.024\pm0.002\%$& $3.9\pm0.6$ & $0.08\pm0.01\%$ & $0.027\pm0.002\%$ & $3.1\pm0.6$ \\
  \citet{siudekValueaddedCatalogPhysical2024} & (57/61\,393) & (147/614\,411) & & (37/43\,644) & (118/435\,959) & \\
  \hline\\[-7pt]
  DL-based & $13.3\pm0.1\%$ & $4.36\pm0.03\%$ & $3.06\pm0.04$ & $11.4\pm0.1\%$ & $3.52\pm0.03\%$ & $3.22\pm0.05$ \\
  \citetalias{Q1-SP015} & (8196/61\,393) & (26\,779/614\,411) & & (4960/43\,644) & (15\,387/435\,959) & \\
  \hline\\[-7pt]
  C75 AllWISE & $0.69\pm 0.03 \%$ & $0.420\pm0.008\%$ & $1.63\pm0.09$ & $4.4 \pm 0.1 \%$ & $2.57\pm 0.02\%$ & $1.69\pm 0.04$ \\
  \citetalias{assefWISEAGNCatalog2018} & (420/61\,393) & (2578/614\,411) & & (1902/43\,644) & (11\,195/435\,959) & \\
  \hline\\[-7pt]
  R90 AllWISE & $0.18\pm0.02\%$ & $0.042\pm 0.003 \%$ & $4.2\pm 0.5$ & $0.58\pm 0.04\%$ & $0.127\pm0.005\%$ & $4.5\pm 0.3$ \\
  \citetalias{assefWISEAGNCatalog2018} & (109/61\,393) & (258/614\,411) & & (252/43\,644) & (555/435\,959) & \\
  \hline
  \end{tabular}
  \label{tab:agn_freq}
  \tablefoot{
  M (NM) indicates mergers (non-mergers). Fractions and relative errors are calculated using bootstrapping with resampling (1000 samples for each population). In brackets, we provide the number of AGN for each type, over the total number of mergers and non-merger controls. 
}
\end{table*}

The frequency of AGN in mergers and respective non-merger controls, per AGN type, is reported in Table \ref{tab:agn_freq} and shown in Fig.~\ref{fig:agn_freq}. 
The frequencies and relative { statistical} uncertainties are estimated using bootstrapping with resampling (1000 samples for each population). In both classes, the AGN frequency is defined as the ratio of identified AGN in the merger class to the total number of objects in the merger class:
\begin{align}\label{eq:agn_freq}
    & {\rm AGN\,\,frequency} = \frac{N_{\rm AGN}}{N_{\rm all}}\, .
\end{align}

For all AGN types, we observed a higher frequency of AGN in mergers than non-merger controls in both $z$ bins, demonstrating that mergers are a viable method to fuel accretion onto SMBHs. To show it more clearly, we calculated the AGN excess, defined as the ratio of the AGN frequency in mergers relative to non-mergers. The AGN excess is reported in Table~\ref{tab:agn_freq} and the lower panel of Fig.~\ref{fig:agn_freq}.
X-ray AGN show the same excess (2.2) relative to controls in both redshift bins. Similarly, DL-based, MIR C75 and R90 AGN have consistent excess in both $z$ bins, showing no clear signs of redshift evolution. On the contrary, DESI AGN go from an AGN excess of $3.9$ at $z<0.9$, to $3.1$ at $z>0.9$. However, this AGN selection shows larger uncertainties. Therefore, one must be cautious in inferring any redshift trends. Interestingly, the two MIR AGN selections exhibit completely different AGN excesses, with a much higher excess in the purer R90 selection. This could indicate that the C75 selection is highly contaminated by non-AGN galaxies. 
While these results indicate a clear excess of AGN activity in mergers, it is important to consider the potential impact of classification uncertainties inherent to automated methods, which we rigorously assess using detailed Monte Carlo (MC) simulations in Sect.~\ref{sect:MC}.

Our results are in agreement with previous studies that adopted the same AGN excess definition. The optical AGN excess we observed is consistent with the 3.7 AGN excess reported by \citet{bickleyAGNsPostmergersUltraviolet2023} for similar AGN in post-mergers, but it is much higher than the upper bound of 1.5 for the optical AGN excess found by \citet{gaoMergersTriggerActive2020} (sample selections in both works, $z<0.3$ and $M_{\star}>10^9\,M_{\odot}$). Regarding the X-ray AGN, the excess we found is comparable to the 1.8 excess found by \citet[][$z<0.3$ and $M_{\star}>10^8\,M_{\odot}$]{bickleyXrayAGNsSRG2024}, the 1.9 excess found by \citet[][at $0.5< z <0.8$ and $M_{\star}>10^9\,M_{\odot}$]{lamarcaDustPowerUnravelling2024}, and the 2.2 excess reported by \citet[][$0.5<z<1$ and $M_{\star}>10^8\,M_{\odot}$]{lacknerLateStageGalaxyMergers2014}. Nevertheless, \citet{lamarcaDustPowerUnravelling2024} observed a much lower X-ray AGN excess of 1.3 at $z<0.5$, while \citet{secrestXrayViewMergerinduced2020} found no statistically significant evidence for an X-ray AGN excess in post-mergers at $z<0.2$ and $M_{\star}>10^{9.5}\,M_{\odot}$. However, the latter work showed a much larger excess for MIR-selected AGN, suggesting that AGN in post-mergers are more likely to be heavily obscured. In fact, several other studies reported a larger excess of MIR AGN in mergers compared to non-merger controls, reaching a factor of 3--7 \citep[][]{gouldingGalaxyInteractionsTrigger2018, bickleyAGNsPostmergersUltraviolet2023, lamarcaDustPowerUnravelling2024}, which is in agreement with our results of the more reliable MIR AGN selection (R90). 

Our results and previous studies allow for some robust conclusions and some speculation. These findings robustly imply that major mergers trigger and fuel AGN, independently of AGN selection and the redshift. Considering that the purest MIR AGN show a larger excess than other AGN selections, we could speculate that mergers are more strongly connected to the triggering of dust-obscured AGN. For example, a major merger could redistribute gas and dust within a galaxy, increasing the dust obscuration surrounding the central active nucleus. This obscuration, if particularly heavy, might also make the detection of optical and soft X-ray AGN more challenging, partially explaining the lower excesses for these AGN.

\subsubsection{Merger fraction in AGN and non-AGN}

\begin{table*}[h]
\caption{Merger fraction ($f_{\rm merg}$) in active galaxies (AGN) and non-active galaxies (non-AGN controls) for different AGN selections divided into two redshift bins. }
\small
  \centering
  \begin{tabular}{lcccc}
  \hline\hline
  & \multicolumn{2}{c}{$0.5\leq z < 0.9$} & \multicolumn{2}{c}{$0.9\leq z \leq 2.0$} \\
  AGN type & $f_{\rm merg}$(AGN) & $f_{\rm merg}$(non-AGN controls) & $f_{\rm merg}$(AGN) & $f_{\rm merg}$(non-AGN controls) \\
  \hline\\[-7pt]
  X-ray & $51\pm3\%$ & $24.7\pm0.8\%$ & $44\pm4\%$ & $27\pm1\%$ \\
  \citetalias{Q1-SP003} & (146/288) & (726/2935) & (66/149) & (386/1435) \\
  \hline\\[-7pt]
  DESI & $59\pm5\%$ & $26\pm1\%$ & $59\pm6\%$ & $26\pm2\%$ \\
  \citet{siudekValueaddedCatalogPhysical2024} & (57/97) & (257/982) & (37/63) & (161/618) \\
  \hline\\[-7pt]
  DL-based & $57.3\pm 0.4\%$ & $27.9\pm0.1\%$ & $54.9\pm0.5\%$ & $26.4\pm0.1\%$ \\
  \citetalias{Q1-SP015} & (8196/14\,313) & (39\,886/143\,223) & (4906/9025) & (23\,777/90\,157) \\
  \hline\\[-7pt]
  C75 AllWISE & $40\pm2\%$ & $27.4\pm0.4\%$ & $40.7\pm0.7\%$ & $26.9\pm0.2\%$ \\
  \citetalias{assefWISEAGNCatalog2018} & (420/1041) & (2848/10\,404) & (1902/4671) & (12\,558/46\,716) \\
  \hline\\[-7pt]
  R90 AllWISE & $64\pm4\%$ & $27\pm1\%$ & $65\pm2\%$ & $27.3\pm0.7\%$ \\
  \citetalias{assefWISEAGNCatalog2018} & (109/170) & (454/1705) & (252/386) & (1052/3855) \\
  \hline
  \end{tabular}
  \label{tab:merg_frac}
  \tablefoot{
  Fractions and relative errors are calculated using bootstrapping with resampling (1000 samples for each population). The numbers of AGN for each type, relative to the total number of mergers and non-merger controls in each $z$-bin, are provided in brackets. 
}
\end{table*}

\begin{figure}[h]
    \centering
    \includegraphics[width=0.49\textwidth]{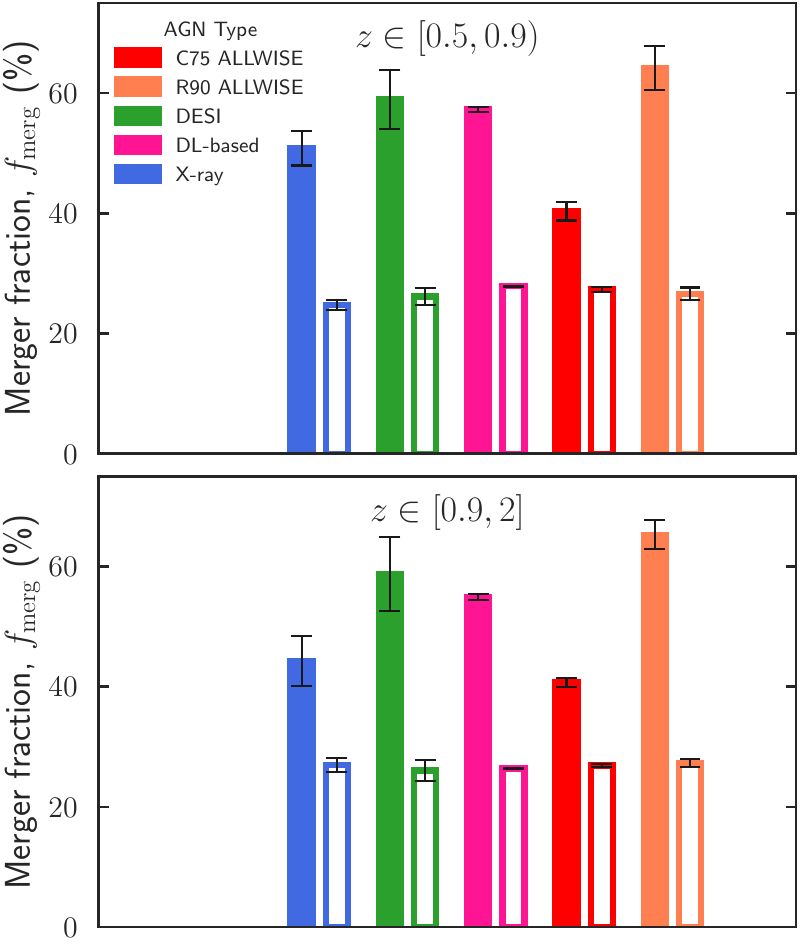}
    \caption{Merger fraction for all AGN types (filled bars) and relative non-AGN controls (empty bars), divided into two redshift bins. The fraction of mergers is higher in the AGN samples than in the non-AGN controls for all AGN types in both bins. }
    \label{fig:merg_frac}
\end{figure}

Table~\ref{tab:merg_frac} and Fig.~\ref{fig:merg_frac} report the merger fraction ($f_{\rm merg}$, Eq.~\ref{eq:f_merg}) in AGN and corresponding non-AGN controls, divided by AGN type.
In both $z$ bins, and for each AGN type, the fraction of mergers is higher for AGN than for non-AGN controls, reinforcing the merger-AGN connection. All AGN types, excluding the X-ray-selected ones, do not show signs of redshift evolution for the $f_{\rm merg}$ in AGN and non-AGN. The difference in $f_{\rm merg}$ for X-ray AGN at $z<0.9$ and $z\geq 0.9$ is within $2\sigma$ uncertainty. Across the entire redshift range, X-ray, DESI, DL-based, and R90 MIR AGN predominantly inhabit merging galaxies, with $f_{\rm merg}$ ranging from 44\% to 65\%. Only in the case of C75 MIR AGN, we reported a merger fraction of 40$\%$, which might indicate a possibly higher contamination degree in this selection. On the other hand, non-AGN controls are classified as mergers in 25--28$\%$ of the cases, about a factor of two less frequently than for the AGN host galaxies. 
To confirm that these findings are not an artefact of imperfections in the merger classification process, we evaluate the influence of classification uncertainties on these merger fractions through the MC analysis detailed in Sect.~\ref{sect:MC}.

Low-redshift studies have found that the fraction of mergers in the MIR-selected AGN is a factor of 1.5--2.3 larger than that of non-AGN controls, in agreement with our results \citep[$z<0.8$ and $M_{\star}>10^9\,M_{\odot}$;][]{ellisonDefinitiveMergerAGNConnection2019, gaoMergersTriggerActive2020, lamarcaDustPowerUnravelling2024}. Likewise, \citet{donleyEvidenceMergerdrivenGrowth2018} found that IR-only AGN out to $z=5$ are more likely to be classified as irregular, asymmetric, or interacting than as regular galaxies. For optically selected AGN, \citet{ellisonDefinitiveMergerAGNConnection2019} found a merger fraction in AGN twice as large as that in non-AGN, similar to what we observe for the DESI AGN. In contrast, \citet{gaoMergersTriggerActive2020} reported an excess of $f_{\rm merg}$ in optical AGN of a factor below 1.5. For X-ray-selected AGN, \citet{bickleyXrayAGNsSRG2024} reported an $f_{\rm merg}$ excess of a factor of 2, while \citet{lamarcaDustPowerUnravelling2024} found an excess of $1.3$ at $z\leq0.5$ and of $1.8$ at $0.5<z<0.8$, close to our findings. At higher redshift ($1\leq z \leq 2$), other studies uncovered only a marginally higher fraction of mergers in X-ray AGN compared to non-AGN, comparable with no excess at all \citep[$M_{\star}\gtrsim 10^{9.5}\,M_{\odot}$;][]{cisternasBulkBlackHole2011, kocevskiCANDELSConstrainingAGNMerger2012, marianMajorMergersAre2019}, although these samples are limited to intermediate AGN luminosity (X-ray luminosity $10^{42}< L_{\rm X} < 10^{44}\,{\rm erg\,s^{-1}}$). Recently, \citet{villforthCompleteCatalogueMerger2023} reviewed several studies in the literature about the merger fraction in AGN and non-AGN controls. They concluded that $f_{\rm merg}$ in X-ray-selected AGN are consistent with no excess over controls, in contrast with our findings, while for optically selected AGN, there is an excess over control samples, in agreement with our results for DESI AGN. 

Although the elevated merger fractions observed in AGN hosts point to a connection between mergers and nuclear activity, caution is needed when interpreting the role of mergers as a primary triggering mechanism. First, a non-negligible merger fraction (25--28\%) is also observed in the non-AGN control samples. Second, our analysis is limited to classified galaxies and does not include sources below our detection and classification thresholds, which may introduce biases and incompleteness. Therefore, our results do not allow us to quantify the exact contribution of mergers relative to other triggering channels.

\subsection{The merger and AGN connection using continuous parameters}\label{sc:continuous}

In this second set of experiments, we examined the merger-AGN connection using continuous parameters, which characterise either the relative or the absolute AGN power. Specifically, we first analysed the PSF fraction, $f_{\rm{PSF}}$, which assesses the power of an AGN relative to its host galaxy. Then, we concentrated on the AGN luminosity for studying the absolute AGN power.

\subsubsection{Dependence on the relative AGN power}

\begin{figure}
    \centering
    \includegraphics[width=0.49\textwidth]{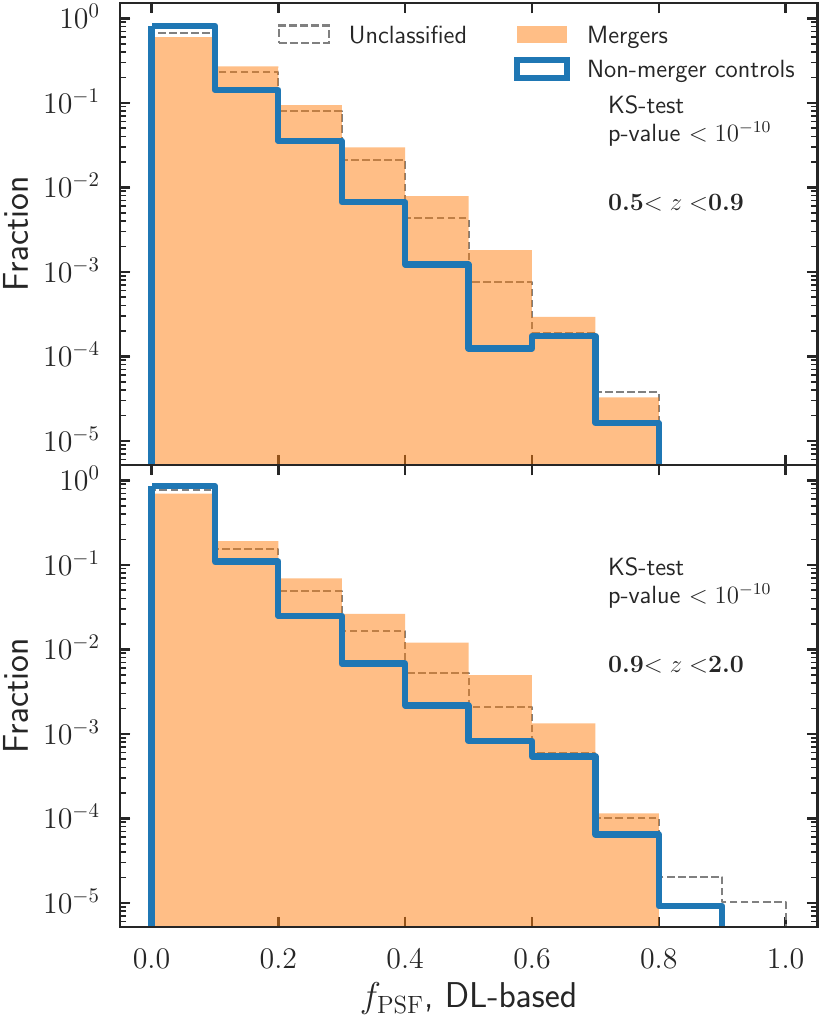}
    \caption{Normalised distributions of the PSF fraction for mergers and non-mergers, in the two redshift bins. The results of a two-sample KS test are reported in each panel. The $f_{\rm{PSF}}$-normalised distribution for unclassified galaxies is overlaid as a comparison. }
    \label{fig:fagn_hist}
\end{figure}

\begin{figure}
    \centering
    \includegraphics[width=.49\textwidth]{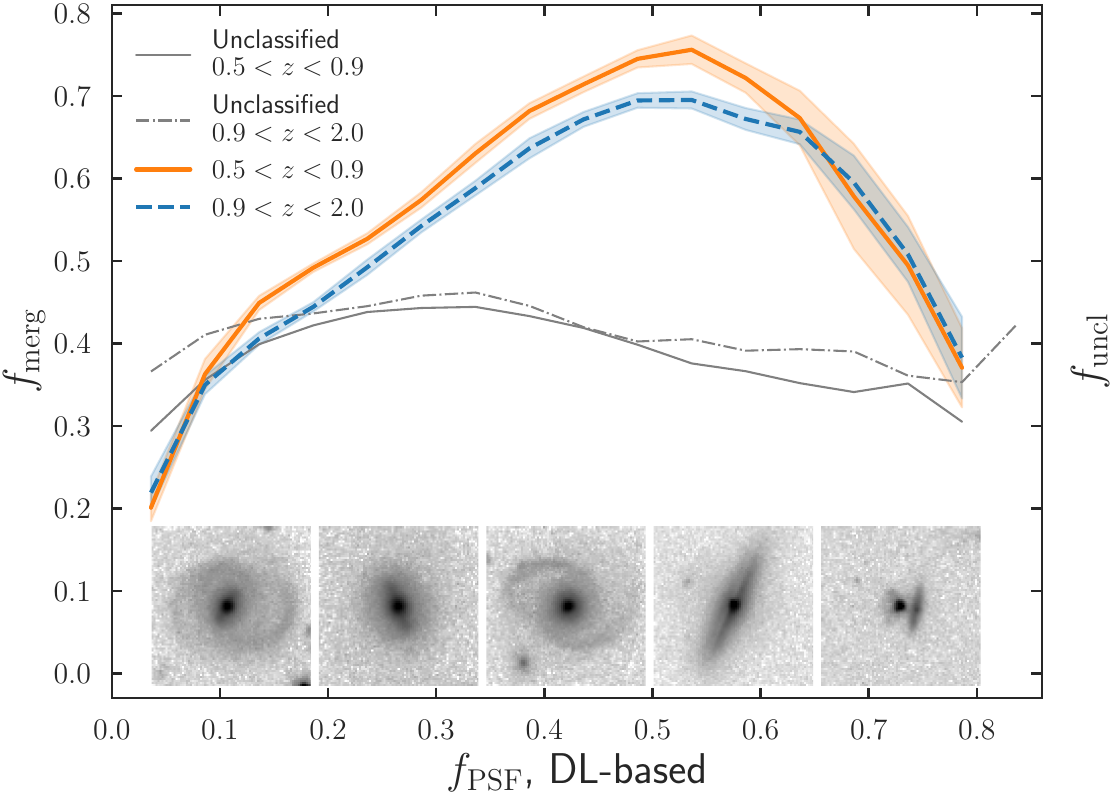}
    \caption{Merger fraction and PSF contribution fraction relationship for the two redshift bins considered. All AGN are included. Trend lines represent the running median, while the shaded areas are one standard deviation. Examples of \Euclid images of galaxies with increasing levels of $f_{\rm{PSF}}$ are shown at the bottom. Cutouts are $8\arcsec \times 8\arcsec$, log-scaled in the $1{\rm st}$--$99{\rm th}$ percentile range. Grey lines indicate the fraction of unclassified objects as a function of $f_{\rm{PSF}}$. 
    }
    \label{fig:fmerg_fagn}
\end{figure}

Here we analysed the connection between mergers and the PSF fraction parameter, $f_{\rm{PSF}}$, which measures the relative nuclear power. We reported the $f_{\rm{PSF}}$ normalised distributions for mergers and relative non-merger control galaxies in Fig.~\ref{fig:fagn_hist}. 
Mergers show a larger fraction of galaxies in the range $0.1 \leq f_{\rm PSF}<0.8$ than non-merger controls, in both redshift bins. The only exception is represented by the $f_{\rm PSF}>0.8$ galaxies at $z\geq0.9$.
We show the results of KS tests in each panel of Fig.~\ref{fig:fagn_hist}. The KS test output strongly excludes the null hypothesis, that is, the difference between the $f_{\rm{PSF}}$ distribution for mergers and non-merger controls is statistically significant, in both $z$ bins. This hints towards a scenario where mergers fuel the accretion onto the SMBH, enhancing its accretion rate and, consequently, the point-source luminosity and contribution to the total galaxy light. 

We present the merger fraction versus $f_{\rm{PSF}}$ relationship for all galaxies in Fig.~\ref{fig:fmerg_fagn}, divided into redshift bins.  We calculated the merger fraction in $N$ $f_{\rm{PSF}}$ bins, logarithmically spaced in the range 0--0.86 (the maximum $f_{\rm{PSF}}$ in our sample). The number of bins $N$ is randomly sampled between 6 and 20. Bootstrapping with resampling is used (1000 samples for each population). The trends reported represent the running median of all outcomes and the respective $1\sigma$ uncertainties, for each population. 
A clear trend emerges for both redshift bins. From $f_{\rm{PSF}} = 0$ up to $f_{\rm{PSF}} = 0.55$, the fraction of mergers monotonically increases, from $f_{\rm merg}=0.2$ to $f_{\rm merg}\simeq 0.7$. After this peak value, the merger fraction declines with increasing $f_{\rm{PSF}}$, down to $f_{\rm merg} = 0.4$ at $f_{\rm{PSF}}=0.8$. Uncertainties become larger with increasing $f_{\rm{PSF}}$, mostly due to fewer galaxies in those bins. In the range $0.2 \leq f_{\rm{PSF}} \leq 0.75$ mergers appear to be the dominant mechanism to trigger AGN ($f_{\rm merg}>0.5$).

The trend inversion for $f_{\rm PSF}=0.55$, where $f_{\rm merg}$ begins to decline despite increasing PSF dominance, presents a complex interpretive challenge. We specifically investigated if this decline is primarily due to the dominant PSF outshining the host galaxy's morphological features, thereby making merger identification more difficult. To test this, we performed an experiment using a randomly selected sample of $\sim 1500$ galaxies without an injected PSF from our test set. Focusing on the dominant regime ($f_{\rm PSF}>0.55$), we re-classified simulated galaxies after injecting a prominent PSF component into the host galaxies from this sample. Our analysis revealed that only approximately 4\% of simulated galaxies originally classified as mergers were re-labelled as non-mergers or unclassified after a dominant PSF was added. This indicates that while a dominant PSF can marginally hinder morphological classification, it is not the primary driver for the observed steep decline in $f_{\rm merg}$ at the highest $f_{\rm PSF}$ values. 
An alternative explanation might lie in the fact that $f_{\rm PSF}$ is a relative quantity. Although extremely dominant, a point source could be faint in absolute terms, in which case mergers might play a minor role, as we show in the next section. 

To investigate possible differences among the various AGN selections, we analysed in Appendix~\ref{app:AGN_analysis} the $f_{\rm merg}$ versus $f_{\rm{PSF}}$ relation for individual AGN types. The C75 MIR AGN show a trend similar to the whole galaxy population, while the trend is less clear in the case of the DESI and X-ray AGN, probably due to the lower number statistics. R90 MIR AGN have a very high $f_{\rm merg}$ ($>60\%$) for the whole $f_{\rm PSF}$ range.

In \citet{lamarcaDustPowerUnravelling2024}, the authors estimated the relative AGN power, the AGN fraction parameter ($f_{\rm{AGN}}$), through SED fitting. This $f_{\rm{AGN}}$ is the fraction of light emitted by the AGN component over the total galaxy light, in the wavelength range 3--30\,\micron. 
This particular wavelength range was chosen as it robustly probes the re-emission from warm dust in the AGN torus, making it a reliable indicator of AGN activity, particularly for obscured sources \citep{hickoxObscuredActiveGalactic2018}.
They presented an $f_{\rm merg}$ versus $f_{\rm{AGN}}$ relation with two regimes, for all AGN types considered: $f_{\rm merg}$ is rather flat as a function of $f_{\rm{AGN}}$ for relatively subdominant AGN, then it steeply rises above $50\%$ for the most dominant AGN ($f_{\rm{AGN}}\geq 0.8$). Although we estimated the AGN relative contribution through photometry, there are some similarities. First, for less dominant AGN ($f_{\rm{PSF}}\leq 0.2$), mergers are not the main AGN triggering mechanism. Second, major mergers are the principal pathway to fuel more dominant AGN. Yet, some differences exist. The $f_{\rm merg}$ versus $f_{\rm{PSF}}$ relation does not show any flat regime, but rather $f_{\rm merg}$ constantly increases, and subsequentially decreases, as a function of $f_{\rm{PSF}}$. 
Overall, these results support the idea that mergers can enhance AGN fuelling and are the prevailing mechanism for producing dominant AGN with respect to their host galaxy. In contrast, less dominant AGN may be primarily fuelled by other mechanisms, such as secular processes { \citep[e.g.,][]{cisternasBulkBlackHole2011, schawinskiHSTObservations2011, treisterMajorGalaxyMergers2012}}.

\subsubsection{Dependence on the absolute AGN power}\label{sect:LPSF}

\begin{figure}
    \centering
    \includegraphics[width=0.49\textwidth]{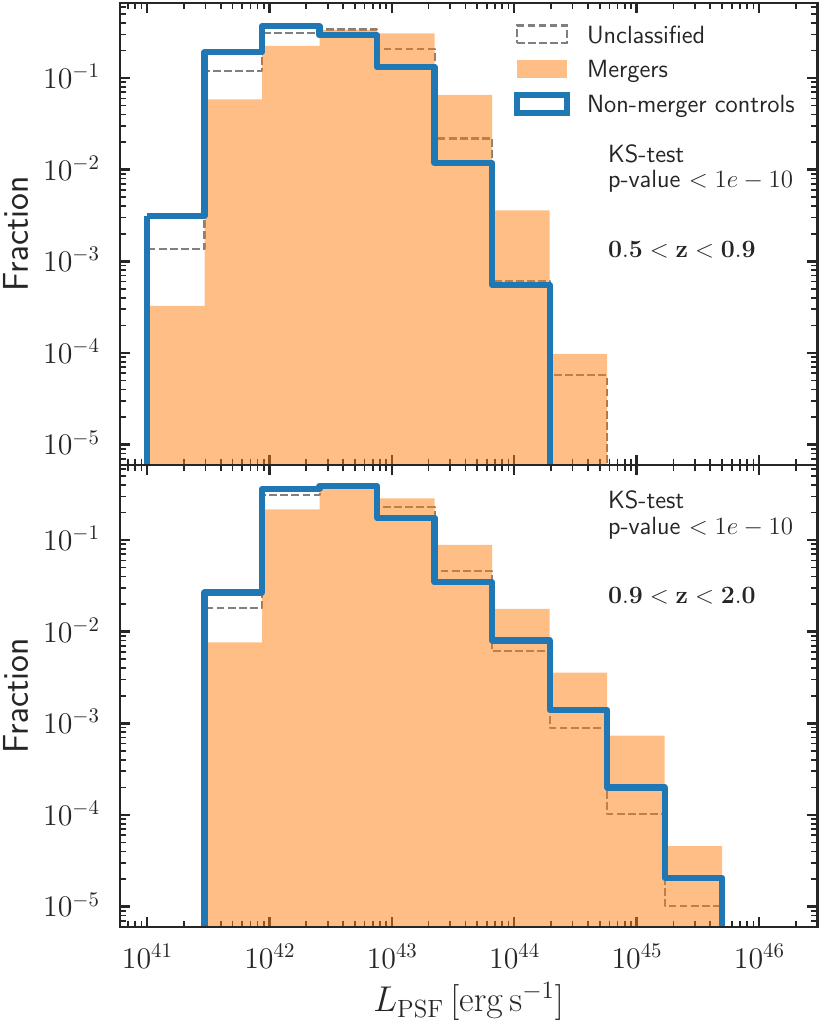}
    \caption{Normalised distributions of the PSF luminosity for mergers and non-mergers, in the two redshift bins. The results of a two-sample KS test are reported in each panel. The $f_{\rm{PSF}}$-normalised distribution for unclassified galaxies is overlaid as a comparison.
    }
    \label{fig:LPSF_hist}
\end{figure}

\begin{figure*}
    \sidecaption
    \includegraphics[width=0.63\textwidth
    ]{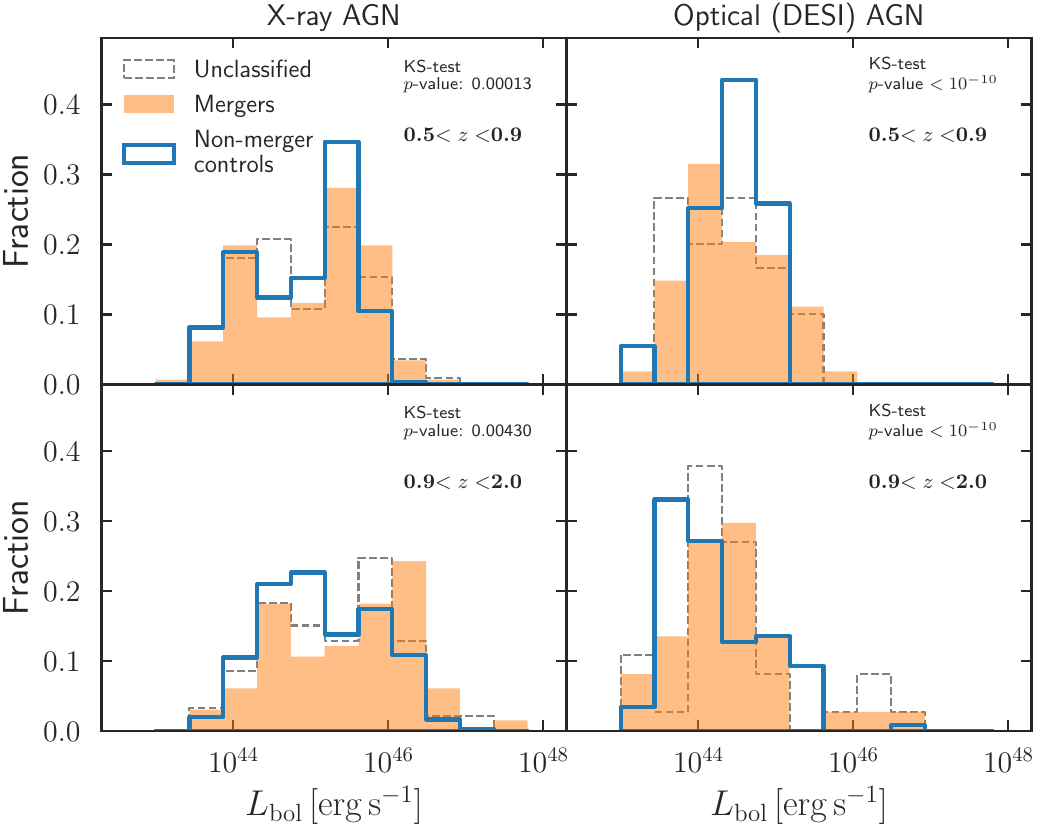}
    \caption{Normalised distributions of the AGN bolometric luminosity for mergers and non-merger controls in the two redshift bins for the X-ray AGN (\emph{left} column) and the DESI-selected AGN (\emph{right} column). 
    We report the results of a two-sample KS test in each panel. The $f_{\rm{PSF}}$ normalised distribution for unclassified galaxies is overlaid as a comparison.
    }
    \label{fig:Lbol_hist}
\end{figure*}

Next, we analysed the AGN luminosity parameters, which trace the absolute AGN power. Specifically, we focused on the point source luminosity, $L_{\rm{PSF}}$, and the bolometric luminosity, $L_{\rm bol}$, where the latter is available only for the X-ray and DESI AGN (see Sect.~\ref{sc:Method} for details on the derivation of $L_{\rm bol}$). 

We present the normalised distributions of $L_{\rm{PSF}}$ for mergers and non-merger controls in Fig.~\ref{fig:LPSF_hist}. In both redshift bins, we observed a higher fraction of mergers at $L_{\rm{PSF}}>10^{43}\,{\rm erg}\,{\rm s}^{-1}$ compared to non-merger controls. Therefore, mergers are more likely to harbour a bright AGN than the relative non-merger control galaxies. We show the normalised $L_{\rm bol}$ distribution for mergers and non-merger controls in Fig.~\ref{fig:Lbol_hist}. In the case of the X-ray AGN, mergers and non-mergers have similar $L_{\rm bol}$ distributions, with some differences at the very bright end in both $z$ bins. Indeed, a larger fraction of mergers host a bright AGN ($L_{\rm bol}\geq 10^{46}\,{\rm erg}\,{\rm s}^{-1}$) compared to non-mergers. The KS test confirms such a difference in both $z$ bins. In the case of the DESI AGN, this difference emerges at lower luminosities, at $L_{\rm bol}\geq 10^{45}\,{\rm erg}\,{\rm s}^{-1}$ for the $0.5\leq z < 0.9$ bin, and at $L_{\rm bol}\geq 10^{44}\,{\rm erg}\,{\rm s}^{-1}$ for the $0.9\leq z \leq 2.0$ bin. For both X-ray and DESI AGN, in both redshift bins, the brightest AGN seem to inhabit almost exclusively interacting galaxies, hinting towards a picture where major mergers are responsible for fuelling the most powerful AGN. 
We evaluated the significance of this overabundance by performing a two-proportion z-test, comparing the fraction of bright X-ray ($L_{\rm bol}>5\times10^{45}\,{\rm erg\,s^{-1}}$) and DESI ($L_{\rm bol}>10^{45}\,{\rm erg\,s^{-1}}$) AGN in mergers and non-merger controls, in both redshift bins. The results showed that these differences are statistically significant ($p$-value$<0.05$) in all cases, except for the DESI AGN in the $0.9\leq z \leq 2.0$ bin ($p$-value$=0.43$). 

\begin{figure}
    \centering
    \includegraphics[width=0.495\textwidth]{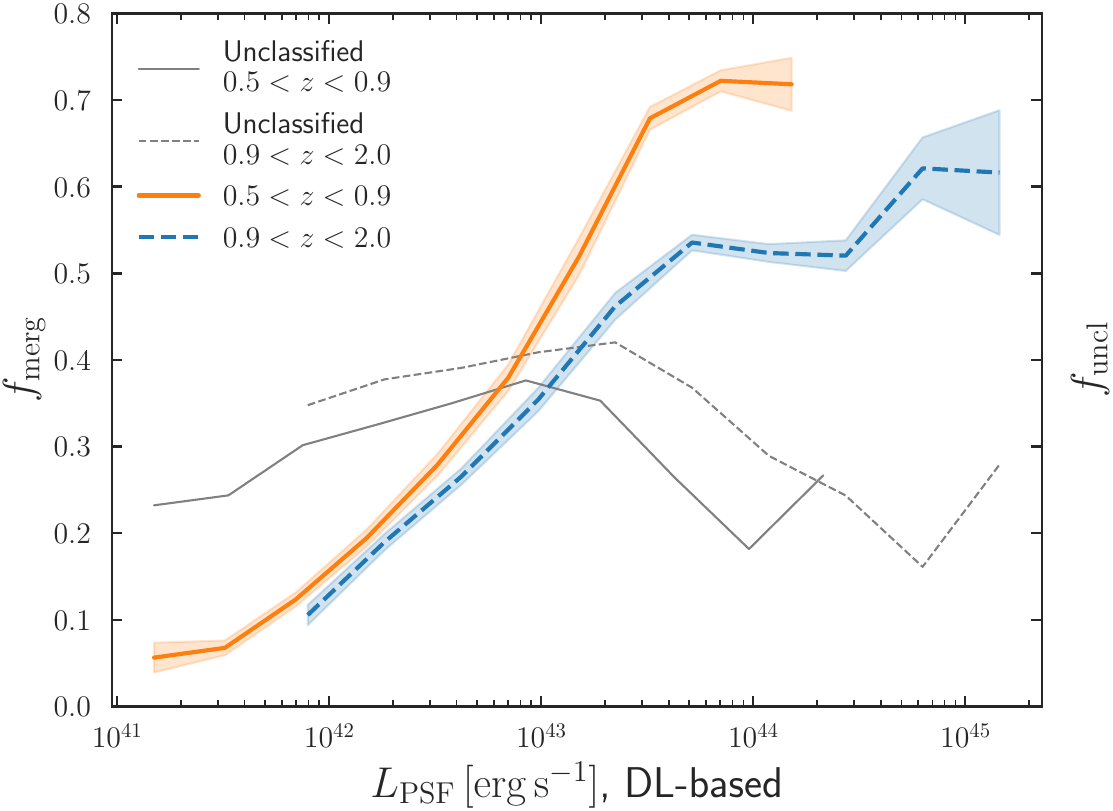}
    \caption{Merger fraction as a function of the PSF luminosity, $L_{\rm{PSF}}$, measured through photometry. The shaded areas show the uncertainties obtained through bootstrapping. The solid and dashed grey lines indicate the fraction of unclassified objects as a function of $L_{\rm PSF}$.
    }
    \label{fig:fmerg_Lagn_phot}
\end{figure}

\begin{figure}
    \centering
    \includegraphics[width=0.48\textwidth]{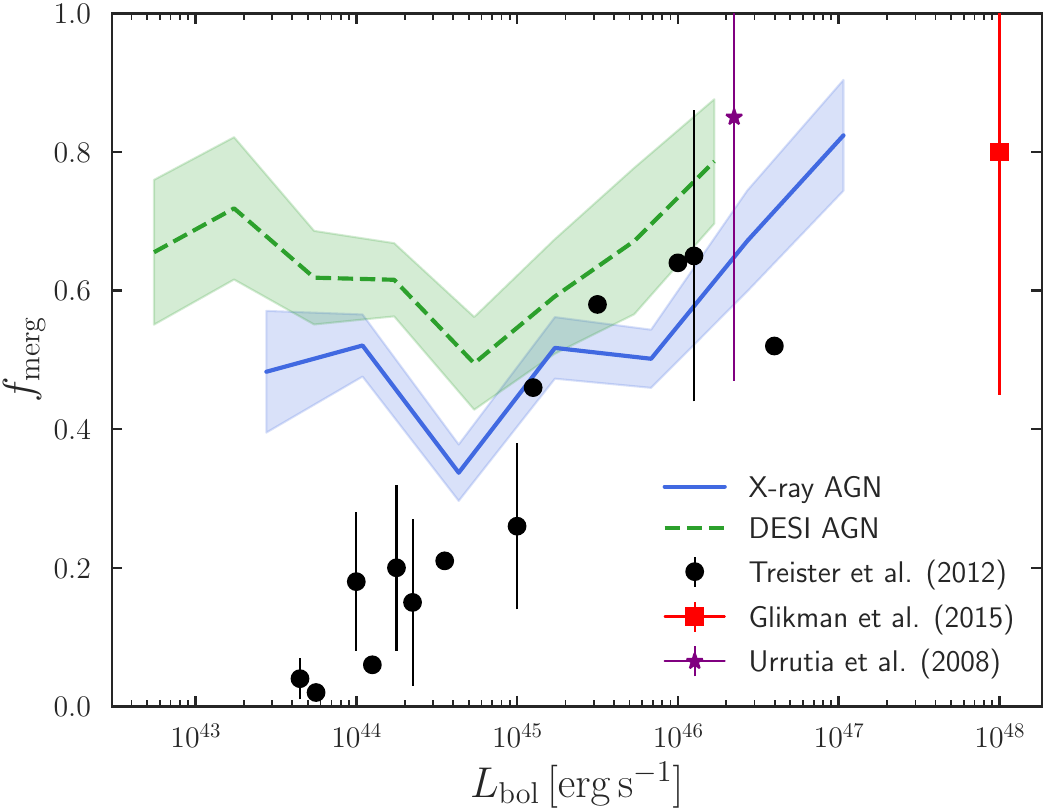}
    \caption{Merger fraction and bolometric AGN luminosity relation for the X-ray and DESI-selected AGN. Trend lines show the running median, and shaded areas are one standard deviation.
    We include data from \citet[][black circles]{treisterMajorGalaxyMergers2012}, \citet[][purple star]{urrutiaEvidenceQuasarActivity2008}, and \citet[][red square]{glikmanMajorMergersHost2015}, with the associated $f_{\rm merg}$ uncertainties, if available.
    }
    \label{fig:fmerg_Lbol}
\end{figure}

We plot the merger fraction as a function of $L_{\rm{PSF}}$ in Fig.~\ref{fig:fmerg_Lagn_phot}, { using the same methodology as for the $f_{\rm PSF}$-$f_{\rm merg}$ relation}. For both redshift bins, $f_{\rm merg}$ increases as a function $L_{\rm{PSF}}$. At $z<0.9$, $f_{\rm merg}$ show a steeper monotonic rise, with most of the galaxies being in mergers at $L_{\rm{PSF}} \simeq 10^{43.5}\,{\rm erg}\,{\rm s}^{-1}$. This happens towards higher luminosities at $0.9\leq z \leq 2.0$, when we observe a flat $f_{\rm merg}$ in the range $10^{43}$--$5\times 10^{44}\,{\rm erg}\,{\rm s}^{-1}$, and mergers become prevalent only for the very bright end of the AGN population, $L_{\rm{PSF}}>10^{45}\,{\rm erg}\,{\rm s}^{-1}$. This might indicate that at higher redshift, when larger gas supplies are available within galaxies \citep{tacconiHighMolecularGas2010}, major mergers are less important in fuelling bright AGN. At the same time, at $z<0.9$, when less gas is available, mergers might be the sole viable path to fuel such powerful AGN. We perform the same analysis but for individual AGN selections in Appendix~\ref{fig:fmerg_Lagn_agn_type}.

To quantify the statistical significance of these apparent redshift differences in the $f_{\rm merg}$ versus $L_{\rm PSF}$ relation, we performed two-proportion z-tests comparing the merger fractions in four equally spaced $L_{\rm PSF}$ bins spanning the range $10^{42}$ to $10^{44}\, {\rm erg\,s^{-1}}$, where the data from both redshift samples overlap. In all four luminosity bins tested, we find that the merger fraction in the lower redshift bin ($0.5 \leq z < 0.9$) is significantly higher than in the higher redshift bin ($0.9 \leq z \leq 2.0$), with $p$-values $< 0.001$. This statistically confirms the visual impression from Fig.~\ref{fig:fmerg_Lagn_phot} that, at a given $L_{\rm PSF}$, mergers are more prevalent at lower redshifts in our sample. However, it is important to note that when the uncertainties arising from the merger classification process are considered (as detailed in our MC simulations in Sect.~\ref{sect:MC}), the shaded error regions for the two redshift trends show considerable overlap. This suggests that the true underlying difference might be less pronounced once the full impact of potential misclassifications is taken into account.

Finally, we present the merger fraction versus AGN bolometric luminosity relationship for DESI and X-ray AGN, and previous literature results \citep{urrutiaEvidenceQuasarActivity2008, treisterMajorGalaxyMergers2012, glikmanMajorMergersHost2015}, in Fig.~\ref{fig:fmerg_Lbol}. As for the $f_{\rm merg}$ and $f_{\rm{PSF}}$ relation, we calculated the $f_{\rm merg}$ in $N$ $L_{\rm bol}$-bins, randomly sampled in the range $10^{42}$--$10^{47}\,{\rm erg}\,{\rm s}^{-1}$. Bootstrapping with resampling was used to estimate uncertainties. In both cases, given that there are only a few numbers in each bin, we report large uncertainties, of the order $f_{\rm merg}=0.1-0.15$. These large uncertainties do not allow for strong conclusions to be drawn. X-ray AGN show a clear trend: the fraction of mergers increases with increasing luminosity. DESI AGN show a less clear trend, having an initially ($L_{\rm bol}\leq10^{45}\,{\rm erg}\,{\rm s}^{-1}$) decreasing merger fraction followed by a steady increase with increasing $L_{\rm bol}$. In both cases, major mergers appear as the dominant triggering mechanism of the most luminous AGN. 

Our finding that the $f_{\rm merg}$ increases with bolometric AGN luminosity, particularly for the X-ray AGN sample, and that $f_{\rm merg}$ is particularly high for the most luminous AGN ($L_{\rm bol}\gtrsim 10^{45.5}\,{\rm erg\,s^{-1}}$) is consistent with previous work. For instance, \citet{urrutiaEvidenceQuasarActivity2008} and \citet{glikmanMajorMergersHost2015} found very high merger fractions, $f_{\rm merg}>80\%$, for luminous, dust-reddened quasars. Similarly, \citet{treisterMajorGalaxyMergers2012} found similar results and argued that major mergers are essential for fuelling the most luminous AGN. 
While direct comparison of absolute merger fractions is challenging due to different merger identification techniques, AGN selection methods, and redshift ranges, the qualitative trend of mergers playing an increasingly dominant role at higher AGN luminosities is a common theme \citep[see also][]{donleyEvidenceMergerdrivenGrowth2018, ellisonDefinitiveMergerAGNConnection2019, lamarcaDustPowerUnravelling2024}.
For producing such powerful emissions, a large amount of matter must be fed to the central SMBH, and major mergers are an efficient way of bringing large amounts of gas to the centres of galaxies \citep{blumenthalGoFlowUnderstanding2018}.


\section{Caveats discussion}\label{sc:caveats}
In this section, we investigate the main factors that might influence the relationship between mergers and the AGN properties characterised by the continuous parameters $f_{\rm PSF}$ and $L_{\rm PSF}$, potentially affecting our results and conclusions. We discuss the role of unclassified galaxies, the dependency on stellar mass, and the systematics in our classification.

\subsection{The unclassified galaxies}
First, we examined the impact of the unclassified galaxies on the observed trends. Figure~\ref{fig:fagn_hist} presents the normalised $f_{\rm PSF}$ distributions for the unclassified galaxies, which lie between the distributions of the non-merger controls and the mergers. This intermediate positioning likely reflects their mixed composition. However, mergers remain dominant over unclassified galaxies, with significantly higher fractions in the range $f_{\rm PSF}=0.1$--$0.8$.
To test whether unclassified galaxies influence the observed $f_{\rm merg}$-$f_{\rm PSF}$ trends, we analysed how the fraction of unclassified galaxies ($f_{\rm uncl}$) varies with $f_{\rm PSF}$. The results, shown in Fig.~\ref{fig:fmerg_fagn}, indicate that $f_{\rm uncl}$ remains roughly constant in both redshift bins, varying very mildly within the range $f_{\rm uncl}=0.3$--$0.4$. These findings suggest that unclassified objects do not significantly impact the relationship between major mergers and $f_{\rm PSF}$. Indeed, this result strengthens our overall conclusion that mergers are predominantly associated with relatively bright central point sources, serving as the primary mechanism for fuelling dominant AGN.

We also investigated the role of unclassified galaxies in the mergers and AGN luminosity relation. Figures~\ref{fig:LPSF_hist} and \ref{fig:Lbol_hist} overlay the normalised $L_{\rm PSF}$ and $L_{\rm bol}$ distributions for unclassified sources. These comparisons indicate that major mergers primarily trigger the most luminous AGN, since they exhibit a significant excess compared to both non-merger controls and unclassified galaxies. Furthermore, we computed $f_{\rm uncl}$ as a function of $L_{\rm PSF}$ for both redshift bins (Fig.~\ref{fig:fmerg_Lagn_phot}). The fraction remains relatively stable at 0.25--0.35 up to $L_{\rm  PSF}=10^{43}\,\rm{erg\,s^{-1}}$, before decreasing to 0.2 for brighter AGN. These marginal variations do not alter the main finding that mergers play an increasingly significant role in fuelling the most luminous AGN.

\subsection{The effect of stellar mass}

\begin{figure}
    \centering
    \includegraphics[width=0.495\textwidth]{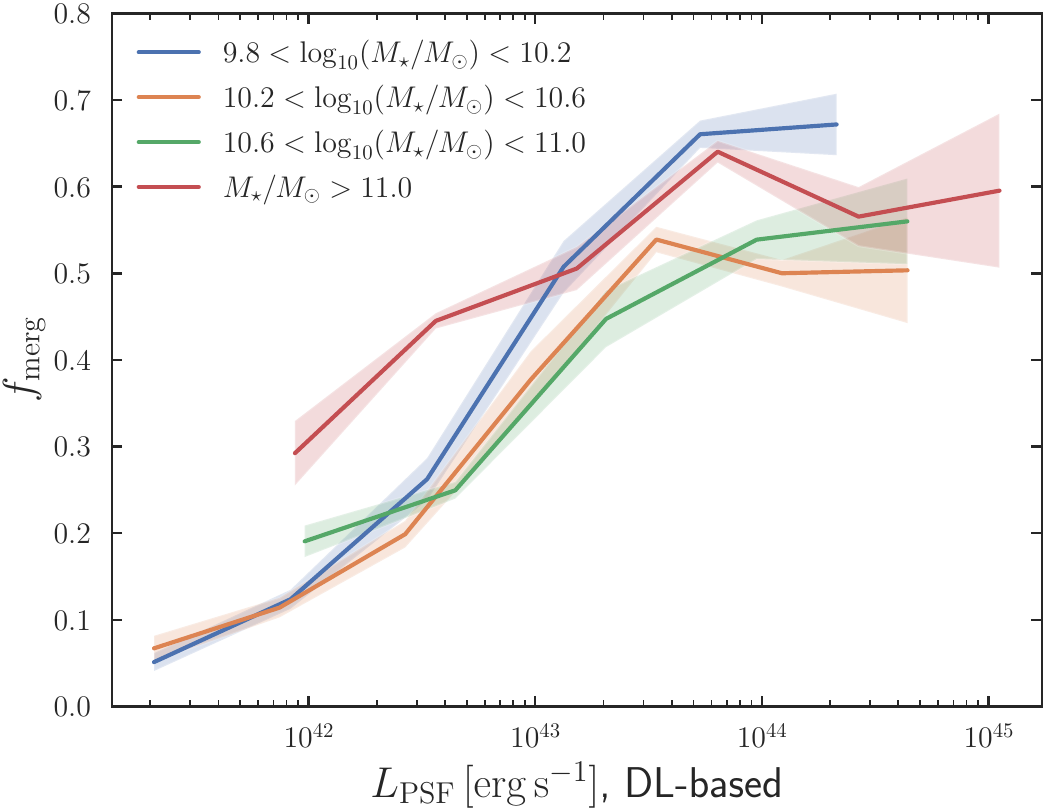}
    \caption{Similar to Fig.~\ref{fig:fmerg_Lagn_phot} but with the merger fraction and PSF luminosity relation divided in stellar mass bins.}
    \label{fig:fmerg_Lagn_mass}
\end{figure}

Another potential concern is whether brighter, and consequently more massive, galaxies are more likely to be classified as mergers. To assess whether the observed trends primarily arise from the galaxy stellar mass, we examined the $f_{\rm merg}$-$L_{\rm PSF}$ relation in four stellar mass bins, each containing a similar number of galaxies. The results (Fig.~\ref{fig:fmerg_Lagn_mass}) confirm the general trend observed in Fig.~\ref{fig:fmerg_Lagn_phot}: the merger fraction increases with $L_{\rm PSF}$ and then flattens for the most luminous point sources ($L_{\rm PSF}>10^{43.5}\,\rm{erg\,s^{-1}}$), where mergers constitute the majority of the population. The most massive galaxies ($M_{\star}>10^{11}\,M_{\odot}$) exhibit the highest $f_{\rm merg}$ on average, consistent with recent studies reporting a positive correlation between $f_{\rm merg}$ and stellar mass \citep[e.g.,][]{nevinDecliningMajorMerger2023}. Thus, we conclude that stellar mass is not the primary driver of the $f_{\rm merg}$-$L_{\rm PSF}$ relation.

\subsection{Systematics in the classification}\label{sect:MC}

A key aspect of this study relies on the automated classification of galaxies into mergers and non-mergers using a CNN. As detailed in Sect.~\ref{sc:CNN} and Table~\ref{tab:performance}, our classifier achieves performance levels comparable to contemporary studies, with a precision of 0.80 and recall of 0.68 for the merger class, and a precision of 0.72 and recall of 0.83 for the non-merger class. While these metrics are robust, they inevitably imply that our classified samples contain non-negligible fractions of misclassified objects and are incomplete. Throughout our manuscript, we relied on statistical uncertainties under the assumption that our classifier perfectly distinguishes mergers from non-mergers. It is therefore crucial to assess the impact of these classification uncertainties on our main scientific findings, derived from both the binary comparisons (Sect.~\ref{sc:binary}) and the analysis of continuous AGN properties (Sect.~\ref{sc:continuous}).

To quantitatively evaluate the robustness of our results against these misclassifications, we performed a detailed MC simulation based directly on the classifier's performance metrics.
\footnote{
Quantitatively assessing and propagating classification systematics (contamination and incompleteness) has historically been challenging in merger studies, due to the absence of precise performance metrics. Our detailed MC simulation represents a significant step forward in rigorously assessing the impact of classification uncertainties on large statistical samples.}
The core simulation procedure involved 1000 independent iterations. In each iteration, we did the following:
\begin{enumerate}
    \item We simulated the effect of contamination (finite precision). Based on the measured precision for mergers ($p_{\rm merger}=0.80$), a fraction ($1-p_{\rm merger}=20\%$) of galaxies initially classified by our CNN as mergers were randomly selected and temporarily relabelled as non-mergers. Similarly, based on the precision for non-mergers ($p_{\rm non-merger}=0.72$), a fraction ($1-p_{non-merger}=28\%$) of galaxies initially classified as non-mergers were randomly selected and temporarily assigned to the merger class. This step yielded temporary `corrected' classifications for all galaxies within that iteration.

    \item We accounted for the classifier's incompleteness (finite recall). To estimate fractions relative to the total underlying population in any given subsample (e.g., AGN hosts, specific luminosity bins), rather than just the classified population, we applied weights based on the recall values. Any galaxy temporarily labelled as `merger' in the iteration received a weight $w_{\rm M} = 1 / {\rm recall_{M}} = 1 / 0.68$. Any galaxy labelled as `non-merger' received a weight $w_{\rm NM} = 1 / {\rm recall_{NM}} = 1 / 0.83$. This weighting statistically corrects the counts for the classifier's detection efficiency.
\end{enumerate}
Using these temporary weighted classifications from each MC iteration, we recalculated our key metrics presented in Tables~\ref{tab:agn_freq} and \ref{tab:merg_frac} and recomputed the $f_{\rm merg}$-$L_{\rm PSF}$ relation (Fig.~\ref{fig:fmerg_Lagn_phot}).

\subsubsection{Systematic uncertainties on the binary experiments}

\begin{figure}
    \centering
    \includegraphics[width=0.45\textwidth]{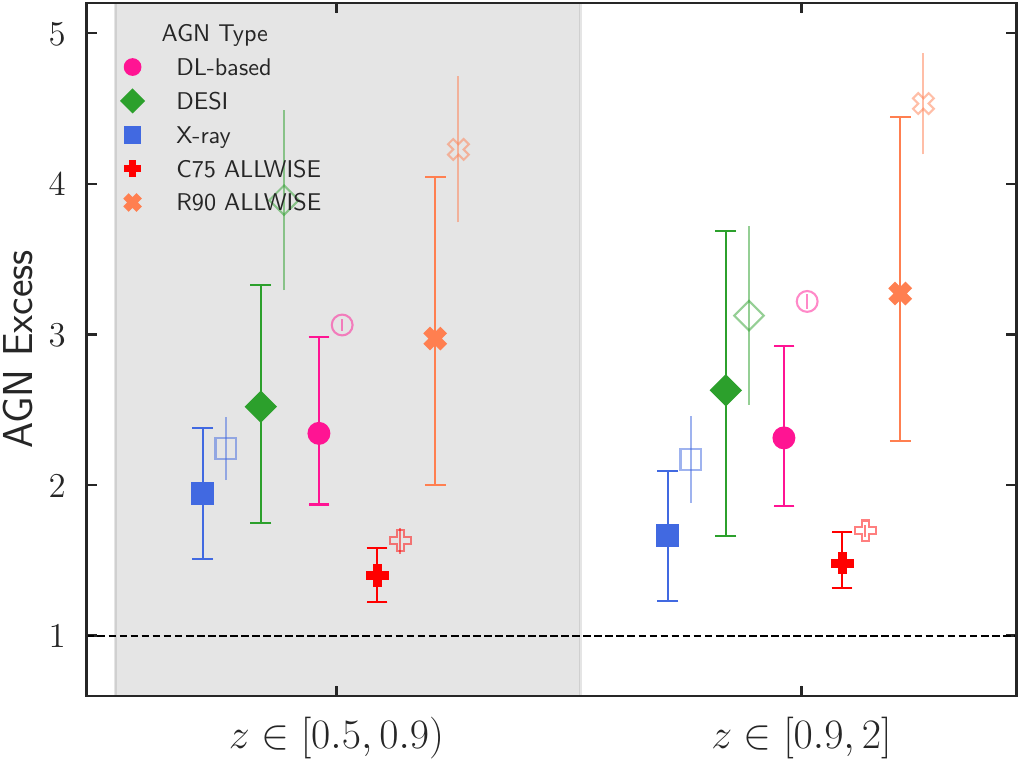}
    \caption{
    Monte Carlo simulation outcomes for the AGN excess in mergers compared to non-merger controls. The symbols represent the median value of each AGN excess distribution, while the error bars cover the 2.5th-97.5th percentile range. Empty, transparent symbols represent the results presented in Fig.~\ref{fig:agn_freq}.}
    \label{fig:agn_excess_MC}
\end{figure}

For each MC iteration, we recalculated the AGN frequency in mergers and non-mergers presented in Table~\ref{tab:agn_freq}, for each AGN type and in both redshift bins, using the relabelled outcomes. The ratio of the two frequencies above was calculated for each iteration, producing a distribution for the AGN excess. The results of these simulations confirm the robustness of our binary analysis findings.
Across the 1000 MC iterations, the AGN frequency was consistently found to be higher in the simulated merger populations compared to the non-merger controls for all AGN types and redshift bins considered.
We plot in Fig.~\ref{fig:agn_excess_MC} the median AGN excess derived from the 1000 iterations, alongside the 2.5th to 97.5th percentile for each AGN type and redshift bin. This range represents the central 95\% interval of the simulated outcomes, effectively illustrating the statistically dominant parameter space explored when accounting for potential misclassifications. The median values are generally lower than the corresponding AGN excesses presented in Fig.~\ref{fig:agn_freq}, but still within the parameter space covered. 
As evident in Fig.~\ref{fig:agn_excess_MC}, even considering the full extent of these uncertainties, the lower bound of the simulated AGN excess consistently remains well above unity, confirming that the observed excess is statistically significant and not merely an artefact of misclassification.

\begin{figure}
    \centering
    \includegraphics[width=0.45\textwidth]{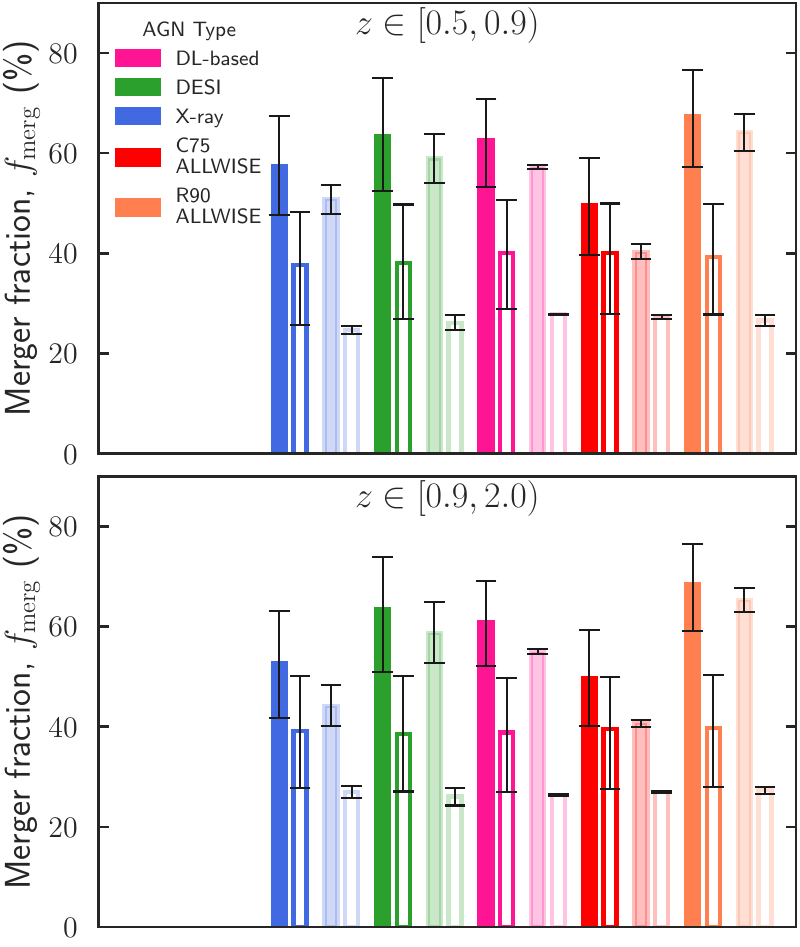}
    \caption{ 
    Monte Carlo simulation outcomes for the merger fractions in AGN and non-AGN control sample (empty bars). The bars show the median value of the 1000 MC simulations, while the error bars display the 2.5th-97.5th percentile range. Results from Fig.~\ref{fig:merg_frac} are reported as transparent bars. 
    }
    \label{fig:f_merg_MC}
\end{figure}

Similarly, we calculated the weighted sum of temporary `mergers' within the AGN host sample and divided it by the total weighted sum of the AGN host sample. The same was done for the non-AGN control sample, generating distributions for the merger fraction in both populations. Figure~\ref{fig:f_merg_MC} illustrates the median $f_{\rm merg}$ from each distribution obtained through the MC simulation, alongside the 2.5th to 97.5th percentile range, divided into two redshift bins. On average, the merger fraction was found to be slightly higher in the AGN samples compared to the non-AGN controls across the MC iterations. However, the parameter space ranges for AGN and non-AGN controls overlap in most cases.

\subsubsection{Systematic uncertainties on the $f_{\rm merg}$-$L_{\rm PSF}$ relation}

\begin{figure}
    \centering
    \includegraphics[width=0.495\textwidth]{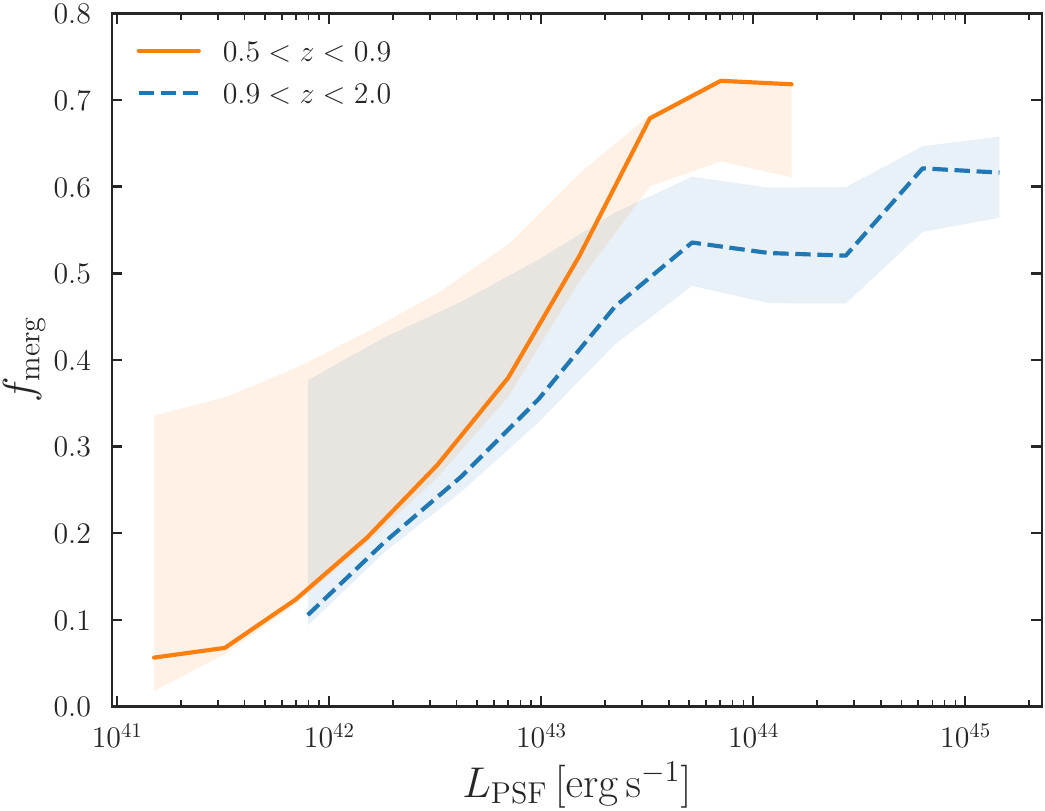}
    \caption{Monte Carlo simulation of the merger fraction and PSF luminosity relation. The { solid line} trends reported are the same as in Fig.~\ref{fig:fmerg_Lagn_phot}, while the shaded areas represent the full parameter space covered (0th-100th percentile range) by the MC simulation results.}
    \label{fig:fmerg_Lagn_MC}
\end{figure}

For each MC iteration, we recalculated $f_{\rm merg}$ in each $L_{\rm PSF}$ bin using the weighted counts. Figure~\ref{fig:fmerg_Lagn_MC} shows that,  as expected, the MC simulations yield a broader range of $f_{\rm merg}$ values compared to the bootstrapping uncertainties. The median trend across the 1000 MC iterations closely follows the trend derived directly from the initial classification (solid lines), while the shaded areas (0th-100th percentile range) illustrate the propagated uncertainty. The MC results demonstrate that while uncertainties introduce larger scatter, particularly at lower luminosities ($L_{\rm PSF}<10^{43}\,\rm{erg\,s^{-1}}$), the rising trend of $f_{\rm merg}$ with increasing $L_{\rm PSF}$ remains robust. The conclusion that mergers dominate ($f_{\rm merg} > 50\%$) among the most luminous AGN holds true across the vast majority of the MC realisations.

In summary, our detailed MC simulations, incorporating the measured precision and recall, indicate that our classifier is sufficiently robust to establish the primary qualitative conclusions presented in Sect.~\ref{sc:Results}. Both the enhanced presence of AGN in mergers found in the binary analysis and the trend of increasing merger importance for more luminous AGN seen in the continuous analysis hold even when accounting for realistic levels of misclassification inherent to automated methods in deep surveys.

\section{\label{sc:Conc} Summary and conclusions}

In this paper, we have performed the first detection of major mergers in the \Euclid VIS \IE-band imaging data and examined the merger and AGN connection at $0.5\leq z \leq 2.0$ in the Q1 EDFs. We constructed a stellar mass-complete sample of galaxies ($M_{\star}>10^{9.8}\,M_{\odot}$) and employed a CNN trained on mock \Euclid observations generated from Illustris-TNG simulations to identify merging galaxies. We defined mergers in Illustris-TNG galaxies with a major merger event (a stellar mass ratio $\leq 4$) in the time interval spanning 800 Myr before to 300 Myr after coalescence. We exploited the rich multi-wavelength datasets for selecting AGN using four different diagnostics to select AGN via X-ray detections, optical spectroscopy (DESI data), two different MIR colour selections, and a DL-based imaging decomposition technique. We analysed the role of mergers in triggering AGN using a binary approach and a more refined approach that focuses on continuous AGN parameters.
Our key findings are the following.
\begin{enumerate}
\renewcommand{\labelenumi}{\roman{enumi})}
    \item A larger fraction of AGN in mergers than in non-merger controls, which results in an excess of AGN in mergers, regardless of the AGN selection used. X-ray and DL-based AGN show a factor of two to three excess in mergers across the whole redshift range. DESI-selected AGN show a larger excess (3.9) at $z\geq 0.9$ than at $z<0.9$ (3.1). MIR AGN show an excess that depends on the criterion adopted. For the more reliable selection, R90, the excess is much larger (a factor of 4.2--4.5) than that of the more complete selection, C75 (1.7). This indicates that mergers can trigger all AGN types but are likely to be more connected with dust-obscured AGN. 
    
    \item A higher merger fraction ($f_{\rm merg}$) in active galaxies, with a larger fraction in AGN by 15--25 percentage points compared to non-AGN controls, for all AGN types. However, we cannot conclude with certainty whether mergers are a primary triggering mechanism. 
    
    \item A rising trend in the $f_{\rm merg}$ as a function of the PSF relative contribution $f_{\rm{PSF}}$, measured in the \IE-band up to $f_{\rm{PSF}}\simeq 0.55$ followed by a decline. This trend is independent of the redshift. In the range $f_{\rm{PSF}} = 0.3$--0.75, most galaxies are classified as mergers, which hints towards a scenario where mergers are the prevalent fuelling mechanism in relatively dominant AGN ($f_{\rm PSF}>0.5$). 
    
    \item A positive correlation between $f_{\rm merg}$ and the PSF luminosity, $L_{\rm{PSF}}$, where mergers represent more than $50\%$ of the galaxies at $L_{\rm{PSF}}> 10^{43.5}\,{\rm erg}\,{\rm s}^{-1}$ for $z<0.9$ and at $10^{45}\,{\rm erg}\,{\rm s}^{-1}$ for $z\geq 0.9$. This confirms the idea that mergers are the main channel to fuel the brightest AGN. 
    
\end{enumerate}
Moreover, we performed detailed MC simulations to assess the impact of potential misclassification and incompleteness from our merger identification pipeline. While showing much larger uncertainties, these tests demonstrated that our primary conclusions are qualitatively robust and not driven by classification systematics.

In conclusion, our results prove that mergers are closely linked to relatively dominant and bright AGN. 
Moreover, larger merger fractions and AGN excesses are observed for MIR AGN, which are usually linked to the dust-obscured phase of AGN lives.
This suggests that mergers efficiently funnel gas to the central regions of galaxies, driving rapid accretion onto the SMBH, possibly obscuring it with dust, and making AGN more detectable in the MIR. For less dominant AGN, other fuelling mechanisms may play a more important role. Although mergers appear to be the primary -- if not the sole -- trigger for the most luminous AGN at $z<0.9$, their influence may decline at higher redshifts, where galaxies typically have larger gas reservoirs capable of sustaining AGN activity without external triggers. 

A key limitation of this study is the reliance on CNN-based merger classification, which inherently has accuracy constraints despite being trained on cosmological simulations. While the main trends remain robust, some level of misclassification is unavoidable. Future improvements in classification techniques will be essential to refining merger identification. This study, although based on only 63 ${\rm deg}^2$, highlights the statistical power of \Euclid in probing mergers and AGN fuelling. With upcoming \Euclid releases, the sample size will increase dramatically, marking a transition from being limited by statistical uncertainties to a regime dominated by systematics, which must be understood to advance the field. Complementary datasets from XMM-{\it Newton}, eROSITA, and JWST, as well as ancillary far-IR and radio observations, will further expand the AGN sample, allowing for a more precise assessment of the role of mergers in AGN evolution. Crucially, these data will allow us to map the merger and AGN connection in a multi-dimensional space and simultaneously analyse it as a function of key galaxy properties such as stellar mass, redshift, gas content, star formation rate, and environment.

%
%

\begin{acknowledgements}
\AckQone
\AckEC  
Based on data from UNIONS, a scientific collaboration using
three Hawaii-based telescopes: CFHT, Pan-STARRS, and Subaru
\url{www.skysurvey.cc}\,. Based on data from the Dark Energy Camera (DECam) on the Blanco 4-m Telescope at CTIO in Chile \url{https://www.darkenergysurvey.org}\,.
This publication is part of the project `Clash of the titans: deciphering the enigmatic role of cosmic collisions' (with project number VI.Vidi.193.113 of the research programme Vidi which is (partly) financed by the Dutch Research Council (NWO).
This research makes use of ESA Datalabs (datalabs.esa.int), an initiative by ESA’s Data Science and Archives Division in the Science and Operations Department, Directorate of Science.
We thank SURF (\url{www.surf.nl}) for the support in using the National Supercomputer Snellius.
This research was supported by the International Space Science Institute (ISSI) in Bern, through ISSI International Team project \#23-573 `Active Galactic Nuclei in Next Generation Surveys'.
This work has benefited from the support of Royal Society Research Grant RGS{\textbackslash}R1\textbackslash231450.
FR, VA acknowledge the support from the INAF Large Grant `AGN \& Euclid: a close entanglement' Ob. Fu. 01.05.23.01.14.
ALM is grateful for the indispensable discussion with Christopher Boettner and Scott C. Trager.
\end{acknowledgements}

%
%

\bibliography{Euclid_Q1_mergers}

%

\begin{appendix}

\section{Comparison with {\tt Zoobot} classification}\label{app:Zoobot}

We compared the predictions of the model trained in this work with the classification given by the {\tt Zoobot} model \citep{Q1-SP047} for the subsample of galaxies in common. We used two {\tt Zoobot} catalogue columns to identify mergers, the \texttt{merging merger fraction} and the \texttt{merging major-disturbance fraction}, which allow us to select both pair galaxies and highly disturbed post-merging galaxies. As a first step, we removed possible artefacts from the {\tt Zoobot} catalogue by setting \texttt{problem artifact fraction} $< 0.01$ \texttt{AND} \texttt{problem star fraction} $< 0.01$ \texttt{AND} \texttt{problem zoom fraction} $< 0.01$. Then, we defined the mergers as those galaxies with \texttt{merging merger fraction} $>0.7$ \texttt{OR} \texttt{merging major-disturbance fraction} $>0.5$, and the non-mergers as \texttt{merging merger fraction} $<0.2$ \texttt{AND} \texttt{merging major-disturbance fraction} $<0.1$. We chose these criteria in order to obtain pure samples of mergers and non-mergers. In total, we found 40\,847 galaxies in common. Of these, 27.8\% are unclassified according to our model classification, which we removed to compare merger and non-merger classifications. 

\begin{figure}
    \centering
    \includegraphics[width=0.48\textwidth]{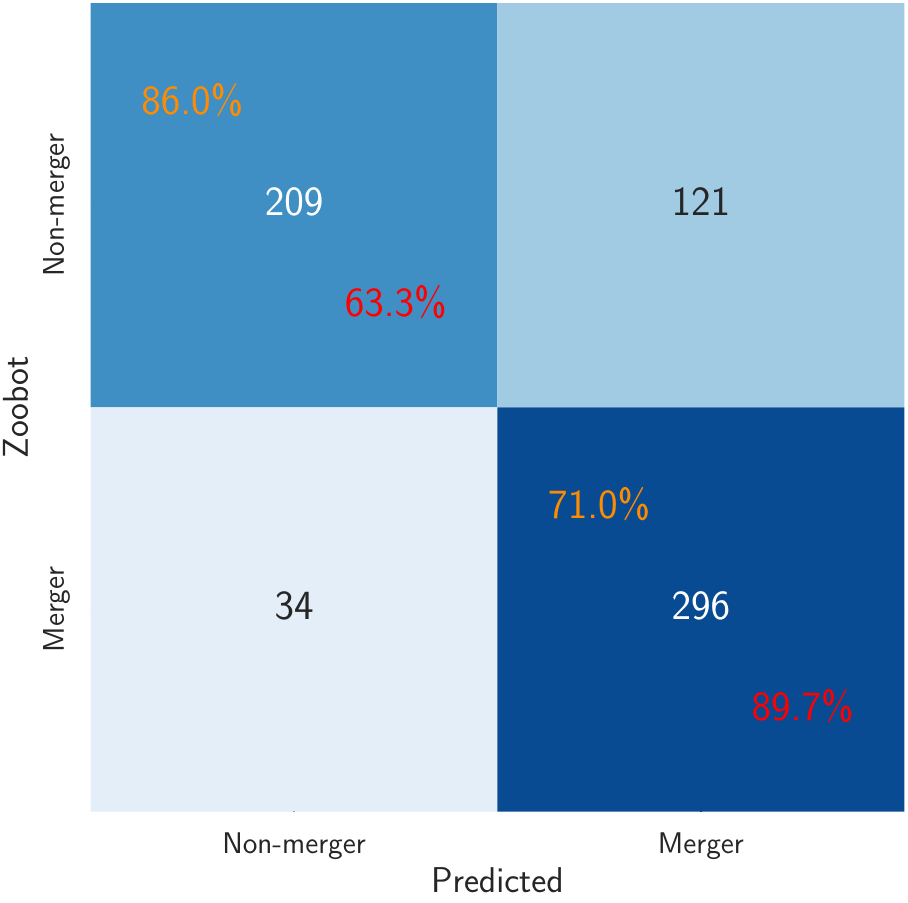}
    \caption{Confusion matrix comparing our model predictions with the {\tt Zoobot} classifications. Along the diagonal, we report the precision (orange) and the recall (red) of each class. In black, the number of galaxies in each cell. Results are averaged over ten different balanced sets.}
    \label{fig:CM_Zoobot}
\end{figure}

We created balanced samples of the {\tt Zoobot} mergers and non-mergers by randomly selecting the same number of mergers among the available non-mergers. This operation was repeated ten times, and we report the average results in Fig.~\ref{fig:CM_Zoobot}. Our model has a precision of 71\% and 86\% for mergers and non-mergers, respectively, when compared to the {\tt Zoobot} labels. Compared to the performance on the TNG test set, we observed a lower precision for the merger class, but an improved precision for the non-mergers. At the same time, our model classification is highly complete with respect to {\tt Zoobot} mergers, with a recall of 90\%, but has a much lower recall for the non-mergers, 63\%. Overall, the F1-scores for both classes are the same as for the TNG test set. Considering all {\tt Zoobot} mergers in the common subsample, our model classifies as mergers 90\% of them (75\% if we do not exclude unclassified galaxies), demonstrating good agreement between our classification and labels obtained from a model trained on visual classification. 

We visually inspected the cases where {\tt Zoobot} and our CNN disagree. We observed that sometimes {\tt Zoobot} misclassified mergers picked up by our CNN, but also the opposite is true (i.e., our CNN misclassified mergers correctly labelled by {\tt Zoobot}). Nevertheless, we note that the comparison between our classifier and {\tt Zoobot} is inherently dependent on the choice of classification thresholds in both approaches. Variations in these thresholds can significantly affect the reported merger fractions and the relative performance of the methods.

The performance of our CNN, when compared to Zoobot labels as shown in Fig.~\ref{fig:CM_Zoobot}, yields precision and recall values for the merger class broadly consistent with the performance achieved by our CNN on the TNG test set (Table~\ref{tab:performance}). These figures, are also comparable to the typical performance levels reported for other state-of-the-art DL methods applied to merger classification in similar large, deep surveys \citep{margalef-bentabolGalaxyMergerChallenge2024}. This consistency suggests that the level of accuracy, and the associated inherent uncertainties (as discussed in Sect.~\ref{sect:MC}), are characteristic of current methodologies rather than specific shortcomings of our individual classifier.

\section{Example of unclassified galaxies}\label{app:uncl}

\begin{figure*}
    \centering
    \includegraphics[width=0.99\textwidth]{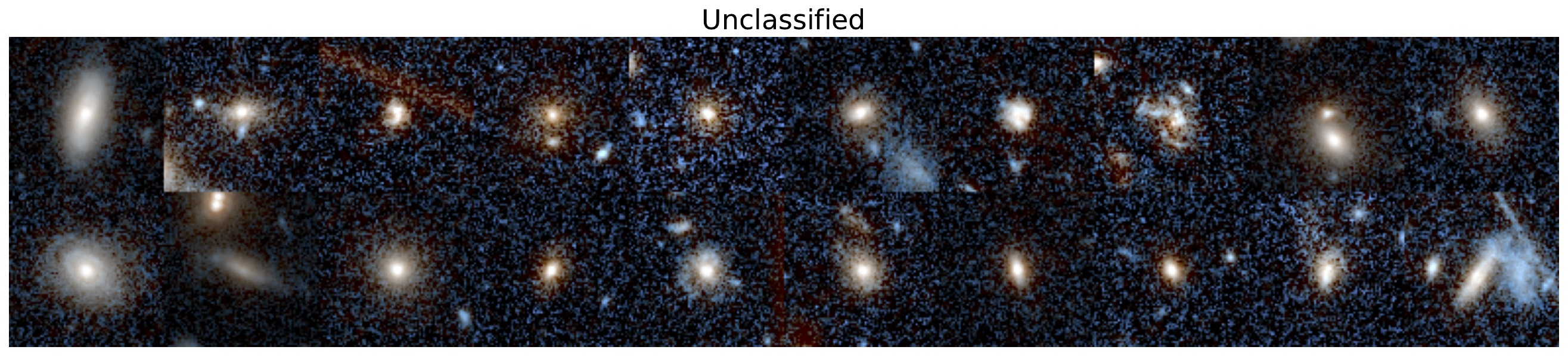}
    \caption{
    Examples of galaxies that we label as unclassified. The cutouts are \Euclid RGB composite images, $8\arcsec\times8\arcsec$, generated as in Fig.~\ref{fig:examples}. 
    }
    \label{fig:uncl_example}
\end{figure*}

We show some randomly sampled examples of unclassified galaxies in Fig.~\ref{fig:uncl_example}. Unclassified galaxies are those objects with a predicted score between 0.35 and 0.59, inclusive. These unclassified objects appear as intermediate between mergers and non-mergers. While some look isolated and undisturbed, others appear to have close neighbours and an irregular morphology.

\section{AGN sample: Additional information}\label{app:AGN_sel}

\begin{figure*}
    \sidecaption
    \includegraphics[width=0.67\textwidth]{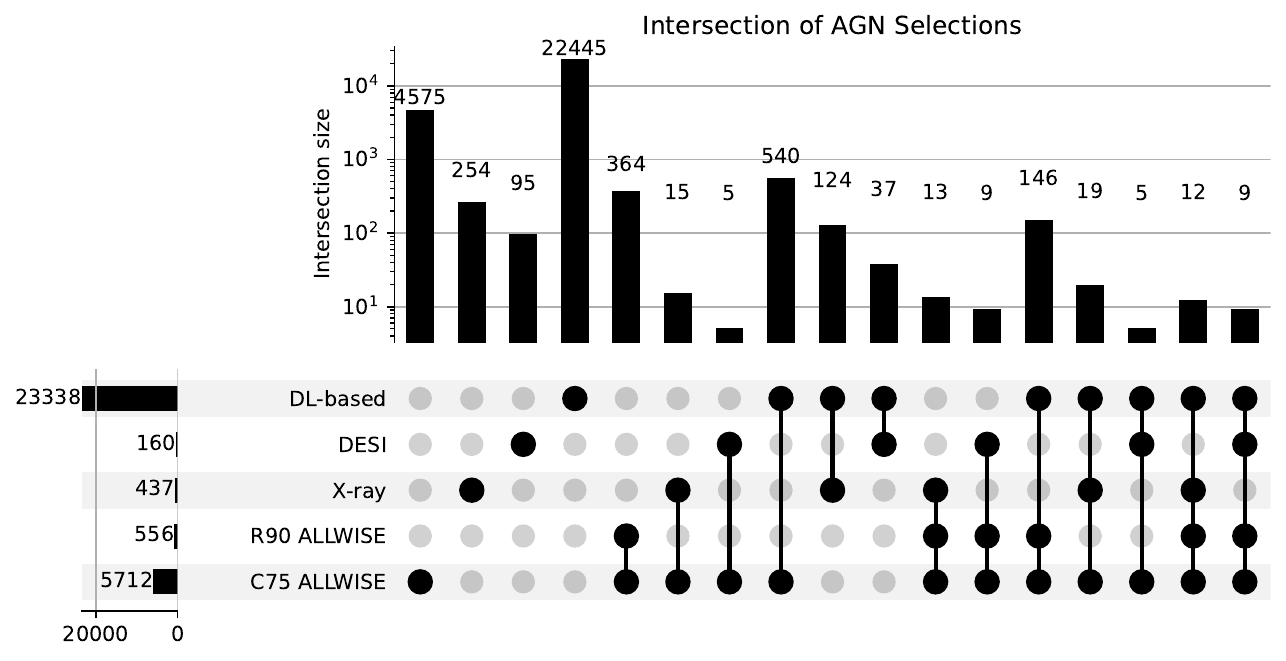}
    \caption{UpSet plot showing the intersections of all AGN selection methods employed. Rows correspond to the AGN selections, while columns correspond to the intersections. Numbers of each selection and intersection are displayed as bar charts. Intersections with fewer than five elements are not shown to facilitate readability.}
    \label{fig:upset}
\end{figure*}

\begin{figure*}
    \sidecaption
    \includegraphics[width=0.67\textwidth]{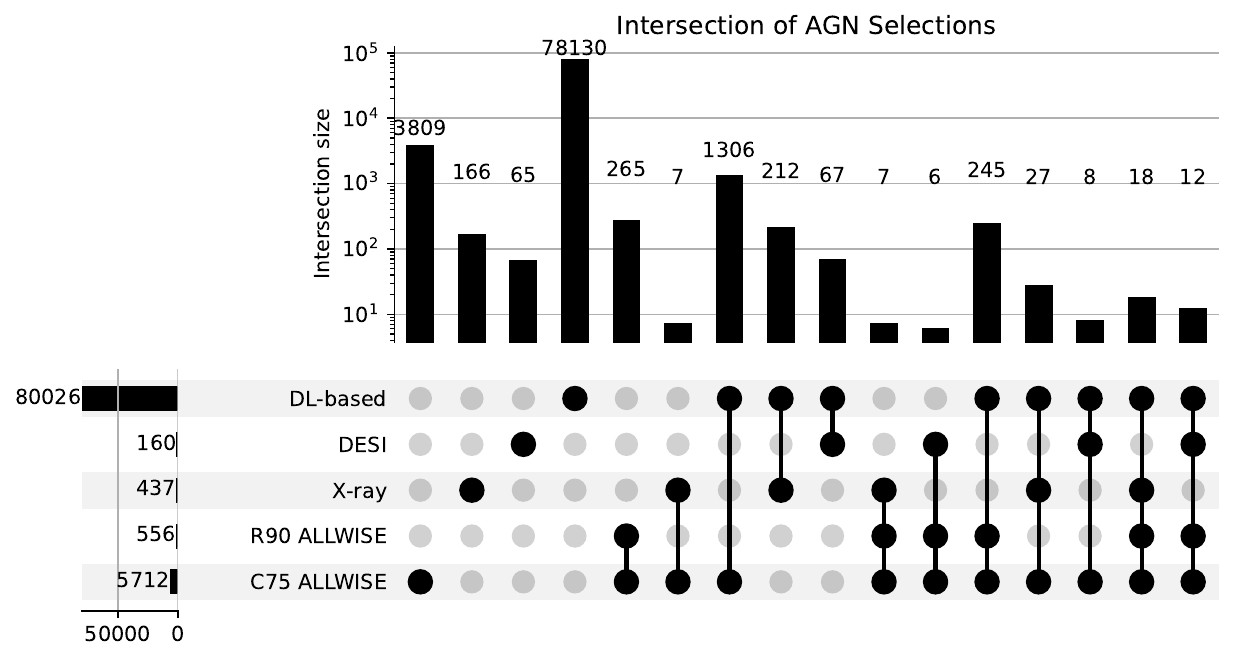}
    \caption{UpSet plot showing the intersections of all AGN selection methods employed but with DL-based AGN defined as $f_{\rm{PSF}}\geq 0.1$, as in Fig.~\ref{fig:upset}.}
    \label{fig:upset_low_fPSF}
\end{figure*}

We show the various intersections of all AGN selection methods used in this work in Fig.~\ref{fig:upset} as an UpSet plot. This plot displays intersections in a matrix, with rows corresponding to the AGN selections and columns representing the intersections between these sets. The size of the sets and the intersections are shown as bar charts. The DL-based method correctly identifies about 25--30\% of DESI, X-ray, and R90 MIR-selected AGN, while only about 10\% of C75 MIR AGN. Relaxing the $f_{\rm PSF} \geq 0.2$ constraint to 0.1, the DL-based model recognises many more AGN (Fig.~\ref{fig:upset_low_fPSF}): it correctly identifies  $>50\%$ of DESI, X-ray and R90 AGN, and about 25\% of C75 AGN. It is not surprising that the C75 selection method has the lowest identification ratio, as this diagnostic is also the most contaminated one.

\begin{figure*}
    \centering
    \includegraphics[width=0.99\textwidth]{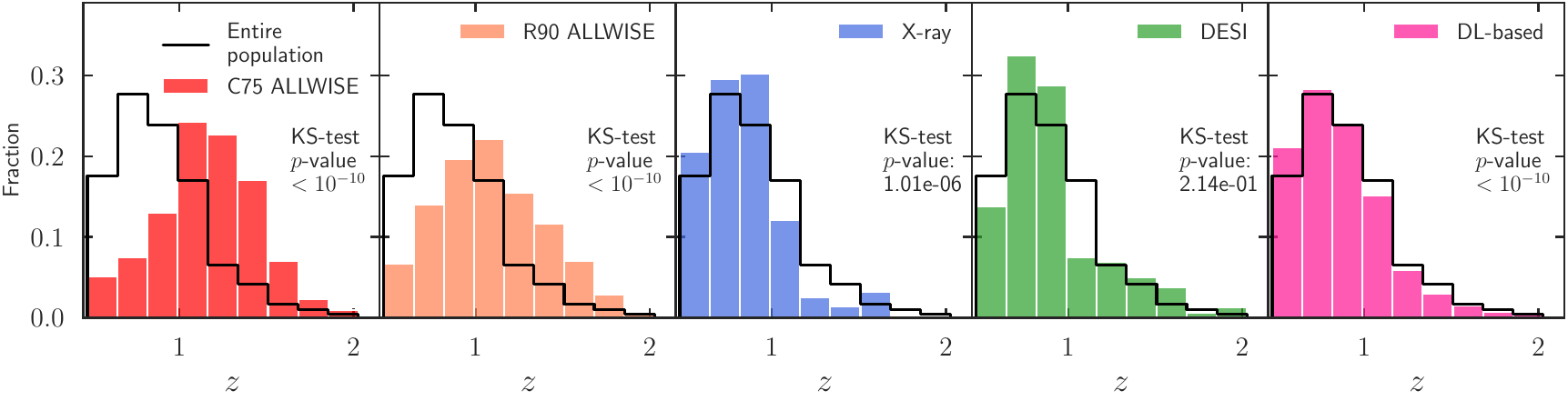}
    \caption{
    Normalised redshift distribution for each AGN population. We overlay the distribution for the entire sample of galaxies and report the results of a KS test in each panel.}
    \label{fig:z_hist}
\end{figure*}

Figure~\ref{fig:z_hist} shows the normalised redshift distributions of all AGN types. X-ray and DESI AGN mostly inhabit $z<1$ galaxies, with very few individuals at higher redshift. MIR AGN, both the C75 and the R90 selections, on average have higher redshifts, with their distributions peaking at $z\simeq 1$. DL-selected AGN mostly follow the same $z$ distribution of the full galaxy sample, which has its maximum at $z\simeq 0.7$ and then monotonically decreases towards higher redshifts. This behaviour is expected because the DL-based AGN population strongly depends on the original population of galaxies. 

\section{$f_{\rm merg}$ as a function of $f_{\rm{PSF}}$ and $L_{\rm{PSF}}$ for individual AGN selections}\label{app:AGN_analysis}

\begin{figure}
    \centering
    \includegraphics[width=0.49\textwidth]{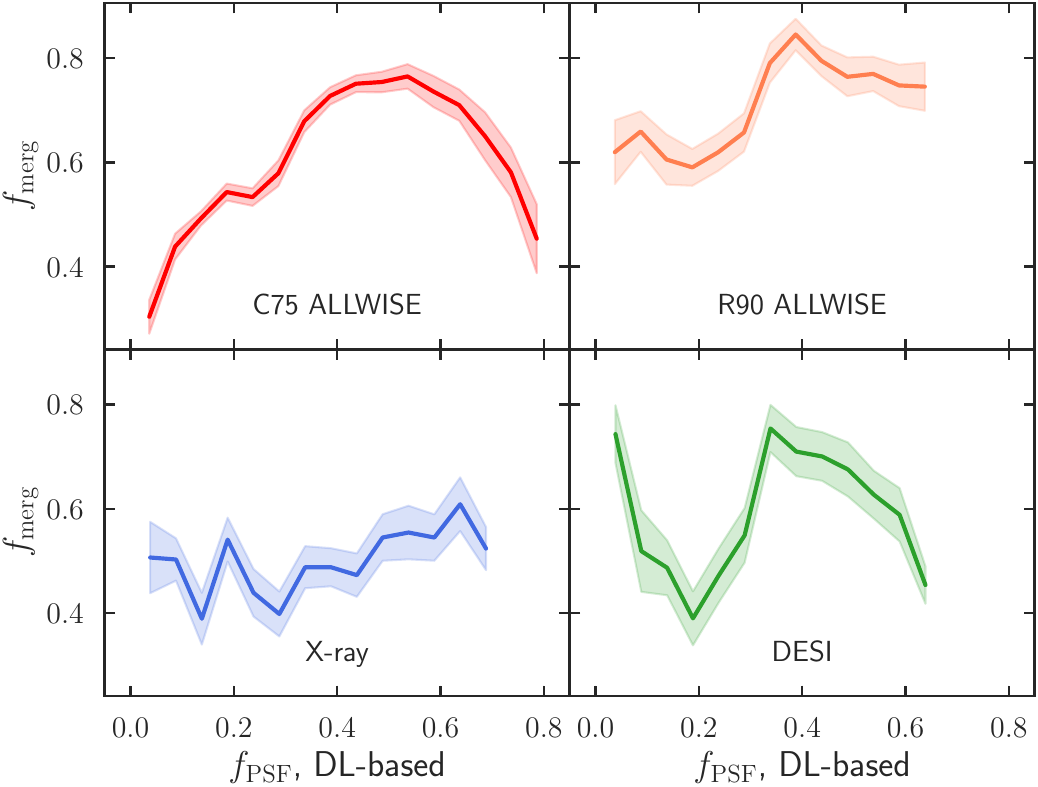}
    \caption{Merger fraction and PSF fraction relationship for each AGN type. Trend lines represent the running median, while shaded areas are one standard deviation. The AGN type is indicated in each panel. }
    \label{fig:fmerg_fagn_agn_type}
\end{figure}

Here, we analyse the relation of the merger fraction with the point source fraction and luminosity for individual AGN selections, to study possible differences. We show the $f_{\rm merg}$ and $f_{\rm{PSF}}$ relation for X-ray, DESI, and MIR colour selections in Fig.~\ref{fig:fmerg_fagn_agn_type}. X-ray and DESI AGN show larger fluctuations, mainly due to fewer objects compared to MIR and DL-based AGN selections. X-ray AGN show a rather flat trend, with a mild increase of $f_{\rm merg}$ as a function of $f_{\rm{PSF}}$. Regarding the MIR AGN, the R90-selected objects show a rising $f_{\rm merg}$ trend with increasing $f_{\rm{PSF}}$, centred at very high $f_{\rm merg}$ values ($>0.6$). The C75-selected AGN have a trend similar to that of the general population reported in Fig.~\ref{fig:fmerg_fagn}: a sharp rise in $f_{\rm merg}$ up to $f_{\rm{PSF}}\simeq0.55$, followed by a decreasing merger fraction. 

\begin{figure}
    \centering
    \includegraphics[width=0.49\textwidth]{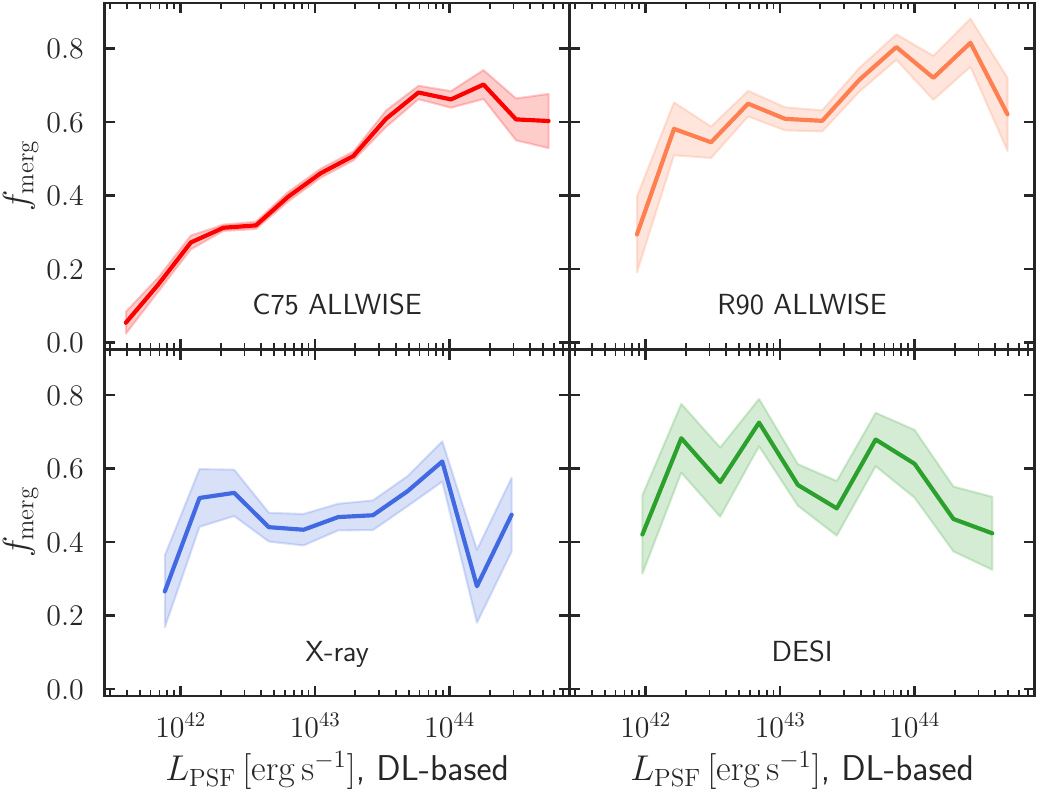}
    \caption{Merger fraction and $L_{\rm{PSF}}$ relation for each AGN type. Trend lines represent the running median, while shaded areas are one standard deviation. The AGN type is indicated in each panel. }
    \label{fig:fmerg_Lagn_agn_type}
    \label{LastPage}
\end{figure}

Likewise, we show the merger fraction and point source luminosity relations for the individual AGN selections in Fig.~\ref{fig:fmerg_Lagn_agn_type}. X-ray and DESI AGN have a rather flat trend, around $f_{\rm merg}\simeq 0.5$ and $f_{\rm merg}\simeq 0.6$, respectively. Although their trends are significantly different from those in Fig.~\ref{fig:fmerg_Lagn_phot}, mergers appear as a dominant fuelling mechanism for both selections. On the other hand, both MIR colour selections show monotonic rising $f_{\rm merg}$ as a function of $L_{\rm PSF}$, with mergers becoming dominant for the brightest point sources ($L_{\rm PSF}> 10^{43.5}\,\rm{erg\, s^{-1}}$). In this case, the trends reported are similar to those of the general AGN population in Fig.~\ref{fig:fmerg_Lagn_phot}.

\end{appendix}

\end{document}